\documentclass[%
 reprint,
superscriptaddress,
nofootinbib,
 amsmath,amssymb,
 prd,
showkeys]{revtex4-2}

\usepackage[margin=5pt,caption=false]{subfig}
\usepackage{slantsc}
\usepackage{comment}
\usepackage{longtable} 
\usepackage[online]{threeparttablex}	
\usepackage{threeparttable}
\usepackage{booktabs}
\usepackage[figuresright]{rotating}
\usepackage{textcase}
\usepackage{multirow}
\newcommand{\be}{\begin{equation}}
\newcommand{\ee}{\end{equation}}
\newcommand{\bea}{\begin{eqnarray}}
\newcommand{\eea}{\end{eqnarray}}
\newcommand{\hunit}{$\rm{km \ s^{-1} \ Mpc^{-1}}$}
\newcommand{\lcdm}{$\Lambda$CDM}
\newcommand{\pcdm}{$\phi$CDM}

\usepackage{array}

\newcommand{\hii}{H\,\textsc{ii}}
\newcommand{\Om}{\Omega_{m0}}
\newcommand{\Ok}{\Omega_{k0}}
\newcommand{\om}{$\Omega_{m0}$}
\newcommand{\ok}{$\Omega_{k0}$}
\newcommand{\wx}{$w_{\rm X}$}
\newcommand{\wX}{w_{\rm X}}

\newcommand{\mii}{Mg\,\textsc{ii}}

\newcommand{\civ}{C\,\textsc{iv}}

\newcommand{\obh}{\Omega_{b}h^2}
\newcommand{\och}{\Omega_{c}h^2}
\newcommand{\onh}{\Omega_{\nu}h^2}
\newcommand{\obhs}{$\Omega_{b}h^2$}
\newcommand{\ochs}{$\Omega_{c}h^2$}

\usepackage[T1]{fontenc}
\usepackage{scalerel}
\usepackage{tikz}
\usetikzlibrary{svg.path}

\definecolor{orcidlogocol}{HTML}{A6CE39}
\tikzset{
  orcidlogo/.pic={
    \fill[orcidlogocol] svg{M256,128c0,70.7-57.3,128-128,128C57.3,256,0,198.7,0,128C0,57.3,57.3,0,128,0C198.7,0,256,57.3,256,128z};
    \fill[white] svg{M86.3,186.2H70.9V79.1h15.4v48.4V186.2z}
                 svg{M108.9,79.1h41.6c39.6,0,57,28.3,57,53.6c0,27.5-21.5,53.6-56.8,53.6h-41.8V79.1z M124.3,172.4h24.5c34.9,0,42.9-26.5,42.9-39.7c0-21.5-13.7-39.7-43.7-39.7h-23.7V172.4z}
                 svg{M88.7,56.8c0,5.5-4.5,10.1-10.1,10.1c-5.6,0-10.1-4.6-10.1-10.1c0-5.6,4.5-10.1,10.1-10.1C84.2,46.7,88.7,51.3,88.7,56.8z};
  }
}

\newcommand\orcidicon[1]{\href{https://orcid.org/#1}{\mbox{\scalerel*{
\begin{tikzpicture}[yscale=-1,transform shape]
\pic{orcidlogo};
\end{tikzpicture}
}{|}}}}

\usepackage{graphicx}	% Including figure files
\usepackage{amsmath,bm}	% Advanced maths commands
\usepackage{amssymb}	% Extra maths symbols
\usepackage{fnpct}
\usepackage[colorlinks=true,allcolors=blue]{hyperref}
\begin{document}

\preprint{APS/123-QED}

\title{Standardizing reverberation-mapped H$\alpha$ and H$\beta$ active galactic nuclei using radius--luminosity relations involving monochromatic and broad H$\alpha$ luminosities }

\author{Shulei Cao$^{\orcidicon{0000-0003-2421-7071}}$}
\email{shuleic@mail.smu.edu}
\affiliation{Department of Physics, Kansas State University, 116 Cardwell Hall, Manhattan, KS 66506, USA}
\affiliation{Department of Physics, Southern Methodist University, 3215 Daniel Ave Fondren Science Building, Dallas, TX 75205, USA}%
\author{Amit Kumar Mandal$^{\orcidicon{0000-0001-9957-6349}}$}
\affiliation{Center for Theoretical Physics, Polish Academy of Sciences, Al. Lotnik\'{o}w 32/46, 02-668 Warsaw, Poland}%
\author{Michal Zaja\v{c}ek$^{\orcidicon{0000-0001-6450-1187}}$}
\affiliation{Department of Theoretical Physics and Astrophysics, Faculty of Science, Masaryk University, Kotlá\v{r}ská 2, 611 37 Brno, Czech Republic}%
\author{Bo\.zena Czerny$^{\orcidicon{0000-0001-5848-4333}}$}%
%\email{bcz@cft.edu.pl}
\affiliation{Center for Theoretical Physics, Polish Academy of Sciences, Al. Lotnik\'{o}w 32/46, 02-668 Warsaw, Poland}%
\author{Bharat Ratra$^{\orcidicon{0000-0002-7307-0726}}$}%
\email{ratra@phys.ksu.edu}
\affiliation{Department of Physics, Kansas State University, 116 Cardwell Hall, Manhattan, KS 66506, USA}%
% \affiliation{ 
% Department of Physics, Kansas State University, 116 Cardwell Hall, Manhattan, KS 66506, USA%\\This line break forced with \textbackslash\textbackslash
% }%

\date{\today}% It is always \today, today,

\begin{abstract}

We test the standardizability of a homogeneous sample of 41 lower-redshift ($0.00415\leq z \leq 0.474$) active galactic nuclei (AGNs) reverberation-mapped (RM) using the broad H$\alpha$ and H$\beta$ emission lines. We find that these sources can be standardized using four radius--luminosity ($R-L$) relations incorporating H$\alpha$ and H$\beta$ time delays and monochromatic and broad H$\alpha$ luminosities. Although the $R-L$ relation parameters are well constrained and independent of the six cosmological models considered, the resulting cosmological constraints are weak. The measured $R-L$ relations exhibit slightly steeper slopes than predicted by a simple photoionization model and steeper than those from previous higher-redshift H$\beta$ analyses based on larger datasets. These differences likely reflect the absence of high-accreting sources in our smaller, lower-redshift sample, which primarily comprises lower-accreting AGNs. The inferred cosmological parameters are consistent within 2$\sigma$ (or better) with those from better-established cosmological probes. This contrasts with our earlier findings using a larger, heterogeneous sample of 118 H$\beta$ AGNs, which yielded cosmological constraints differing by $\gtrsim 2\sigma$ from better-established cosmological probes. Our analysis demonstrates that sample homogeneity---specifically, the use of a consistent time-lag determination method---is crucial for developing RM AGNs as a cosmological probe.
\end{abstract}

 %\keywords{quasars: emission lines; reverberation mapping; cosmological parameters; cosmology: observations; dark energy}

\maketitle

%\tableofcontents
\section{Introduction}

The standard spatially-flat $\Lambda$CDM cosmological model \citep{peeb84}, with current cosmological energy budget dominated by dark energy in the form of a cosmological constant $\Lambda$ and dark matter in the form of non-relativistic cold dark matter (CDM), passes most observational tests. However, there are some measurements that are difficult to reconcile with the predictions of this model, \citep{PerivolaropoulosSkara2021, Morescoetal2022, Abdallaetal2022, Hu:2023jqc}. 

Some of these tensions may be real, in which case the flat $\Lambda$CDM model will have to be replaced by a new standard cosmological model, or they may be due to the underestimation of systematic errors. Cosmological probes can have biases. For example, higher redshift probes are located in the earlier Universe, therefore biases due to the evolution of the probes themselves and of their surroundings (metallicity, extinction, etc.) may be hard to properly account for. To test for biases we need numerous probes, as it is likely that different probes suffer from different biases, and the hope is that by systematically comparing results from different probes we will be able to identify and correct potential biases. 

In this paper, we focus on developing an active galactic nuclei (AGNs) cosmological probe that uses reverberation-mapped (RM) time delay measurements of the broad emission lines with respect to the variable photoionizing continuum emission. These time delays reflect the intrinsic absolute luminosity of each source via the so-called radius--luminosity ($R-L$) relation. We previously obtained useful results along this path \citep{Khadkaetal_2021a,Cao:2022pdv} that showed that higher-redshift (up to $z = 3.4$) quasars (QSOs) are standardizable through the $R-L$ relation. This QSO standardizability holds, within the errors, for moderate- and higher-redshift QSOs. Quasars, or more properly AGNs, are also present at lower redshifts; however, previously available time delay measurements for such samples were not of a satisfactory quality \citep{Khadkaetal2021c}. In particular, AGNs that were reverberation-mapped using the broad component of the H$\beta$ line yielded weak cosmological constraints that implied currently decelerated cosmological expansion and were generally in $\sim 2\sigma$ tension with cosmological constraints determined from better-established probes \citep{Khadkaetal2021c}. 

Here we perform an analysis of a new homogeneous, but small, sample of AGNs covering the lower-redshift range $0.00415 \leq z \leq 0.474$. These 41 AGNs have H$\beta$ and H$\alpha$ time delays determined rather homogeneously using the interpolated cross-correlation function \citep{Choetal2023}. We also correct the source redshifts for peculiar velocities. Additionally, in our analysis we account for uncertainties in cosmological parameters. Further, instead of restricting our analysis to a single cosmological model, we examine these AGNs across half a dozen different cosmological models, to determine whether or not the $R-L$ relation depends on the assumed cosmological model: if it is independent of the assumed cosmological model then these sources are standardizable. We test whether the quality of this new sample is high enough so that it can later be combined with other samples, based on \mii\ and \civ\ lines, covering redshifts up to $\sim 3.4$ \citep[e.g.][]{Caoetal2024} to more fully cover the redshift range of AGN/QSO cosmological probes.

Using both the monochromatic luminosity at 5100\,\AA\, ($L_{5100}$) and the broad H$\alpha$ luminosity ($L_{\text{H}\alpha}$), we test H$\alpha$ and H$\beta$ AGN standardizability using four $R-L$ relations. By simultaneously determining $R-L$ relations and cosmological model parameters in a number of different cosmological models, we show that these AGNs are standardizable, i.e. the $R-L$ relation parameters do not depend on the cosmological model. The $R-L$ relations we find are slightly steeper with respect to the slope of 0.5 predicted by a simple photoionization model, as well as with respect to the slopes found in earlier H$\beta$ $R-L$ studies \citep[see][for reviews]{2013peag.book.....N, 2021bhns.confE...1K, Czerny:2022xfj, 2024SSRv..220...29Z}. This is most likely due to the homogeneous and relatively small, low-redshift sample analyzed here, which lacks highly accreting sources that are known to exhibit shortened time delays \citep{2018ApJ...856....6D, MartinezAldama2019, 2020ApJ...896..146Z}, most likely caused by intrinsic changes in the accretion flow at accretion rates close to the Eddington limit \citep{2014ApJ...797...65W}. As for the cosmological model parameters, we obtain only weak constraints that are overall consistent with the standard flat $\Lambda$CDM model, i.e. we do find significant constraints against the standard model, in contrast to what we found earlier from a larger but less homogeneous H$\beta$ sample \citep{Khadkaetal2021c}.

Our paper is structured as follows. In Sec.~\ref{sec:model}, we introduce the cosmological models whose parameters are inferred in this study. The datasets we use are described in Sec.~\ref{sec:data}. The methods that are applied to test the standardizability of QSOs and to infer cosmological model parameters are summarized in Sec.~\ref{sec:analysis}. Results are presented in Sec.~\ref{sec:results}. The determined radius--luminosity relations, their intrinsic scatter, the determined $L_{\text{H}\alpha} - L_{5100}$ relations, and the determined cosmological parameters are discussed in Sec.~\ref{sec:discussion}. Finally, we summarize the main conclusions in Sec.~\ref{sec:conclusion}.

\section{Cosmological models}
\label{sec:model}

We simultaneously constrain cosmological model parameters and the $R-L$ relation parameters in six spatially flat and nonflat relativistic dark energy models, to test whether the H$\alpha$ and H$\beta$ RM AGN datasets are standardizable through the $R-L$ relation. If the derived $R-L$ relation parameters are independent of the assumed cosmological model then the AGN dataset can be assumed to be standardizable. This technique \cite{KhadkaRatra2020c, Caoetal_2021} allows one to get around the circularity problem associated with such data, i.e., that these data must simultaneously determine the cosmological model parameters and the $R-L$ relation parameters.\footnote{This technique has been used to determine which gamma-ray burst data compilations are standardizable \cite{Khadkaetal_2021b, CaoKhadkaRatra2022, CaoDainottiRatra2022, CaoDainottiRatra2022b, CaoRatra2024b, CaoRatra2025}, to show that the \citet{Lussoetal2020} $L_X-L_{UV}$ quasar dataset is not standardizable and so cannot be used for cosmological purposes \cite{KhadkaRatra2021, KhadkaRatra2022, Petrosian:2022tlp, Khadka:2022aeg, Zajaceketal2024}, and to show that the \hii\ starburst galaxy dataset of \citet{Gonzalez-Moran:2021drc} is not standardizable and so cannot be used for cosmological purposes \cite{CaoRatra2024a, Melnick:2024ywi}.} 

Each cosmological model predicts the luminosity distance $D_L(z)$, through the dimensionless Hubble parameter $E(z) = H(z)/H_0$, as a function of redshift $z$ and cosmological model parameters,
\begin{equation}
  \label{eq:DL}
\resizebox{0.475\textwidth}{!}{%
    $D_L(z) = 
    \begin{cases}
    \frac{c(1+z)}{H_0\sqrt{\Omega_{\rm k0}}}\sinh\left[\frac{H_0\sqrt{\Omega_{\rm k0}}}{c}D_C(z)\right] & \text{if}\ \Omega_{\rm k0} > 0, \\
    \vspace{1mm}
    (1+z)D_C(z) & \text{if}\ \Omega_{\rm k0} = 0,\\
    \vspace{1mm}
    \frac{c(1+z)}{H_0\sqrt{|\Omega_{\rm k0}|}}\sin\left[\frac{H_0\sqrt{|\Omega_{\rm k0}|}}{c}D_C(z)\right] & \text{if}\ \Omega_{\rm k0} < 0,
    \end{cases}$%
    }
\end{equation}
where the comoving distance is
\begin{equation}
\label{eq:gz}
   D_C(z) = \frac{c}{H_0}\int^z_0 \frac{dz'}{E(z')}.
\end{equation}
Here, $c$ is the speed of light, $H_0$ is the Hubble constant, the present value of the Hubble parameter $H(z)$, and \ok\ is the current value of the spatial curvature density parameter.\footnote{For recent discussions on spatial curvature, refer to Refs.\ \cite{Oobaetal2018b, ParkRatra2019b, KhadkaRatra2020a, DiValentinoetal2021a, CaoRyanRatra2020, CaoRyanRatra2021, ArjonaNesseris2021, Dhawanetal2021, Renzietal2021, Gengetal2022, MukherjeeBanerjee2022, Glanvilleetal2022, Wuetal2023, deCruzPerezetal2023, DahiyaJain2022, Stevensetal2023,Favaleetal2023, Qietal2023, deCruzPerez:2024shj, Shimon:2024mbm, Wu:2024faw}.} We assume one massive and two massless neutrino species, setting the effective number of relativistic neutrino species to $N_{\rm eff} = 3.046$ and the total neutrino mass to $\sum m_{\nu} = 0.06$ eV. The current physical energy density parameter of nonrelativistic neutrinos is then fixed as $\onh=\sum m_{\nu}/(93.14\ \rm eV)$, where $h$ represents $H_0$ in units of 100 \hunit. The nonrelativistic matter density parameter today, $\Om = (\onh + \obh + \och)/{h^2}$, incorporates contributions from neutrinos, baryonic matter, and cold dark matter, with radiation neglected due to its negligible influence at late times.

In the \lcdm\ models, dark energy is a time- and space-independent cosmological constant, $\Lambda$, with the dark energy equation of state parameter $w_{\rm DE}=-1$, where $w_{\rm DE}=p_{\rm DE}/\rho_{\rm DE}$ with $p_{\rm DE}$ and $\rho_{\rm DE}$ being the pressure and energy density of the dark energy, respectively. Whereas in the XCDM parametrizations, time-dependent but space-independent dynamical dark energy is modeled as an ideal, spatially homogeneous X-fluid with a constant $w_{\rm DE}$ that can deviate from $-1$. The XCDM parametrizations are physically incomplete because they lack a full description of spatial inhomogeneities. In \lcdm\ and XCDM, the dimensionless Hubble parameter is
\be
\label{eq:EzL}
\resizebox{0.475\textwidth}{!}{%
    $E(z) = \sqrt{\Om\left(1 + z\right)^3 + \Ok\left(1 + z\right)^2 + \Omega_{\rm DE0}\left(1+z\right)^{1+w_{\rm DE}}},$%
    }
\ee
where $\Omega_{\rm DE0} = 1 - \Om - \Ok$ denotes the current value of the dark energy density parameter. Given that H$\alpha$ and H$\beta$ RM AGN data do not constrain $H_0$ and $\Omega_{b}$, we set $H_0=70$ \hunit\ and $\Omega_{b}=0.05$ for simplicity when we analyze the AGN datasets. In these analyses, $\Om$ and $\Ok$ are constrained in nonflat \lcdm, with $\Ok$ excluded in flat \lcdm. $\Om, \wX$, and $\Ok$ are constrained in nonflat XCDM with $\Ok=0$ for flat XCDM. For analyses including $H(z)$ + BAO data, the parameters $H_0, \obh\!, \och\!$, and $\Ok$ are constrained in nonflat \lcdm\ and nonflat XCDM, with $\wX$ further constrained in XCDM, and with $\Ok = 0$ in flat \lcdm\ and flat XCDM. 

For the \pcdm\ models \citep{peebrat88,ratpeeb88,pavlov13}\footnote{Recent constraints on \pcdm\ can be found in Refs.\ \cite{ooba_etal_2018b, ooba_etal_2019, park_ratra_2018, park_ratra_2019b, park_ratra_2020, Singhetal2019, KhadkaRatra2020b, UrenaLopezRoy2020, SinhaBanerjee2021, CaoRyanRatra2022, deCruzetal2021, Xuetal2022, Jesusetal2022, CaoRatra2022, Adiletal2023, Dongetal2023, VanRaamsdonkWaddell2023}.} time-dependent and space-dependent dynamical dark energy is a dynamical scalar field $\phi$ with potential energy density
\be
\label{PE}
V(\phi)=\frac{1}{2}\kappa m_p^2\phi^{-\alpha},
\ee
following an inverse power-law. The dimensionless Hubble parameter
\be
\label{eq:Ezp}
    E(z) = \sqrt{\Om\left(1 + z\right)^3 + \Ok\left(1 + z\right)^2 + \Omega_{\phi}(z,\alpha)},
\ee
with a dynamical scalar field dark energy density parameter 
\be
\label{Op}
\Omega_{\phi}(z,\alpha)=\frac{1}{6H_0^2}\bigg[\frac{1}{2}\dot{\phi}^2+V(\phi)\bigg],
\ee
is the governing Friedmann equation, where $\Omega_{\phi}(z,\alpha)$ is numerically determined by both Eq.\ \eqref{eq:Ezp} and the scalar field equation of motion
\be
\label{em}
\ddot{\phi}+3H\dot{\phi}+V'(\phi)=0.
\ee
In these equations, $\alpha$ is the inverse power-law potential energy density exponent and the \lcdm\ models are recovered when $\alpha = 0$, the Planck mass is denoted as $m_p$, and the time and $\phi$ derivatives are represented by an overdot and a prime, respectively. The inverse power-law potential energy density coefficient $\kappa$ is determined via the shooting method in the Cosmic Linear Anisotropy Solving System (\textsc{class}) code \citep{class}. For RM AGN data, $\Om$, $\Ok$, and $\alpha$ are constrained in \pcdm\ with $\Ok=0$ in the spatially flat case. For analyses including $H(z)$ + BAO data, $H_0, \obh\!, \och\!$, and $\Ok$ are constrained in \pcdm\ with $\Ok=0$ in the flat case.

\section{Sample and Data}
\label{sec:data}

In this section we summarize the data selection and characteristics for H$\alpha$ and H$\beta$ AGNs (see Subsection~\ref{subsec_AGNsample}), while in Subsection~\ref{subsec_HzBAO} we introduce the comparison data sample consisting of better-established cosmological probes, in particular a joint sample of expansion rate and baryon acoustic oscillation data, $H(z)$ + BAO.

\subsection{AGN Sample and Data}
\label{subsec_AGNsample}

The AGN sample used in this study is smaller than most previous samples we have analyzed \cite{Khadkaetal_2021a, Khadkaetal2021c, Khadkaetal2022a, Cao:2022pdv, Czerny:2022xfj, Caoetal2024}. However, it is significantly more homogeneous, as all objects in this sample were analyzed using the same methodology. In contrast, using a larger sample with measurements of mixed quality or inconsistent lag measurement methodologies does not yield improved results, as demonstrated by \citet{Caoetal2024}. This issue is particularly pronounced at lower redshifts, where previously available H$\beta$ lag measurements lack a standardized approach due to data inhomogeneity \citep{Mehrabietal2021, Khadkaetal2021c}.

\begin{figure*}
\centering
\includegraphics[width=\columnwidth]{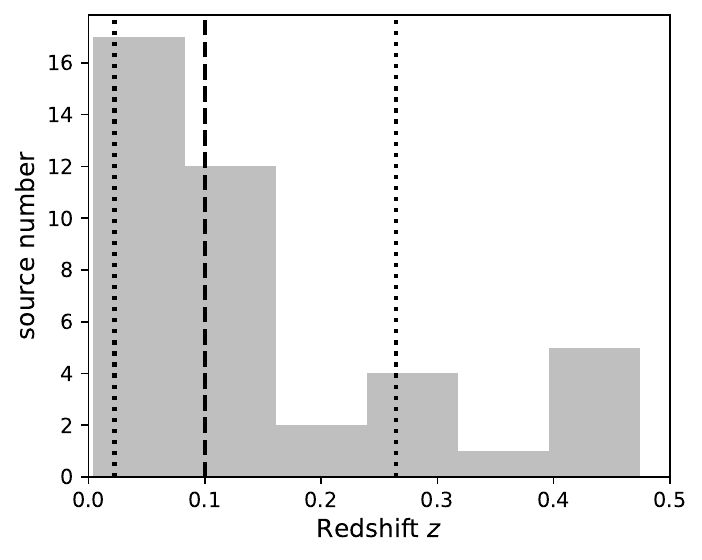}
\includegraphics[width=\columnwidth]{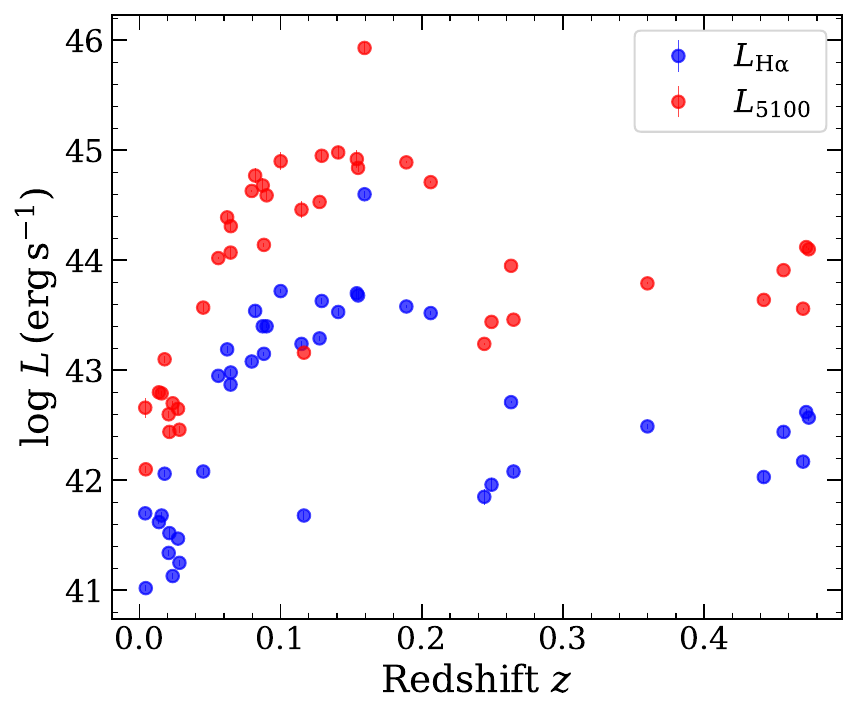}
\caption{Characteristics for the sample of 41 H$\alpha$ and H$\beta$ AGNs. \textit{Left panel:} Distribution of source redshifts. The bin width used, $\Delta z\sim 0.078$, is inferred from Knuth's rule. The dashed vertical line shows the sample median redshift, while the dotted vertical lines represent 16\%- and 84\%-percentiles. \textit{Right panel:} Distribution of AGN H$\alpha$ luminosities (blue) and continuum luminosities at 5100 {\AA} (red) as a function of redshift ($z$) for our final sample. The H$\alpha$ luminosities are measured from the broad H$\alpha$ fluxes, while the continuum luminosities at 5100 {\AA} are derived from the monochromatic flux at 5100 {\AA}, assuming a flat $\Lambda$CDM cosmological model with $H_0=72\,{\rm km\,s^{-1}\,Mpc^{-1}}$ and $\Omega_{ m0}=0.3$.}
\label{fig:dist}
\end{figure*}

Hence, to constrain cosmological parameters using a uniform and homogeneous dataset, we compiled a sample of Type 1 AGNs with H$\alpha$ and H$\beta$ time lag measurements from RM campaigns reported in the literature. Our primary sample is drawn from the dataset compiled by \citet{Choetal2023}, who provided reliable H$\alpha$ time lags for 5 of 6 AGNs observed in the Seoul National University AGN Monitoring Project \citep[SAMP;][]{2024ApJ...962...67W}. Additionally, to explore the BLR radius--luminosity relation, \citet{Choetal2023} included 42 AGNs from other sources \citep{Kaspietal2000, Bentzetal2010, 2017ApJ...851...21G, 2022ApJ...936...75L, 2017MNRAS.465.1898S, 2011ApJ...732..121B} with reliable H$\alpha$ lag measurements, selected based on lag uncertainties, maximum of cross-correlation coefficients ($\mathrm{r_{max}} > 0.5$), and various quality ratings for the time lags \citep[see][for details]{2017ApJ...851...21G, Choetal2023}.

Of the initial 47 AGNs with H$\alpha$ lag measurements, 42 objects also have H$\beta$ lag measurements. However, one of these targets, NGC 2617, lacks an AGN 5100 {\AA} continuum luminosity ($L_{5100}$) measurement. As a result, we finalized a sample of 41 AGNs with both H$\alpha$ and H$\beta$ lag measurements for our analysis. It is important to note that all lags were consistently derived using the Interpolated Cross-Correlation Function (ICCF) analysis method \citep[ICCF;][]{1986ApJ...305..175G, 1987ApJS...65....1G}.

\begin{turnpage}
\begin{table*}
\centering
\resizebox{2.7\columnwidth}{!}{%
\begin{threeparttable}
\caption{QSO sample of 41 sources with H$\alpha$ and H$\beta$ time delays. From left to right the columns list: object name, redshift, broad H$\alpha$ flux density, broad H$\alpha$ luminosity for the flat $\Lambda$CDM model ($H_0=72\,{\rm km\,s^{-1}\,Mpc^{-1}}$ and $\Omega_{ m0}=0.3$), flux density at 5100\,\AA, monochromatic luminosity at 5100\,\AA\, for the flat $\Lambda$CDM model ($H_0=72\,{\rm km\,s^{-1}\,Mpc^{-1}}$ and $\Omega_{ m0}=0.3$), rest-frame H$\alpha$\ time lag (in days), rest-frame H$\beta$\ time lag (in days), and the original references. Source redshifts are corrected for peculiar velocities which are used to compute the corrected luminosities.}
\label{tab:HalphaQSO}
\setlength{\tabcolsep}{12pt}
\begin{tabular}{lcccccccc}
\toprule\toprule
Object & $z$ & $\log \left(F_{\text{H}\alpha}/{\rm erg}\,{\rm s^{-1}}{\rm cm^{-2}}\right)$ & $\log \left(L_{\text{H}\alpha}/{\rm erg}\,{\rm s^{-1}}\right)$ & $\log \left(F_{5100}/{\rm erg}\,{\rm s^{-1}}{\rm cm^{-2}}\right)$ & $\log \left(L_{5100}/{\rm erg}\,{\rm s^{-1}}\right)$ & $\tau_{\text{H}\alpha}$ (days) & $\tau_{\text{H}\beta}$ (days) & Ref.\tnote{a}\\
\midrule
Mrk1501 & 0.08819 & $-12.13 \pm 0.02$ & $43.14 \pm 0.02$ & $-11.14 \pm 0.02$ & $44.13 \pm 0.02$ & $67^{+24}_{-38}$ & $12^{+8}_{-9}$ & \citep{Choetal2023,2024ApJ...962...67W} \\
PG0026+129 & 0.14083 & $-12.18 \pm 0.04$ & $43.52 \pm 0.04$ & $-10.73 \pm 0.06$ & $44.97 \pm 0.06$ & $116^{+25}_{-27}$ & $109^{+25}_{-32}$ & \citep{Kaspietal2000} \\
PG0052+251 & 0.15399 & $-12.09 \pm 0.05$ & $43.69 \pm 0.05$ & $-10.87 \pm 0.08$ & $44.91 \pm 0.08$ & $183^{+57}_{-38}$ & $86^{+26}_{-27}$ & \citep{Kaspietal2000} \\
J0101+422 & 0.18909 & $-12.41 \pm 0.01$ & $43.58 \pm 0.01$ & $-11.10 \pm 0.01$ & $44.89 \pm 0.01$ & $118^{+17}_{-17}$ & $76^{+13}_{-12}$ & \citep{Choetal2023,2024ApJ...962...67W} \\
PG0804+761 & 0.10006 & $-11.66 \pm 0.02$ & $43.72 \pm 0.02$ & $-10.48 \pm 0.08$ & $44.90 \pm 0.08$ & $175^{+18}_{-15}$ & $137^{+24}_{-22}$ & \citep{Kaspietal2000} \\
PG0844+349 & 0.06473 & $-11.99 \pm 0.03$ & $42.99 \pm 0.03$ & $-10.66 \pm 0.04$ & $44.32 \pm 0.04$ & $37^{+15}_{-15}$ & $12^{+13}_{-10}$ & \citep{Kaspietal2000} \\
PG0947+396 & 0.20633 & $-12.54 \pm 0.02$ & $43.52 \pm 0.02$ & $-11.35 \pm 0.01$ & $44.71 \pm 0.01$ & $71^{+16}_{-35}$ & $37^{+10}_{-11}$ & \citep{Choetal2023,2024ApJ...962...67W} \\
Mrk142 & 0.04521 & $-12.56 \pm 0.03$ & $42.09 \pm 0.03$ & $-11.07 \pm 0.04$ & $43.58 \pm 0.04$ & $2.8^{+1.2}_{-0.9}$ & $2.7^{+0.7}_{-0.8}$ & \citep{Bentzetal2010} \\
SBS1116+583A & 0.02836 & $-12.97 \pm 0.03$ & $41.27 \pm 0.03$ & $-11.76 \pm 0.03$ & $42.48 \pm 0.03$ & $4^{+1.4}_{-1}$ & $2.3^{+0.6}_{-0.5}$ & \citep{Bentzetal2010} \\
Arp151 & 0.02128 & $-12.44 \pm 0.05$ & $41.54 \pm 0.05$ & $-11.52 \pm 0.05$ & $42.46 \pm 0.05$ & $7.8^{+1}_{-1}$ & $4^{+0.5}_{-0.7}$ & \citep{Bentzetal2010} \\
Mrk1310 & 0.02077 & $-12.57 \pm 0.03$ & $41.39 \pm 0.03$ & $-11.31 \pm 0.03$ & $42.65 \pm 0.03$ & $4.5^{+0.7}_{-0.6}$ & $3.7^{+0.6}_{-0.6}$ & \citep{Bentzetal2010} \\
NGC4151 & 0.00415 & $-10.66 \pm 0.02$ & $41.89 \pm 0.02$ & $-9.70 \pm 0.09$ & $42.85 \pm 0.09$ & $7.6^{+1.9}_{-2.6}$ & $6.2^{+1.4}_{-1.1}$ & \citep{2011ApJ...732..121B} \\
PG1211+143 & 0.08201 & $-11.64 \pm 0.05$ & $43.55 \pm 0.05$ & $-10.41 \pm 0.07$ & $44.78 \pm 0.07$ & $107^{+35}_{-42}$ & $95^{+29}_{-41}$ & \citep{Kaspietal2000} \\
Mrk202 & 0.02359 & $-12.93 \pm 0.03$ & $41.15 \pm 0.03$ & $-11.36 \pm 0.02$ & $42.72 \pm 0.02$ & $22^{+1}_{-4}$ & $3.1^{+1.7}_{-1.1}$ & \citep{Bentzetal2010} \\
NGC4253 & 0.01383 & $-11.93 \pm 0.02$ & $41.68 \pm 0.02$ & $-10.75 \pm 0.02$ & $42.86 \pm 0.02$ & $25^{+1}_{-1}$ & $6.2^{+1.6}_{-1.2}$ & \citep{Bentzetal2010} \\
PG1226+023 & 0.15949 & $-11.21 \pm 0.03$ & $44.61 \pm 0.03$ & $-9.88 \pm 0.05$ & $45.94 \pm 0.05$ & $444^{+56}_{-55}$ & $330^{+101}_{-83}$ & \citep{Kaspietal2000} \\
PG1229+204 & 0.06458 & $-12.09 \pm 0.04$ & $42.88 \pm 0.04$ & $-10.89 \pm 0.05$ & $44.08 \pm 0.05$ & $67^{+37}_{-43}$ & $34^{+30}_{-17}$ & \citep{Kaspietal2000} \\
NGC4748 & 0.01575 & $-11.98 \pm 0.03$ & $41.74 \pm 0.03$ & $-10.87 \pm 0.02$ & $42.85 \pm 0.02$ & $7.5^{+3}_{-4.6}$ & $5.5^{+1.6}_{-2.2}$ & \citep{Bentzetal2010} \\
VIIIZw218 & 0.12763 & $-12.31 \pm 0.02$ & $43.30 \pm 0.02$ & $-11.07 \pm 0.01$ & $44.54 \pm 0.01$ & $140^{+26}_{-26}$ & $63^{+16}_{-15}$ & \citep{Choetal2023,2024ApJ...962...67W} \\
PG1307+085 & 0.15487 & $-12.10 \pm 0.04$ & $43.69 \pm 0.04$ & $-10.94 \pm 0.04$ & $44.85 \pm 0.04$ & $155^{+81}_{-126}$ & $94^{+40}_{-100}$ & \citep{Kaspietal2000} \\
SDSSJ140812.09+535303.3\tnote{b} & 0.11655 & $-13.84 \pm 0.04$ & $41.68 \pm 0.04$ & $-12.36 \pm 0.03\tnote{b}$ & $43.16 \pm 0.03\tnote{b}$ & $7.2^{+4.8}_{-5.6}$ & $9^{+5.6}_{-3.8}$ & \citep{2017ApJ...851...21G} \\
SDSSJ141018.04+532937.5 & 0.47008 & $-14.72 \pm 0.04$ & $42.17 \pm 0.04$ & $-13.33 \pm 0.01$ & $43.56 \pm 0.01$ & $23^{+13}_{-8}$ & $14^{+4}_{-6}$ & \citep{2017ApJ...851...21G} \\
SDSSJ141041.25+531849.0 & 0.35985 & $-14.13 \pm 0.02$ & $42.49 \pm 0.02$ & $-12.83 \pm 0.01$ & $43.79 \pm 0.01$ & $12^{+8}_{-7}$ & $11^{+7}_{-7}$ & \citep{2017ApJ...851...21G} \\
SDSSJ141123.42+521331.7 & 0.47232 & $-14.28 \pm 0.03$ & $42.62 \pm 0.03$ & $-12.78 \pm 0.01$ & $44.12 \pm 0.01$ & $13^{+10}_{-14}$ & $6.5^{+8.8}_{-5.4}$ & \citep{2017ApJ...851...21G} \\
PG1411+442 & 0.09012 & $-11.88 \pm 0.02$ & $43.41 \pm 0.02$ & $-10.69 \pm 0.04$ & $44.60 \pm 0.04$ & $95^{+37}_{-34}$ & $108^{+66}_{-65}$ & \citep{Kaspietal2000} \\
SDSSJ141324.28+530527.0\tnote{b} & 0.45626 & $-14.42 \pm 0.05$ & $42.44 \pm 0.05$ & $-12.95 \pm 0.03\tnote{b}$ & $43.91 \pm 0.03\tnote{b}$ & $45^{+14}_{-11}$ & $22^{+11}_{-11}$ & \citep{2017ApJ...851...21G} \\
SDSSJ141625.71+535438.5\tnote{b} & 0.26324 & $-13.59 \pm 0.01$ & $42.71 \pm 0.01$ & $-12.35 \pm 0.03\tnote{b}$ & $43.95 \pm 0.03\tnote{b}$ & $33^{+19}_{-17}$ & $17^{+6}_{-7}$ & \citep{2017ApJ...851...21G} \\
SDSSJ141645.15+542540.8 & 0.24430 & $-14.38 \pm 0.07$ & $41.85 \pm 0.07$ & $-12.99 \pm 0.01$ & $43.24 \pm 0.01$ & $9.6^{+4.5}_{-3}$ & $6.5^{+2.7}_{-1.8}$ & \citep{2017ApJ...851...21G} \\
SDSSJ141645.58+534446.8 & 0.44226 & $-14.80 \pm 0.05$ & $42.03 \pm 0.05$ & $-13.19 \pm 0.01$ & $43.64 \pm 0.01$ & $18^{+7}_{-8}$ & $9.7^{+4}_{-4}$ & \citep{2017ApJ...851...21G} \\
NGC5548 & 0.01788 & $-11.74 \pm 0.03$ & $42.10 \pm 0.03$ & $-10.70 \pm 0.03$ & $43.14 \pm 0.03$ & $11^{+1}_{-1}$ & $4.2^{+0.9}_{-1.3}$ & \citep{Bentzetal2010} \\
SDSSJ142038.52+532416.5\tnote{b} & 0.26500 & $-14.23 \pm 0.03$ & $42.08 \pm 0.03$ & $-12.85 \pm 0.03\tnote{b}$ & $43.46 \pm 0.03\tnote{b}$ & $20^{+15}_{-15}$ & $27^{+8}_{-14}$ & \citep{2017ApJ...851...21G} \\
SDSSJ142039.80+520359.7 & 0.47413 & $-14.33 \pm 0.03$ & $42.57 \pm 0.03$ & $-12.80 \pm 0.01$ & $44.10 \pm 0.01$ & $18^{+6}_{-16}$ & $5.1^{+6.4}_{-8.5}$ & \citep{2017ApJ...851...21G} \\
SDSSJ142135.90+523138.9\tnote{b} & 0.24941 & $-14.29 \pm 0.06$ & $41.96 \pm 0.06$ & $-12.81 \pm 0.03\tnote{b}$ & $43.44 \pm 0.03\tnote{b}$ & $7.2^{+3.4}_{-5.6}$ & $1^{+3.8}_{-4.2}$ & \citep{2017ApJ...851...21G} \\
PG1426+015 & 0.08737 & $-11.85 \pm 0.03$ & $43.41 \pm 0.03$ & $-10.57 \pm 0.07$ & $44.69 \pm 0.07$ & $83^{+42}_{-48}$ & $106^{+45}_{-63}$ & \citep{Kaspietal2000} \\
PG1440+356\tnote{c} & 0.07958 & $-12.08 \pm 0.03$ & $43.09 \pm 0.03$ & $-10.53 \pm 0.01\tnote{c}$ & $44.64 \pm 0.01\tnote{c}$ & $80^{+63}_{-30}$ & $51^{+17}_{-21}$ & \citep{Choetal2023,2024ApJ...962...67W} \\
PG1613+658 & 0.12916 & $-11.99 \pm 0.03$ & $43.63 \pm 0.03$ & $-10.67 \pm 0.05$ & $44.95 \pm 0.05$ & $38^{+35}_{-19}$ & $39^{+18}_{-20}$ & \citep{Kaspietal2000} \\
PG1617+175 & 0.11482 & $-12.27 \pm 0.03$ & $43.24 \pm 0.03$ & $-11.05 \pm 0.08$ & $44.46 \pm 0.08$ & $100^{+28}_{-33}$ & $70^{+27}_{-37}$ & \citep{Kaspietal2000} \\
3C390.3 & 0.05590 & $-11.90 \pm 0.02$ & $42.95 \pm 0.02$ & $-10.83 \pm 0.01$ & $44.02 \pm 0.01$ & $153^{+14}_{-14}$ & $84^{+8}_{-8}$ & \citep{Kaspietal2000} \\
Zw229-015\tnote{b} & 0.02735 & $-12.76 \pm 0.03\tnote{b}$ & $41.45 \pm 0.03\tnote{b}$ & $-11.58 \pm 0.05$ & $42.63 \pm 0.05$ & $5.1^{+0.8}_{-1.1}$ & $3.9^{+0.7}_{-0.9}$ & \citep{Kaspietal2000} \\
NGC6814 & 0.00451 & $-11.74 \pm 0.03$ & $40.89 \pm 0.03$ & $-10.66 \pm 0.03$ & $41.97 \pm 0.03$ & $9.5^{+1.9}_{-1.6}$ & $6.6^{+0.9}_{-0.9}$ & \citep{Bentzetal2010} \\
PG2130+099 & 0.06217 & $-11.77 \pm 0.03$ & $43.17 \pm 0.03$ & $-10.57 \pm 0.04$ & $44.37 \pm 0.04$ & $223^{+50}_{-26}$ & $177^{+128}_{-25}$ & \citep{Kaspietal2000} \\

\bottomrule\bottomrule
\end{tabular}
\begin{tablenotes}
\item [a] Two references: H$\alpha$ lag from the first and H$\beta$ lag from the second. One reference: all lags from the same reference.
\item [b] Zero uncertainties are replaced by median uncertainty of $0.03$.
\item [c] The uncertainty of $\log L_{5100}$ is updated with that reported by Ref.\ \citet{2024ApJ...962...67W}.
\end{tablenotes}
\end{threeparttable}%
}
\end{table*}
\end{turnpage}

For our sample, we adopted AGN continuum luminosity at 5100 {\AA}, $L_{5100}$, and broad H$\alpha$ luminosity, $L_{\mathrm{H}\alpha}$, numerical values from \citet{Choetal2023}. These were computed from the flux measurements\footnote{The \citet{Choetal2023} fluxes were derived through spectral decomposition, which separates the AGN continuum from the stellar continuum, and corrected for galactic extinction. Note that, the reported values of $L_{5100}$ and $L_{\mathrm{H}\alpha}$ were retrieved from various literature sources, making them inherently heterogeneous. Nonetheless, \citet{Choetal2023} confirmed that this heterogeneity has a negligible impact on any systematic deviations from the radius--luminosity relations.} assuming a flat $\Lambda$CDM cosmological model with $H_0=72\,{\rm km\,s^{-1}\,Mpc^{-1}}$ and $\Omega_{ m0}=0.3$. We used the same cosmological parameter values to convert the \citet{Choetal2023} luminosities back to fluxes. These fluxes are listed in the third and fifth columns of Table \ref{tab:HalphaQSO}.

Usually redshifts of AGNs determined from spectra are measured in the heliocentric reference frame. However, additional corrections for the Milky Way's peculiar velocity are often required. While this effect is minimal for AGNs at high redshift, it becomes significant for those at low redshift, necessitating corrections for peculiar velocities. Additionally, the motion of an AGN relative to the cosmic microwave background (CMB) frame varies depending on the source's location in the sky. To address these factors, we used the NED Velocity Correction Calculator\footnote{\url{https://ned.ipac.caltech.edu/help/velc_help.html}}, which corrects for Galactic rotation, the peculiar motion of the Galaxy within the Local Group, the `infall' of the Local Group toward the center of the Local Supercluster, and motion in the reference frame established by the CMB radiation. We applied this correction to all AGNs studied in this work, resulting in new redshifts that account for peculiar velocities. These corrected redshifts are listed in the second column of Table \ref{tab:HalphaQSO}. 

Using these corrected redshifts, we computed corrected  flat $\Lambda$CDM cosmological model (with $H_0=72\,{\rm km\,s^{-1}\,Mpc^{-1}}$ and $\Omega_{ m0}=0.3$) $L_{5100}$ and $L_{\mathrm{H}\alpha}$ values from the measured fluxes. These corrected absolute luminosities are listed in the fourth and sixth columns of Table \ref{tab:HalphaQSO}. We emphasize that our analyses use the measured fluxes and not these absolute luminosity values. 

Thus, our final sample consists of 41 AGNs with H$\alpha$ and H$\beta$ rest-frame time lag measurements, taken from \citet{Choetal2023} and listed in the seventh and eighth columns of Table \ref{tab:HalphaQSO}, covering a redshift range from 0.00415 to 0.474, with a median value of 0.100 and 16\%- and 84\%-percentiles of 0.022 and 0.264, respectively. The redshift distribution is shown as a histogram in Fig.~\ref{fig:dist} (left panel). The monochromatic luminosity $L_{5100}$ spans from $10^{42.1}$ to $10^{45.9}$ $\mathrm{erg \, s^{-1}}$, and the broad H$\alpha$ luminosity $L_{\mathrm{H}\alpha}$ is in the range from $10^{40.9}$ to $10^{44.6}$ $\mathrm{erg \, s^{-1}}$. Figure~\ref{fig:dist} (right panel) presents the distribution of $L_{\mathrm{H}\alpha}$ and $L_{5100}$ as a function of redshift for this sample. For subsequent analysis, we designate the sample as H$\alpha$ broad (H$\beta$ broad) and H$\alpha$ mono (H$\beta$ mono), depending on whether broad $L_{\mathrm{H}\alpha}$ or monochromatic $L_{5100}$ are used with H$\alpha$ (H$\beta$) time lags, respectively.

\subsection{$H(z)$ and BAO data}
\label{subsec_HzBAO}

For the analyses including $H(z)$ + BAO data, we use 32 $H(z)$ measurements from cosmic chronometers and 12 baryon acoustic oscillation (BAO) measurements spanning the redshift ranges of $0.070 \leq z \leq 1.965$ and $0.122 \leq z \leq 2.334$, respectively. The detailed data descriptions can be found in Tables 1 and 2, and section III of Ref.\ \cite{CaoRatra2023}.

\section{Data Analysis Methodology}
\label{sec:analysis}

The AGN $R-L$ relation for H$\alpha$ or H$\beta$ AGNs can be expressed as
\begin{equation}
    \label{eq:R-L}
    \log{\left(\frac{\tau}{\rm day}\right)}=\beta+\gamma \log{\left(\frac{L_{\text{broad/mono}}}{10^{44}\,{\rm erg\ s^{-1}}}\right)},
\end{equation}
where $\tau (=R/c)$, $\beta$, and $\gamma$ are the H$\alpha$ or H$\beta$\ rest-frame time lag, the intercept parameter, and the slope parameter, respectively, and the luminosity $L_{\text{broad/mono}}$ is H$\alpha$ broad or monochromatic at 5100\,\AA\
\be
\label{eq:L1350}
    L_{\text{broad/mono}}=4\pi D_L^2F_{\text{broad/mono}},
\ee
with measured AGN flux $F_{\text{broad/mono}}$ in units of $\rm erg\ s^{-1}\ cm^{-2}$ and the luminosity distance $D_L(z)$ is computed from Eq.\ \eqref{eq:DL}.

The natural log of the H$\alpha$ or H$\beta$\ likelihood function \citep{D'Agostini_2005} is
\be
\label{eq:LH}
    \ln\mathcal{L}= -\frac{1}{2}\Bigg[\chi^2+\sum^{N}_{i=1}\ln\left(2\pi\sigma^2_{\mathrm{tot},i}\right)\Bigg],
\ee
where
\be
\label{eq:chi2}
    \chi^2 = \sum^{N}_{i=1}\bigg[\frac{(\log \tau_{\mathrm{obs},i} - \beta - \gamma\log L_{\text{broad/mono},i})^2}{\sigma^2_{\mathrm{tot},i}}\bigg],
\ee
where $\tau_{\mathrm{obs},i}$ is in units of day and $L_{\text{broad/mono},i}$ is in units of ${10^{44}\,{\rm erg\ s^{-1}}}$, with total uncertainty
\be
\label{eq:sigma_civ}
\sigma^2_{\mathrm{tot},i}=\sigma_{\rm int}^2+\sigma_{{\log \tau_{\mathrm{obs},i}}}^2+\gamma^2\sigma_{\log F_{\text{broad/mono},i}}^2,
\ee
where $\sigma_{\rm int}$ is the intrinsic scatter parameter of H$\alpha$ or H$\beta$ AGNs which also contains the unknown systematic uncertainty, $\sigma_{{\log \tau_{\mathrm{obs},i}}}$ and $\sigma_{{\log F_{\text{broad/mono},i}}}$ are the measured time-delay and flux density uncertainties for individual H$\alpha$ or H$\beta$ AGNs, and $N$ is the number of data points. Note that we account for the asymmetric uncertainties in $\tau$ by separately incorporating its upper error ($\sigma_{\tau,+}$) and lower error ($\sigma_{\tau,-}$). Specifically, when the theoretical prediction for $\log{\tau}$ exceeds the observed value, we use $\sigma_{\tau} = \sigma_{\tau,+}$. Conversely, when the prediction is smaller than the observed value, we adopt $\sigma_{\tau} = \sigma_{\tau,-}$.

We also use the same likelihood function \eqref{eq:LH} for the analyses of luminosity-luminosity correlations, but now with
\be
% \label{eq:chi2}
    \chi^2 = \sum^{N}_{i=1}\bigg[\frac{(y_i - \beta - \gamma x_i)^2}{\sigma^2_{\mathrm{tot},i}}\bigg],
\ee
and
\be
\sigma^2_{\mathrm{tot},i} = \sigma_{\rm int}^2+\sigma_{y_i}^2+\gamma^2\sigma_{x_i}^2.
\ee
Here $y$ is $\log \left(L_{\rm broad}/10^{44}\, \rm erg\, s^{-1}\right)$ and $x$ is $\log \left( L_{5100}/10^{42}\, \rm erg\, s^{-1}\right)$ with $D_L$ computed using posterior mean cosmological parameters from the corresponding H$\alpha$ or H$\beta$ $R-L$ relations. 

The likelihood functions of $H(z)$ and BAO data can be found in section IV of Ref.\ \cite{CaoRatra2023} with covariance matrices described in section III of Ref.\ \cite{CaoRatra2023}. 

We use the Akaike information criterion (AIC), the Bayesian information criterion (BIC), and the more reliable deviance information criterion (DIC) to compare the goodness of fit of different cosmological models for a given dataset \cite[for a  detailed description see section IV]{CaoRatra2023}.

We use the Markov chain Monte Carlo (MCMC) code {\footnotesize MontePython} \cite{Brinckmann2019} for the Bayesian inference analyses with flat (uniform) priors of free parameters listed in Table \ref{tab:priors}. The posterior statistics and visualizations are generated using {\footnotesize GetDist} \cite{Lewis_2019}.

\begin{table}[htbp]
\centering
% \resizebox{\columnwidth}{!}{%
\setlength\tabcolsep{3.3pt}
\begin{threeparttable}
\caption{Flat (uniform) priors of the constrained parameters.}
\label{tab:priors}
\begin{tabular}{lcc}
\toprule\toprule
Parameter & & Prior\\
\midrule
 & Cosmological Parameters & \\
\midrule
$H_0$\,\tnote{a} &  & [None, None]\\
\obhs\,\tnote{b} &  & [0, 1]\\
\ochs\,\tnote{b} &  & [0, 1]\\
\ok &  & [$-2$, 2]\\
$\alpha$ &  & [0, 10]\\
\wx &  & [$-5$, 0.33]\\
\om\,\tnote{c} &  & [0.051314766115, 1]\\
% \midrule
\\
 & $R-L$ Relation Parameters & \\
\midrule
$\beta$ &  & [0, 10]\\
$\gamma$ &  & [0, 5]\\
$\sigma_{\mathrm{int}}$ &  & [0, 5]\\
\bottomrule\bottomrule
\end{tabular}
\begin{tablenotes}
\item [a] \hunit. In the RM AGNs only cases, $H_0=70$ \hunit.
\item [b] Analyses involving $H(z)$ + BAO data. In the RM AGNs only cases $\Omega_{b}=0.05$.
\item [c] RM AGNs only, to ensure that $\Omega_{c}$ remains positive.
\end{tablenotes}
\end{threeparttable}%
% }
\end{table}

\begin{figure*}[htbp]
\centering
 \subfloat[Flat \lcdm]{%
    \includegraphics[width=0.45\textwidth,height=0.35\textwidth]{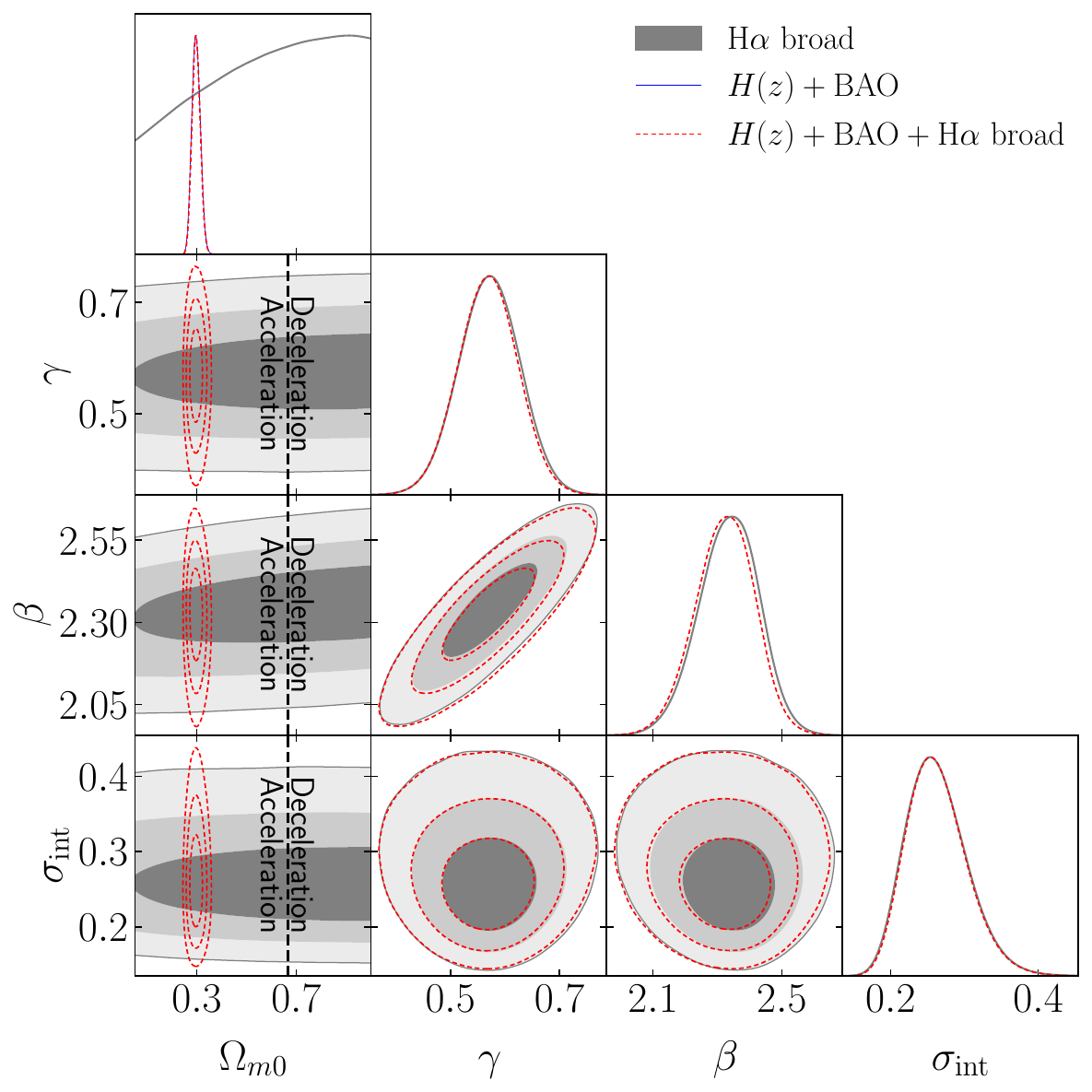}}
 \subfloat[Nonflat \lcdm]{%
    \includegraphics[width=0.45\textwidth,height=0.35\textwidth]{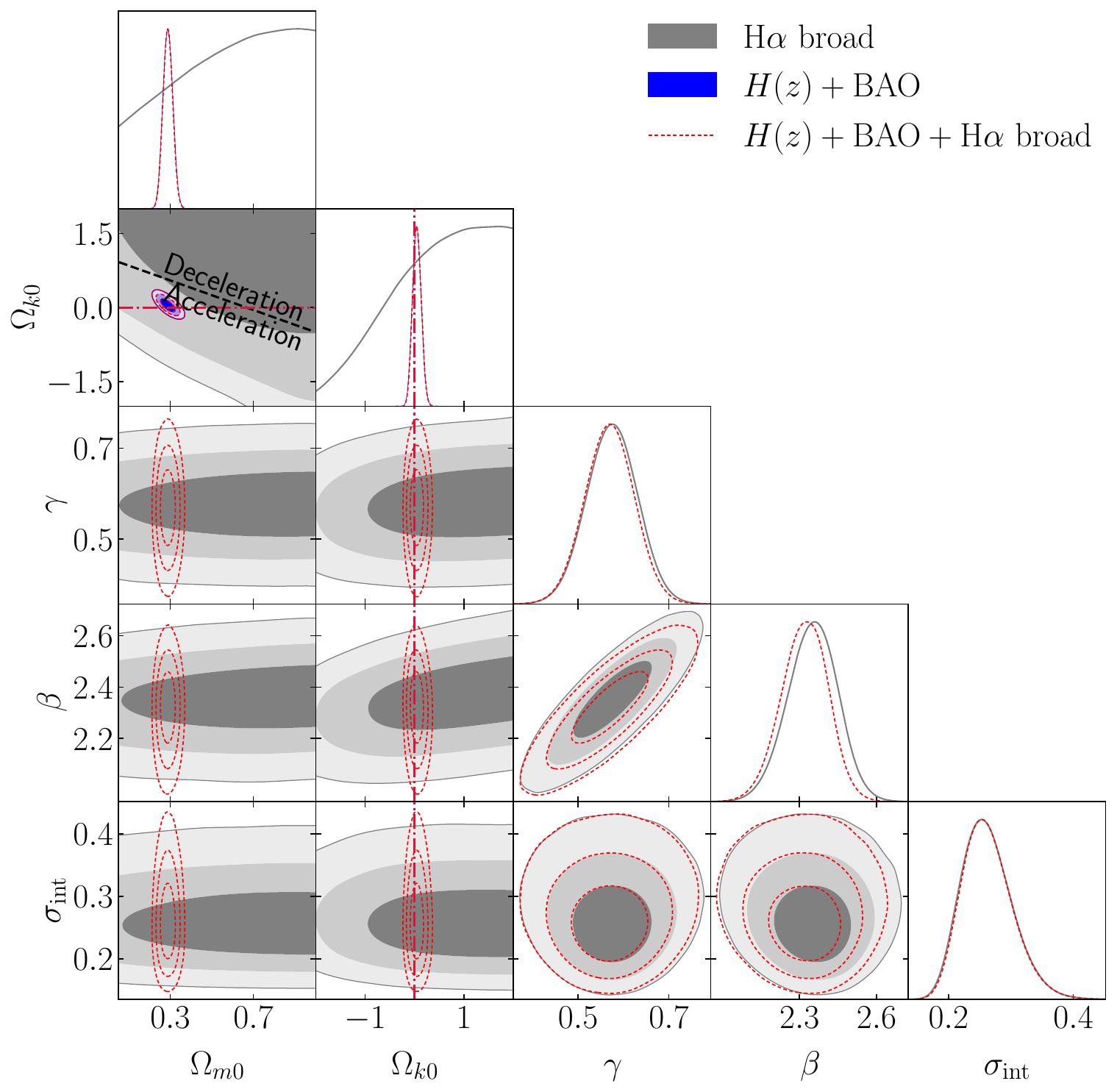}}\\
 \subfloat[Flat XCDM]{%
    \includegraphics[width=0.45\textwidth,height=0.35\textwidth]{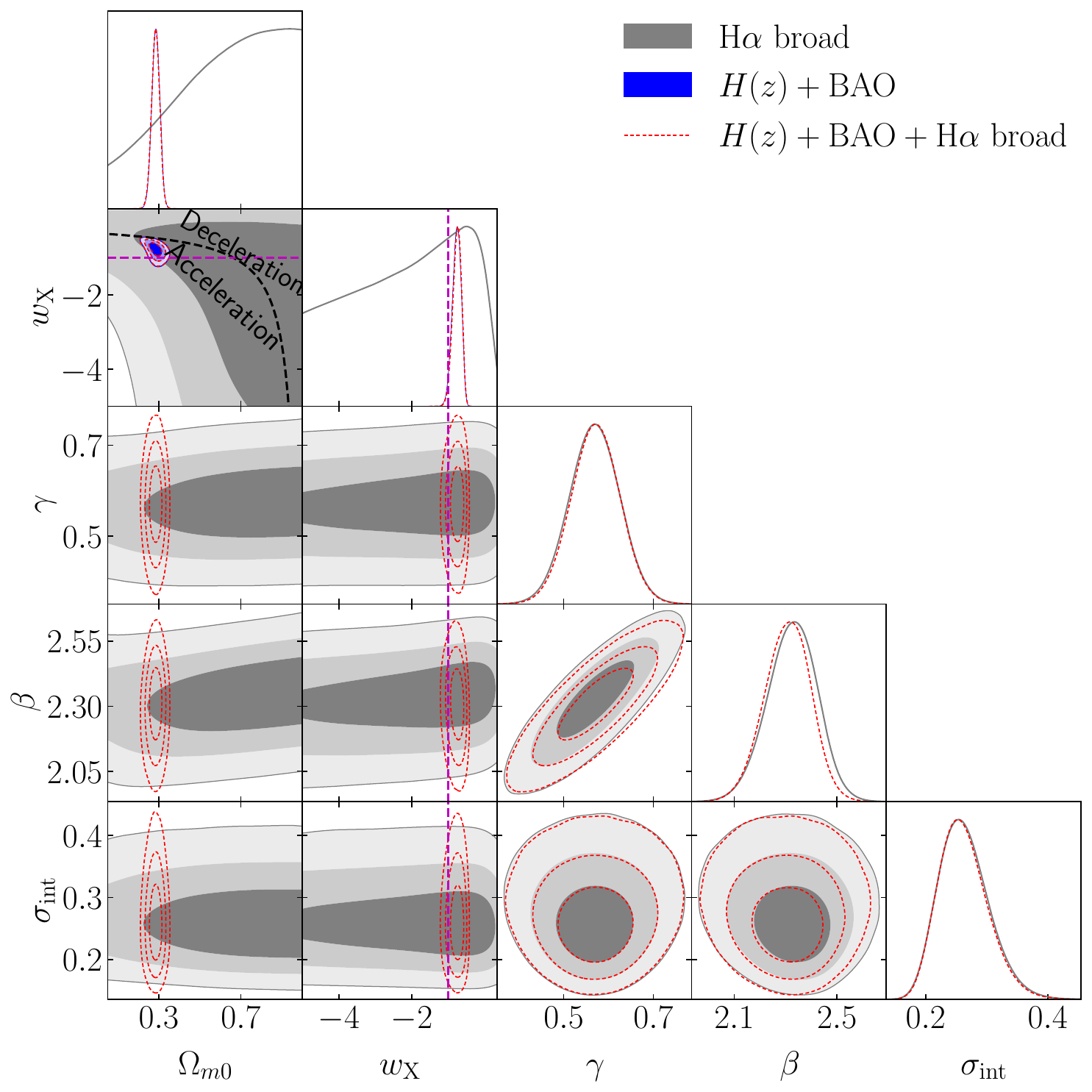}}
 \subfloat[Nonflat XCDM]{%
    \includegraphics[width=0.45\textwidth,height=0.35\textwidth]{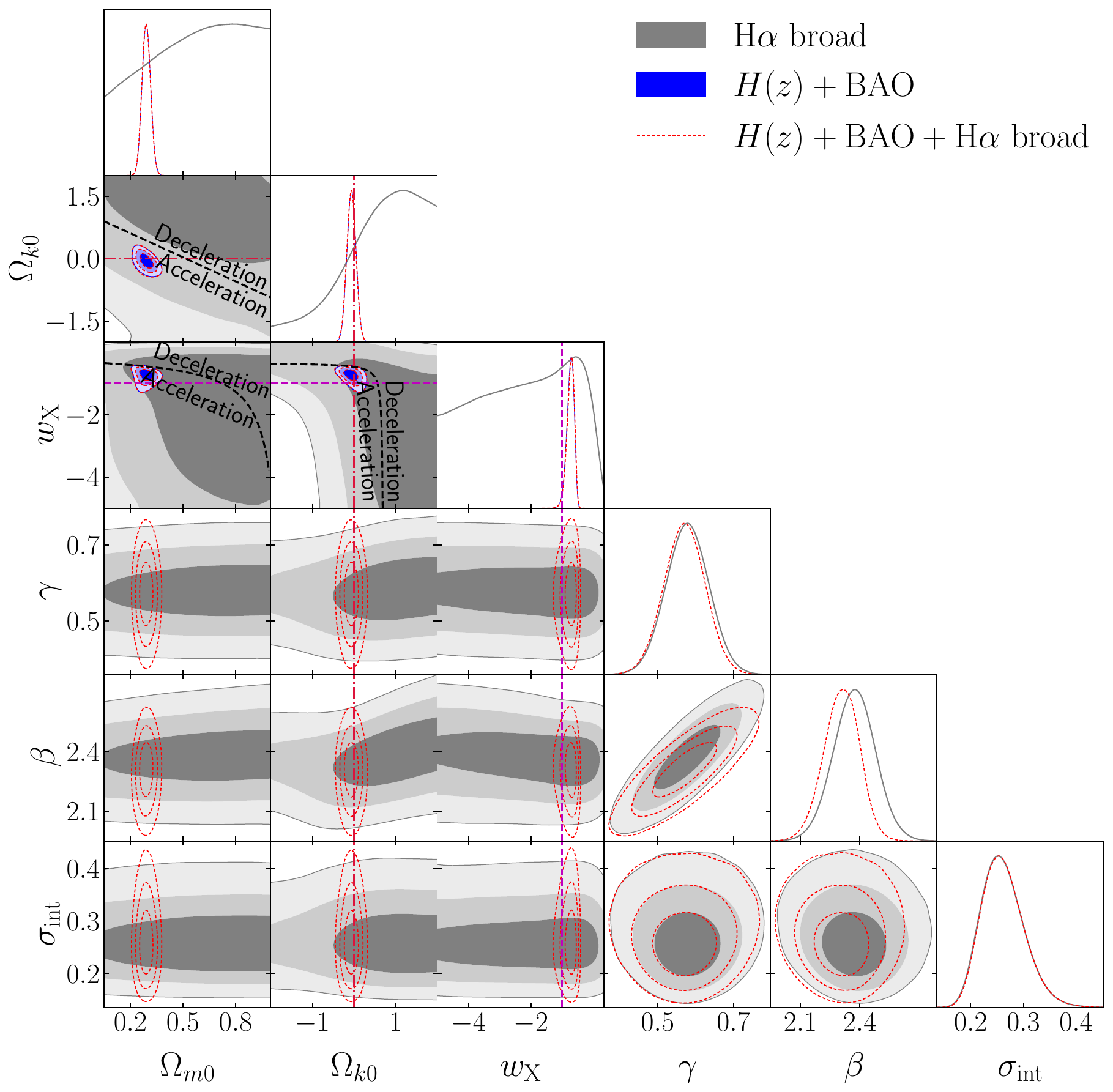}}\\
 \subfloat[Flat \pcdm]{%
    \includegraphics[width=0.45\textwidth,height=0.35\textwidth]{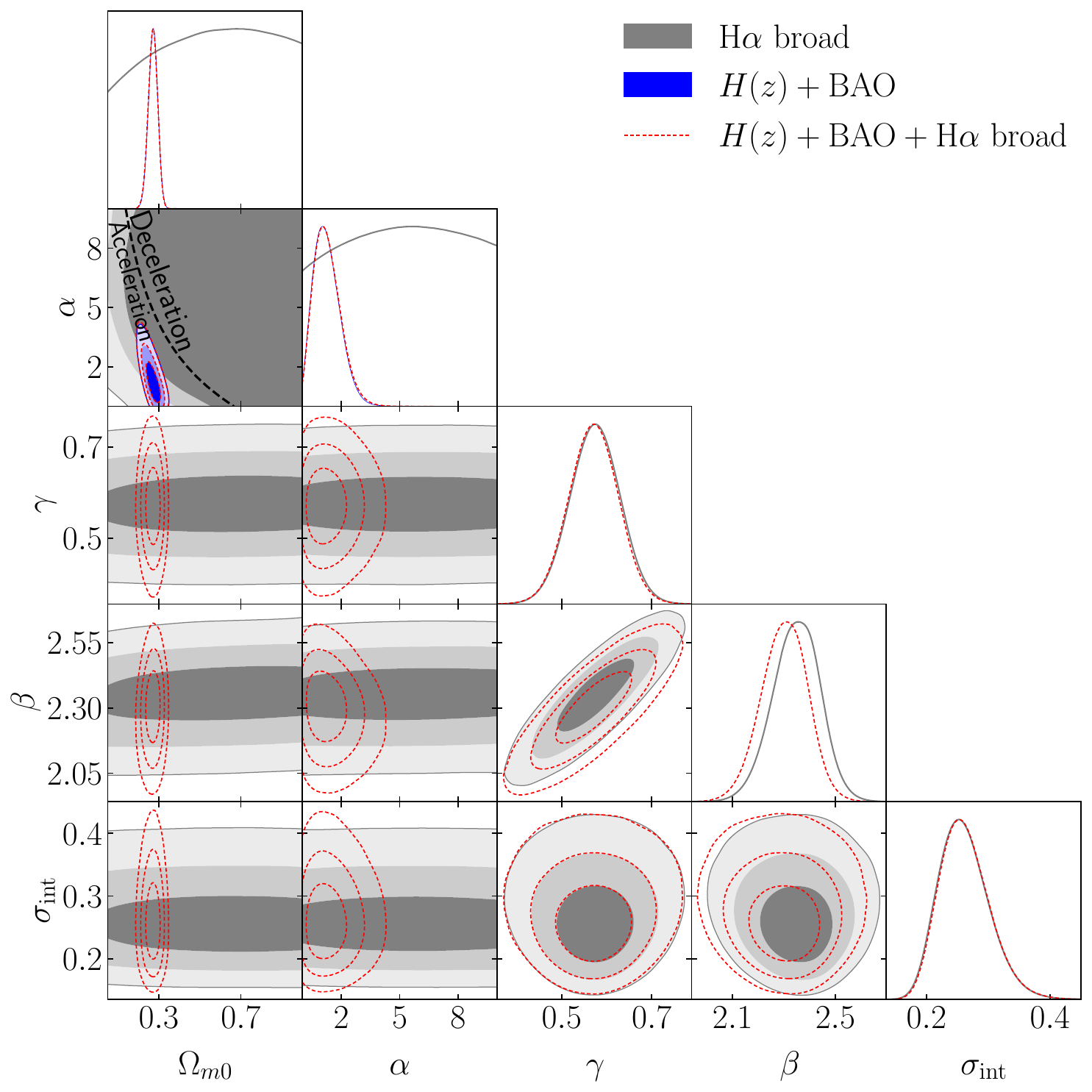}}
 \subfloat[Nonflat \pcdm]{%
    \includegraphics[width=0.45\textwidth,height=0.35\textwidth]{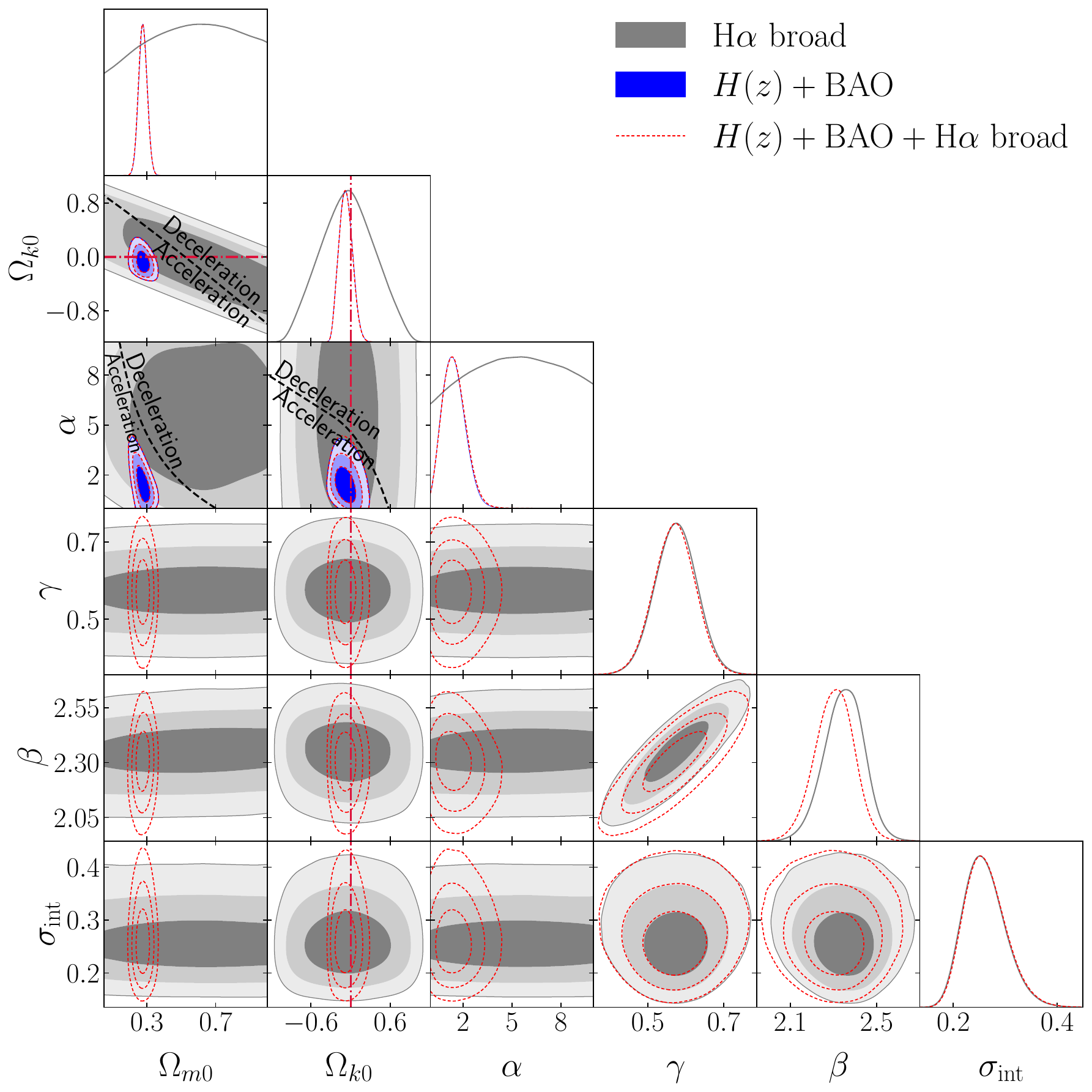}}\\
\caption{One-dimensional likelihoods and 1$\sigma$, 2$\sigma$, and 3$\sigma$ two-dimensional likelihood confidence contours from RM AGN of H$\alpha$ broad (gray), $H(z)$ + BAO (blue), and $H(z)$ + BAO + H$\alpha$ broad (dashed red) data for six different models, with \lcdm, XCDM, and \pcdm\ in the top, middle, and bottom rows, and flat (nonflat) models in the left (right) column. The black dashed zero-acceleration lines, computed for the third cosmological parameter set to the $H(z)$ + BAO data best-fitting values listed in Table \ref{tab:BFP1} in panels (d) and (f), divide the parameter space into regions associated with currently-accelerating (below or below left) and currently-decelerating (above or above right) cosmological expansion. The crimson dash-dot lines represent flat hypersurfaces, with closed spatial hypersurfaces either below or to the left. The magenta lines represent $w_{\rm X}=-1$, i.e.\ flat or nonflat \lcdm\ models. The $\alpha = 0$ axes correspond to flat and nonflat \lcdm\ models in panels (e) and (f), respectively.}
\label{fig1}
\vspace{-50pt}
\end{figure*}

\begin{figure*}
\centering
 \subfloat[Flat \lcdm]{%
    \includegraphics[width=0.4\textwidth,height=0.35\textwidth]{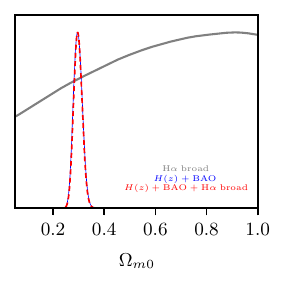}}
 \subfloat[Nonflat \lcdm]{%
    \includegraphics[width=0.4\textwidth,height=0.35\textwidth]{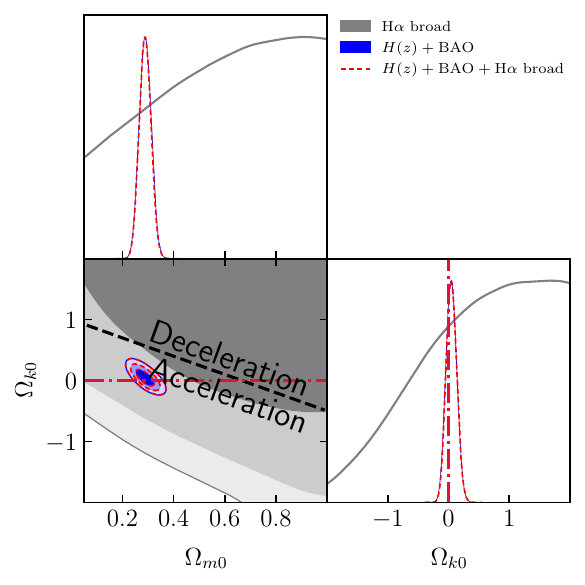}}\\
 \subfloat[Flat XCDM]{%
    \includegraphics[width=0.4\textwidth,height=0.35\textwidth]{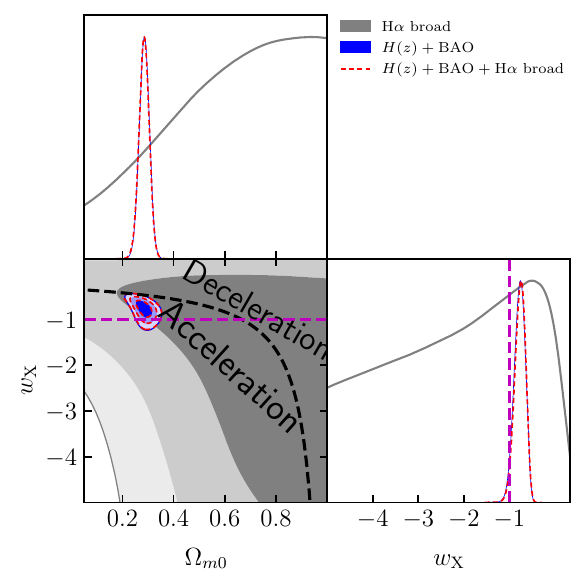}}
 \subfloat[Nonflat XCDM]{%
    \includegraphics[width=0.4\textwidth,height=0.35\textwidth]{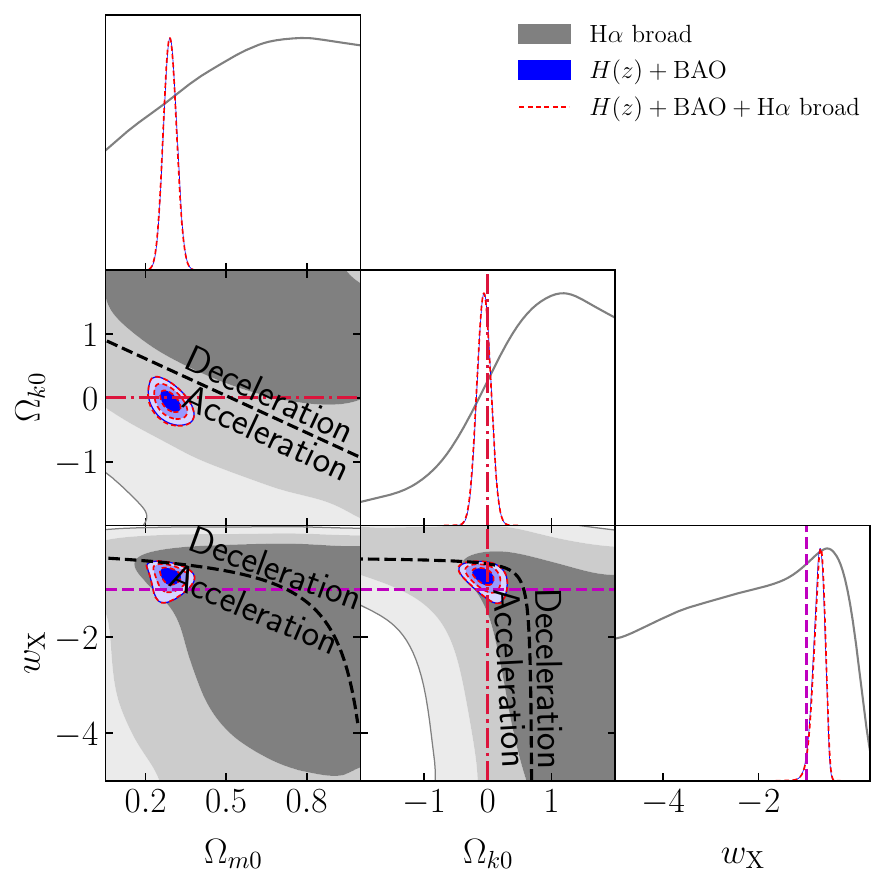}}\\
 \subfloat[Flat \pcdm]{%
    \includegraphics[width=0.4\textwidth,height=0.35\textwidth]{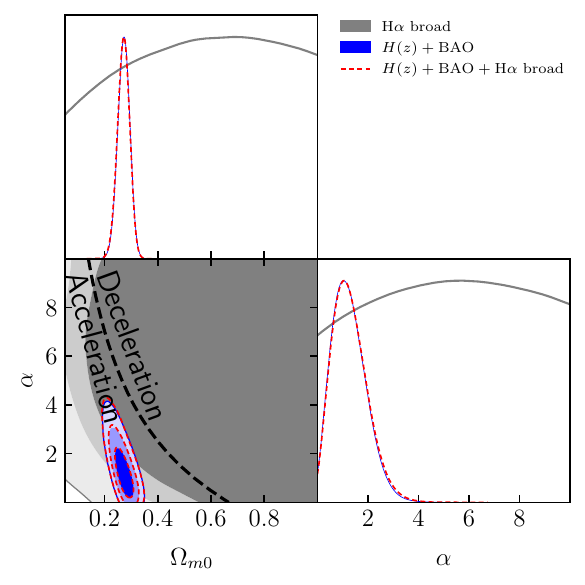}}
 \subfloat[Nonflat \pcdm]{%
    \includegraphics[width=0.4\textwidth,height=0.35\textwidth]{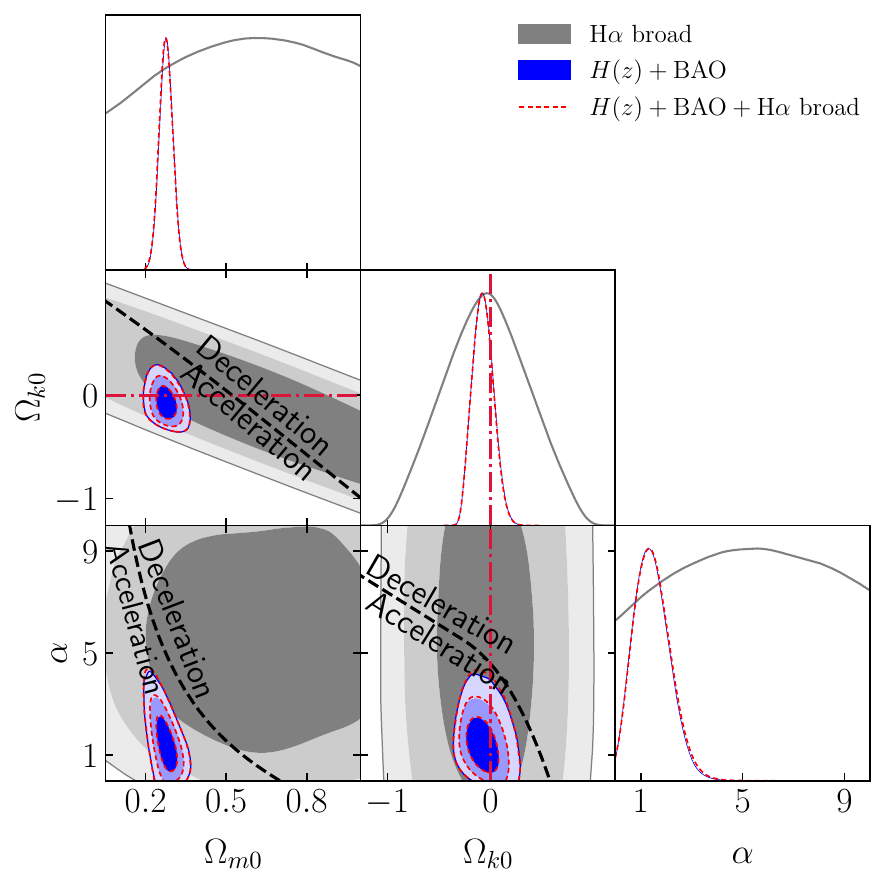}}\\
\caption{Same as Fig.\ \ref{fig1}, but for cosmological parameters only.}
\label{fig2}
\end{figure*}

\begin{turnpage}
\begin{table*}
\centering
% \footnotesize
% \fontsize{8.5pt}{10pt}\selectfont
% \resizebox*{2.7\columnwidth}{1.2\columnwidth}{%
\resizebox{2.7\columnwidth}{!}{%
\begin{threeparttable}
\caption{Unmarginalized best-fitting parameter values for all models from various combinations of data, where H$\alpha$ and H$\beta$ stand for H$\alpha$ and H$\beta$ time delays, respectively, and ``broad'' stands for H$\alpha$ broad luminosity.}\label{tab:BFP1}
\begin{tabular}{lccccccccccccccccc}
\toprule\toprule
Model & dataset & $\Omega_{b}h^2$ & $\Omega_{c}h^2$ & $\Omega_{m0}$ & $\Omega_{k0}$ & $w_{\mathrm{X}}$/$\alpha$\tnote{a} & $H_0$\tnote{b} & $\gamma$ & $\beta$ & $\sigma_{\mathrm{int}}$ & $-2\ln\mathcal{L}_{\mathrm{max}}$ & AIC & BIC & DIC & $\Delta \mathrm{AIC}$ & $\Delta \mathrm{BIC}$ & $\Delta \mathrm{DIC}$ \\
\midrule
 & $H(z)$ + BAO & 0.0254 & 0.1200 & 0.297 & $\cdots$ & $\cdots$ & 70.12 & $\cdots$ & $\cdots$ & $\cdots$ & 30.56 & 36.56 & 41.91 & 37.32 & 0.00 & 0.00 & 0.00\\%36.5576, 41.91016890175479, 37.318529778549525
 & H$\alpha$ broad & $\cdots$ & $\cdots$ & 0.998 & $\cdots$ & $\cdots$ & $\cdots$ & 0.590 & 2.399 & 0.241 & 11.53 & 17.53 & 22.67 & 19.95 & 0.00 & 0.00 & 0.00\\%17.52722, 22.667936200112923, 19.94821112657045
Flat \lcdm & $H(z)$ + BAO + H$\alpha$ broad & 0.0251 & 0.1196 & 0.297 & $\cdots$ & $\cdots$ & 79.97 & 0.590 & 2.361 & 0.241 & 42.99 & 54.99 & 69.64 & 56.56 & 0.00 & 0.00 & 0.00\\%54.9886, 69.6445075389419, 56.56026900850302
 & H$\beta$ broad & $\cdots$ & $\cdots$ & 0.860 & $\cdots$ & $\cdots$ & $\cdots$ & 0.612 & 2.193 & 0.225 & 13.28 & 19.28 & 24.42 & 22.02 & 0.00 & 0.00 & 0.00\\%19.28052, 24.421236200112922, 22.019671458890787
 & $H(z)$ + BAO + H$\beta$ broad & 0.0257 & 0.1204 & 0.297 & $\cdots$ & $\cdots$ & 70.32 & 0.602 & 2.155 & 0.228 & 44.07 & 56.07 & 70.73 & 58.67 & 0.00 & 0.00 & 0.00\\%56.0734, 70.7293075389419, 58.67416989032812
[6pt]
 & $H(z)$ + BAO & 0.0269 & 0.1128 & 0.289 & 0.041 & $\cdots$ & 69.61 & $\cdots$ & $\cdots$ & $\cdots$ & 30.34 & 38.34 & 45.48 & 38.80 & 1.78 & 3.56 & 1.48\\%38.3384, 45.475158535673046, 38.79768870799054
 & H$\alpha$ broad & $\cdots$ & $\cdots$ & 0.969 & 1.846 & $\cdots$ & $\cdots$ & 0.604 & 2.450 & 0.237 & 11.26 & 19.26 & 26.12 & 19.81 & 1.74 & 3.45 & $-0.14$\\%19.2633, 26.11758826681723, 19.805921719151428
Nonflat \lcdm & $H(z)$ + BAO + H$\alpha$ broad & 0.0268 & 0.1109 & 0.288 & 0.048 & $\cdots$ & 69.27 & 0.593 & 2.359 & 0.242 & 42.78 & 56.78 & 73.88 & 58.14 & 1.79 & 4.23 & 1.58\\%56.7808, 73.87935879543221, 58.144048463424085
 & H$\beta$ broad & $\cdots$ & $\cdots$ & 0.329 & 1.870 & $\cdots$ & $\cdots$ & 0.626 & 2.225 & 0.225 & 13.17 & 21.17 & 28.02 & 22.11 & 1.88 & 3.60 & 0.09\\%21.16512, 28.01940826681723, 22.114025600507624
 & $H(z)$ + BAO + H$\beta$ broad & 0.0264 & 0.1125 & 0.290 & 0.038 & $\cdots$ & 69.36 & 0.603 & 2.150 & 0.221 & 43.87 & 57.87 & 74.96 & 60.28 & 1.79 & 4.24 & 1.61\\%57.866, 74.96455879543223, 60.28421758371914
[6pt]
 & $H(z)$ + BAO & 0.0320 & 0.0932 & 0.283 & $\cdots$ & $-0.731$ & 66.69 & $\cdots$ & $\cdots$ & $\cdots$ & 26.57 & 34.57 & 41.71 & 34.52 & $-1.98$ & $-0.20$ & $-2.80$\\%34.5736, 41.710358535673045, 34.51703264021917
 & H$\alpha$ broad & $\cdots$ & $\cdots$ & 0.140 & $\cdots$ & 0.096 & $\cdots$ & 0.592 & 2.403 & 0.241 & 11.44 & 19.44 & 26.30 & 20.29 & 1.92 & 3.63 & 0.35\\%19.44416, 26.298448266817232, 20.294164461239152
Flat XCDM & $H(z)$ + BAO + H$\alpha$ broad & 0.0320 & 0.0901 & 0.281 & $\cdots$ & $-0.714$ & 66.12 & 0.589 & 2.344 & 0.237 & 38.78 & 52.78 & 69.87 & 53.67 & $-2.21$ & 0.23 & $-2.89$\\%52.7754, 69.87395879543222, 53.66856126409743
 & H$\beta$ broad & $\cdots$ & $\cdots$ & 0.089 & $\cdots$ & 0.101 & $\cdots$ & 0.628 & 2.213 & 0.226 & 13.25 & 21.25 & 28.11 & 22.13 & 1.97 & 3.68 & 0.11\\%21.25132, 28.10560826681723, 22.127932119914306
 & $H(z)$ + BAO + H$\beta$ broad & 0.0323 & 0.0932 & 0.283 & $\cdots$ & $-0.726$ & 66.73 & 0.608 & 2.144 & 0.219 & 40.03 & 54.03 & 71.13 & 55.87 & $-2.05$ & 0.40 & $-2.80$\\%54.028, 71.12655879543222, 55.874731623904445
[6pt]
 & $H(z)$ + BAO & 0.0312 & 0.0990 & 0.293 & $-0.085$ & $-0.693$ & 66.84 & $\cdots$ & $\cdots$ & $\cdots$ & 26.00 & 36.00 & 44.92 & 36.17 & $-0.56$ & 3.01 & $-1.15$\\%35.996, 44.91694816959131, 36.17275325321086
 & H$\alpha$ broad & $\cdots$ & $\cdots$ & 0.158 & 1.716 & $-4.752$ & $\cdots$ & 0.616 & 2.488 & 0.238 & 11.23 & 21.23 & 29.80 & 19.36 & 3.70 & 7.13 & $-0.59$\\%21.23008, 29.797940333521538, 19.359387630335448
Nonflat XCDM & $H(z)$ + BAO + H$\alpha$ broad & 0.0326 & 0.0970 & 0.289 & $-0.074$ & $-0.692$ & 67.16 & 0.586 & 2.353 & 0.234 & 38.30 & 54.30 & 73.84 & 54.99 & $-0.69$ & 4.19 & $-1.57$\\%54.2982, 73.83941005192253, 54.98592124209937
 & H$\beta$ broad & $\cdots$ & $\cdots$ & 0.094 & 1.519 & $-2.027$ & $\cdots$ & 0.625 & 2.223 & 0.226 & 13.17 & 23.17 & 31.73 & 22.32 & 3.88 & 7.31 & 0.30\\%23.1653, 31.73316033352154, 22.321314754497074
 & $H(z)$ + BAO + H$\beta$ broad & 0.0311 & 0.0965 & 0.292 & $-0.088$ & $-0.683$ & 66.34 & 0.613 & 2.159 & 0.221 & 39.50 & 55.50 & 75.04 & 57.26 & $-0.58$ & 4.31 & $-1.41$\\%55.4984, 75.03961005192252, 57.259642029708836
[6pt]
 & $H(z)$ + BAO & 0.0337 & 0.0864 & 0.271 & $\cdots$ & 1.169 & 66.78 & $\cdots$ & $\cdots$ & $\cdots$ & 26.50 & 34.50 & 41.64 & 34.01 & $-2.05$ & $-0.27$ & $-3.31$\\%34.5036, 41.640358535673045, 34.00595987367331
 & H$\alpha$ broad & $\cdots$ & $\cdots$ & 0.992 & $\cdots$ & 8.355& $\cdots$ & 0.594 & 2.399 & 0.239 & 11.51 & 19.51 & 26.36 & 19.29 & 1.98 & 3.70 & $-0.66$\\%19.50974, 26.364028266817233, 19.286379353077514
Flat \pcdm & $H(z)$ + BAO + H$\alpha$ broad & 0.0358 & 0.0832 & 0.267 & $\cdots$ & 1.337 & 66.96 & 0.590 & 2.347 & 0.241 & 38.71 & 52.71 & 69.81 & 52.89 & $-2.27$ & 0.17 & $-3.67$\\%52.7144, 69.81295879543222, 52.89155991353178
 & H$\beta$ broad & $\cdots$ & $\cdots$ & 0.833 & $\cdots$ & 0.617 & $\cdots$ & 0.613 & 2.195 & 0.225 & 13.28 & 21.28 & 28.14 & 21.70 & 2.00 & 3.71 & $-0.32$\\%21.2808, 28.13508826681723, 21.70147359229677
 & $H(z)$ + BAO + H$\beta$ broad & 0.0336 & 0.0864 & 0.272 & $\cdots$ & 1.203 & 66.63 & 0.602 & 2.137 & 0.226 & 39.94 & 53.94 & 71.04 & 55.27 & $-2.14$ & 0.31 & $-3.40$\\%53.9368, 71.03535879543222, 55.274265960077166
[6pt]
 & $H(z)$ + BAO & 0.0338 & 0.0878 & 0.273 & $-0.077$ & 1.441 & 66.86 & $\cdots$ & $\cdots$ & $\cdots$ & 25.92 & 35.92 & 44.84 & 35.12 & $-0.64$ & 2.93 & $-2.20$\\%35.9212, 44.842148169591304, 35.12310308752016 D
 & H$\alpha$ broad & $\cdots$ & $\cdots$ & 0.914 & $-0.741$ & 9.649 & $\cdots$ & 0.593 & 2.408 & 0.240 & 11.43 & 21.43 & 30.00 & 19.20 & 3.90 & 7.33 & $-0.74$\\%21.4305, 29.99836033352154, 19.203995446070223 D
Nonflat \pcdm & $H(z)$ + BAO + H$\alpha$ broad & 0.0354 & 0.0861 & 0.270 & $-0.085$ & 1.592 & 67.29 & 0.586 & 2.353 & 0.250 & 38.30 & 54.30 & 73.84 & 53.86 & $-0.69$ & 4.19 & $-2.71$\\%54.2968, 73.83801005192254, 53.85510462548766 D
 & H$\beta$ broad & $\cdots$ & $\cdots$ & 0.969 & $-0.715$ & 8.040 & $\cdots$ & 0.627 & 2.214 & 0.228 & 13.24 & 23.24 & 31.81 & 21.63 & 3.96 & 7.38 & $-0.38$\\%23.23814, 31.80600033352154, 21.634976021830433 D
 & $H(z)$ + BAO + H$\beta$ broad & 0.0336 & 0.0910 & 0.278 & $-0.099$ & 1.474 & 67.06 & 0.610 & 2.144 & 0.221 & 39.41 & 55.41 & 74.95 & 56.25 & $-0.66$ & 4.22 & $-2.42$\\%55.4118, 74.95301005192253, 56.25309505425841 D
\bottomrule\bottomrule
\end{tabular}
%}
\begin{tablenotes}
\item [a] \wx\ corresponds to flat/nonflat XCDM and $\alpha$ corresponds to flat/nonflat \pcdm.
\item [b] \hunit.
% \item [c] $\Omega_b=0.05$ and $H_0=70$ \hunit.
\end{tablenotes}
\end{threeparttable}%
}
\end{table*}
\end{turnpage}

\begin{turnpage}
\begin{table*}
\centering
% \footnotesize
% \resizebox*{2.7\columnwidth}{1.2\columnwidth}{%
\resizebox{2.7\columnwidth}{!}{%
\begin{threeparttable}
\caption{One-dimensional marginalized posterior mean values and uncertainties ($\pm 1\sigma$ error bars or $2\sigma$ limits) of the parameters for all models from various combinations of data, where H$\alpha$ and H$\beta$ stand for H$\alpha$ and H$\beta$ time delays, respectively, and ``broad'' stands for H$\alpha$ broad luminosity.}\label{tab:1d_BFP1}
\begin{tabular}{lcccccccccc}
\toprule\toprule
Model & dataset & $\Omega_{b}h^2$ & $\Omega_{c}h^2$ & $\Omega_{m0}$ & $\Omega_{k0}$ & $w_{\mathrm{X}}$/$\alpha$\tnote{a} & $H_0$\tnote{b} & $\gamma$ & $\beta$ & $\sigma_{\mathrm{int}}$\\
\midrule
 & $H(z)$ + BAO & $0.0260\pm0.0040$ & $0.1213^{+0.0091}_{-0.0103}$ & $0.298^{+0.015}_{-0.018}$ & $\cdots$ & $\cdots$ & $70.51\pm2.72$ & $\cdots$ & $\cdots$ & $\cdots$ \\
 & H$\alpha$ broad & $\cdots$ & $\cdots$ & $>0.424$\tnote{c} & $\cdots$ & $\cdots$ & $\cdots$ & $0.571\pm0.055$ & $2.336^{+0.096}_{-0.086}$ & $0.263^{+0.033}_{-0.046}$ \\
Flat \lcdm & $H(z)$ + BAO + H$\alpha$ broad & $0.0259\pm0.0038$ & $0.1212^{+0.0088}_{-0.0098}$ & $0.298^{+0.015}_{-0.017}$ & $\cdots$ & $\cdots$ & $70.45\pm2.62$ & $0.569\pm0.053$ & $2.323^{+0.095}_{-0.084}$ & $0.263^{+0.032}_{-0.045}$ \\
 & H$\beta$ broad & $\cdots$ & $\cdots$ & $\cdots$ & $\cdots$ & $\cdots$ & $\cdots$ & $0.616\pm0.053$ & $2.184^{+0.095}_{-0.087}$ & $0.248^{+0.032}_{-0.046}$ \\
 & $H(z)$ + BAO + H$\beta$ broad & $0.0259^{+0.0037}_{-0.0040}$ & $0.1212^{+0.0088}_{-0.0099}$ & $0.298^{+0.015}_{-0.017}$ & $\cdots$ & $\cdots$ & $70.46^{+2.62}_{-2.61}$ & $0.612\pm0.051$ & $2.172^{+0.093}_{-0.084}$ & $0.247^{+0.031}_{-0.045}$ \\
[6pt]
 & $H(z)$ + BAO & $0.0275^{+0.0047}_{-0.0053}$ & $0.1132\pm0.0183$ & $0.289\pm0.023$ & $0.047^{+0.083}_{-0.091}$ & $\cdots$ & $69.81\pm2.87$ & $\cdots$ & $\cdots$ & $\cdots$ \\
 & H$\alpha$ broad & $\cdots$ & $\cdots$ & $>0.438$\tnote{c} & $>-1.280$ & $\cdots$ & $\cdots$ & $0.575\pm0.055$ & $2.351^{+0.100}_{-0.091}$ & $0.262^{+0.032}_{-0.046}$ \\
Nonflat \lcdm & $H(z)$ + BAO + H$\alpha$ broad & $0.0275^{+0.0046}_{-0.0052}$ & $0.1128\pm0.0178$ & $0.289\pm0.023$ & $0.048^{+0.081}_{-0.089}$ & $\cdots$ & $69.75\pm2.80$ & $0.569\pm0.053$ & $2.319^{+0.094}_{-0.083}$ & $0.263^{+0.032}_{-0.045}$ \\
 & H$\beta$ broad & $\cdots$ & $\cdots$ & $>0.423$\tnote{c} & $0.323^{+1.496}_{-0.696}$ & $\cdots$ & $\cdots$ & $0.618\pm0.053$ & $2.194^{+0.094}_{-0.096}$ & $0.248^{+0.032}_{-0.046}$ \\
 & $H(z)$ + BAO + H$\beta$ broad & $0.0275^{+0.0046}_{-0.0052}$ & $0.1130\pm0.0179$ & $0.289\pm0.023$ & $0.047^{+0.081}_{-0.089}$ & $\cdots$ & $69.79^{+2.79}_{-2.80}$ & $0.613\pm0.051$ & $2.169^{+0.091}_{-0.083}$ & $0.247^{+0.031}_{-0.045}$ \\
[6pt]
 & $H(z)$ + BAO & $0.0308^{+0.0053}_{-0.0046}$ & $0.0980^{+0.0182}_{-0.0161}$ & $0.286\pm0.019$ & $\cdots$ & $-0.778^{+0.132}_{-0.104}$ & $67.20^{+3.05}_{-3.06}$ & $\cdots$ & $\cdots$ & $\cdots$ \\
 & H$\alpha$ broad & $\cdots$ & $\cdots$ & $>0.501$\tnote{c} & $\cdots$ & $-2.103^{+2.238}_{-0.891}$ & $\cdots$ & $0.569\pm0.055$ & $2.325^{+0.101}_{-0.088}$ & $0.264^{+0.033}_{-0.046}$ \\
Flat XCDM & $H(z)$ + BAO + H$\alpha$ broad & $0.0311^{+0.0054}_{-0.0046}$ & $0.0967^{+0.0183}_{-0.0161}$ & $0.285\pm0.019$ & $\cdots$ & $-0.768^{+0.131}_{-0.100}$ & $67.05\pm2.98$ & $0.571\pm0.053$ & $2.307^{+0.093}_{-0.083}$ & $0.262^{+0.032}_{-0.045}$ \\
 & H$\beta$ broad & $\cdots$ & $\cdots$ & $>0.446$\tnote{c} & $\cdots$ & $-2.251^{+2.362}_{-1.032}$ & $\cdots$ & $0.611\pm0.053$ & $2.166^{+0.100}_{-0.092}$ & $0.248^{+0.032}_{-0.046}$ \\
 & $H(z)$ + BAO + H$\beta$ broad & $0.0310^{+0.0054}_{-0.0046}$ & $0.0971^{+0.0182}_{-0.0160}$ & $0.285\pm0.019$ & $\cdots$ & $-0.771^{+0.131}_{-0.101}$ & $67.10\pm2.98$ & $0.615\pm0.051$ & $2.155^{+0.091}_{-0.082}$ & $0.247^{+0.031}_{-0.045}$ \\
[6pt]
 & $H(z)$ + BAO & $0.0305^{+0.0055}_{-0.0047}$ & $0.1011\pm0.0196$ & $0.292\pm0.024$ & $-0.059\pm0.106$ & $-0.746^{+0.135}_{-0.090}$ & $67.17^{+2.96}_{-2.97}$ & $\cdots$ & $\cdots$ & $\cdots$ \\
 & H$\alpha$ broad & $\cdots$ & $\cdots$ & $>0.425$\tnote{c} & $>-1.174$ & $-2.231^{+2.281}_{-0.939}$ & $\cdots$ & $0.579\pm0.055$ & $2.373\pm0.103$ & $0.262^{+0.032}_{-0.045}$ \\
Nonflat XCDM & $H(z)$ + BAO + H$\alpha$ broad & $0.0307^{+0.0057}_{-0.0047}$ & $0.1000\pm0.0196$ & $0.291\pm0.024$ & $-0.062^{+0.107}_{-0.108}$ & $-0.737^{+0.136}_{-0.088}$ & $67.04^{+2.93}_{-2.94}$ & $0.571\pm0.053$ & $2.308^{+0.092}_{-0.083}$ & $0.262^{+0.032}_{-0.045}$ \\
 & H$\beta$ broad & $\cdots$ & $\cdots$ & $>0.406$\tnote{c} & $0.480^{+1.219}_{-0.657}$ & $-2.135^{+2.163}_{-0.885}$ & $\cdots$ & $0.620\pm0.054$ & $2.208\pm0.106$ & $0.248^{+0.032}_{-0.046}$ \\
 & $H(z)$ + BAO + H$\beta$ broad & $0.0306^{+0.0057}_{-0.0048}$ & $0.1004^{+0.0196}_{-0.0197}$ & $0.291\pm0.024$ & $-0.061\pm0.107$ & $-0.740^{+0.137}_{-0.089}$ & $67.10^{+2.94}_{-2.95}$ & $0.615\pm0.051$ & $2.155^{+0.091}_{-0.082}$ & $0.246^{+0.032}_{-0.044}$ \\
[6pt]
 & $H(z)$ + BAO & $0.0327^{+0.0060}_{-0.0031}$ & $0.0866^{+0.0192}_{-0.0176}$ & $0.272\pm0.022$ & $\cdots$ & $1.261^{+0.494}_{-0.810}$ & $66.23^{+2.87}_{-2.86}$ & $\cdots$ & $\cdots$ & $\cdots$ \\%$1.261^{+1.276}_{-1.206}$ 2$\sigma$
 & H$\alpha$ broad & $\cdots$ & $\cdots$ & $\cdots$ & $\cdots$ & $\cdots$ & $\cdots$ & $0.574\pm0.054$ & $2.348^{+0.094}_{-0.084}$ & $0.262^{+0.032}_{-0.046}$ \\
Flat \pcdm & $H(z)$ + BAO + H$\alpha$ broad & $0.0329^{+0.0069}_{-0.0023}$ & $0.0859^{+0.0195}_{-0.0178}$ & $0.271\pm0.022$ & $\cdots$ & $1.290^{+0.496}_{-0.829}$ & $66.21^{+2.86}_{-2.84}$ & $0.571\pm0.053$ & $2.301^{+0.092}_{-0.083}$ & $0.263^{+0.031}_{-0.045}$ \\
 & H$\beta$ broad & $\cdots$ & $\cdots$ & $\cdots$ & $\cdots$ & $\cdots$ & $\cdots$ & $0.619^{+0.054}_{-0.050}$ & $2.197^{+0.094}_{-0.085}$ & $0.247^{+0.032}_{-0.046}$ \\
 & $H(z)$ + BAO + H$\beta$ broad & $0.0329^{+0.0068}_{-0.0023}$ & $0.0863^{+0.0195}_{-0.0177}$ & $0.272^{+0.023}_{-0.021}$ & $\cdots$ & $1.274^{+0.491}_{-0.827}$ & $66.26^{+2.86}_{-2.85}$ & $0.615\pm0.051$ & $2.149^{+0.091}_{-0.082}$ & $0.247^{+0.031}_{-0.045}$ \\%$0.271547^{+0.023024}_{-0.021147}$
[6pt]
 & $H(z)$ + BAO & $0.0324^{+0.0062}_{-0.0031}$ & $0.0900\pm0.0200$ & $0.277\pm0.025$ & $-0.072^{+0.093}_{-0.107}$ & $1.435^{+0.579}_{-0.788}$ & $66.50\pm2.88$ & $\cdots$ & $\cdots$ & $\cdots$ \\%$1.435^{+1.319}_{-1.290}$
 & H$\alpha$ broad & $\cdots$ & $\cdots$ & $\cdots$ & $-0.041\pm0.411$ & $\cdots$ & $\cdots$ & $0.574\pm0.054$ & $2.348^{+0.094}_{-0.083}$ & $0.261^{+0.032}_{-0.045}$ \\
Nonflat \pcdm & $H(z)$ + BAO + H$\alpha$ broad & $0.0327^{+0.0070}_{-0.0023}$ & $0.0894\pm0.0203$ & $0.276\pm0.025$ & $-0.073^{+0.092}_{-0.107}$ & $1.464^{+0.584}_{-0.811}$ & $66.50^{+2.87}_{-2.86}$ & $0.571\pm0.053$ & $2.304^{+0.091}_{-0.082}$ & $0.262^{+0.032}_{-0.045}$ \\
 & H$\beta$ broad & $\cdots$ & $\cdots$ & $\cdots$ & $-0.027\pm0.410$ & $\cdots$ & $\cdots$ & $0.619^{+0.054}_{-0.050}$ & $2.199^{+0.092}_{-0.085}$ & $0.247^{+0.031}_{-0.046}$ \\
 & $H(z)$ + BAO + H$\beta$ broad & $0.0326^{+0.0071}_{-0.0025}$ & $0.0897^{+0.0202}_{-0.0203}$ & $0.276\pm0.025$ & $-0.073^{+0.092}_{-0.107}$ & $1.450^{+0.582}_{-0.812}$ & $66.52\pm2.86$ & $0.615^{+0.050}_{-0.051}$ & $2.151^{+0.089}_{-0.082}$ & $0.246^{+0.031}_{-0.044}$ \\
\bottomrule\bottomrule
\end{tabular}
\begin{tablenotes}
\item [a] \wx\ corresponds to flat/nonflat XCDM and $\alpha$ corresponds to flat/nonflat \pcdm.
\item [b] \hunit. For the RM AGN alone cases, $\Omega_b=0.05$ and $H_0=70$ \hunit.
\item [c] This is the 1$\sigma$ limit. The 2$\sigma$ limit is set by the prior and not shown here.
\end{tablenotes}
\end{threeparttable}%
}
\end{table*}
\end{turnpage}

\begin{turnpage}
\begin{table*}
\centering
% \footnotesize
% \fontsize{8.5pt}{10pt}\selectfont
% \resizebox*{2.7\columnwidth}{1.2\columnwidth}{%
\resizebox{2.7\columnwidth}{!}{%
\begin{threeparttable}
\caption{Unmarginalized best-fitting parameter values for all models from various combinations of data, where H$\alpha$ and H$\beta$ stand for H$\alpha$ and H$\beta$ time delays, respectively, and ``mono'' stands for monochromatic luminosity at 5100 \AA.}\label{tab:BFP2}
\begin{tabular}{lccccccccccccccccc}
\toprule\toprule
Model & dataset & $\Omega_{b}h^2$ & $\Omega_{c}h^2$ & $\Omega_{m0}$ & $\Omega_{k0}$ & $w_{\mathrm{X}}$/$\alpha$\tnote{a} & $H_0$\tnote{b} & $\gamma$ & $\beta$ & $\sigma_{\mathrm{int}}$ & $-2\ln\mathcal{L}_{\mathrm{max}}$ & AIC & BIC & DIC & $\Delta \mathrm{AIC}$ & $\Delta \mathrm{BIC}$ & $\Delta \mathrm{DIC}$ \\
\midrule
 & $H(z)$ + BAO & 0.0254 & 0.1200 & 0.297 & $\cdots$ & $\cdots$ & 70.12 & $\cdots$ & $\cdots$ & $\cdots$ & 30.56 & 36.56 & 41.91 & 37.32 & 0.00 & 0.00 & 0.00\\%36.5576, 41.91016890175479, 37.318529778549525
 & H$\alpha$ mono & $\cdots$ & $\cdots$ & 0.999 & $\cdots$ & $\cdots$ & $\cdots$ & 0.547 & 1.622 & 0.256 & 17.64 & 23.64 & 28.79 & 26.87 & 0.00 & 0.00 & 0.00\\%23.64474, 28.78545620011292, 26.874259621065924
Flat \lcdm & $H(z)$ + BAO + H$\alpha$ mono & 0.0249 & 0.1191 & 0.298 & $\cdots$ & $\cdots$ & 69.63 & 0.514 & 1.561 & 0.266 & 50.02 & 62.02 & 76.67 & 63.69 & 0.00 & 0.00 & 0.00\\%62.0158, 76.6717075389419, 63.685324988355404
 & H$\beta$ mono & $\cdots$ & $\cdots$ & 0.997 & $\cdots$ & $\cdots$ & $\cdots$ & 0.578 & 1.420 & 0.247 & 19.14 & 25.14 & 30.28 & 27.80 & 0.00 & 0.00 & 0.00\\%25.13752, 30.27823620011292, 27.79852067820409
 & $H(z)$ + BAO + H$\beta$ mono & 0.0254 & 0.1207 & 0.297 & $\cdots$ & $\cdots$ & 70.24 & 0.567 & 1.392 & 0.253 & 50.67 & 62.67 & 77.33 & 64.53 & 0.00 & 0.00 & 0.00\\%62.6696, 77.3255075389419, 64.52560471698241
[6pt]
 & $H(z)$ + BAO & 0.0269 & 0.1128 & 0.289 & 0.041 & $\cdots$ & 69.61 & $\cdots$ & $\cdots$ & $\cdots$ & 30.34 & 38.34 & 45.48 & 38.80 & 1.78 & 3.56 & 1.48\\%38.3384, 45.475158535673046, 38.79768870799054
 & H$\alpha$ mono & $\cdots$ & $\cdots$ & 0.994 & 1.962 & $\cdots$ & $\cdots$ & 0.568 & 1.663 & 0.256 & 16.40 & 24.40 & 31.25 & 26.56 & 0.75 & 2.46 & $-0.32$\\%24.39544, 31.249728266817232, 26.55637191393592
Nonflat \lcdm & $H(z)$ + BAO + H$\alpha$ mono & 0.0258 & 0.1114 & 0.290 & 0.037 & $\cdots$ & 68.97 & 0.510 & 1.559 & 0.275 & 49.85 & 63.85 & 80.95 & 65.11 & 1.83 & 4.27 & 1.43\\%63.8472, 80.94575879543223, 65.11180298178371
 & H$\beta$ mono & $\cdots$ & $\cdots$ & 0.983 & 1.959 & $\cdots$ & $\cdots$ & 0.604 & 1.464 & 0.243 & 18.56 & 26.56 & 33.42 & 27.71 & 1.43 & 3.14 & $-0.08$\\%26.56358, 33.41786826681724, 27.71453305585495
 & $H(z)$ + BAO + H$\beta$ mono & 0.0271 & 0.1107 & 0.288 & 0.052 & $\cdots$ & 69.35 & 0.570 & 1.381 & 0.254 & 50.45 & 64.45 & 81.54 & 66.06 & 1.78 & 4.22 & 1.54\\%64.4456, 81.54415879543222, 66.06090606556982
[6pt]
 & $H(z)$ + BAO & 0.0320 & 0.0932 & 0.283 & $\cdots$ & $-0.731$ & 66.69 & $\cdots$ & $\cdots$ & $\cdots$ & 26.57 & 34.57 & 41.71 & 34.52 & $-1.98$ & $-0.20$ & $-2.80$\\%34.5736, 41.710358535673045, 34.51703264021917
 & H$\alpha$ mono & $\cdots$ & $\cdots$ & 0.071 & $\cdots$ & 0.139 & $\cdots$ & 0.545 & 1.628 & 0.258 & 17.34 & 25.34 & 32.19 & 27.11 & 1.69 & 3.41 & 0.23\\%25.33928, 32.19356826681723, 27.107815813017627
Flat XCDM & $H(z)$ + BAO + H$\alpha$ mono & 0.0333 & 0.0903 & 0.279 & $\cdots$ & $-0.711$ & 66.69 & 0.549 & 1.582 & 0.266 & 45.65 & 59.65 & 76.75 & 60.40 & $-2.37$ & 0.07 & $-3.28$\\%59.6468, 76.74535879543222, 60.40172570406631
 & H$\beta$ mono & $\cdots$ & $\cdots$ & 0.168 & $\cdots$ & 0.142 & $\cdots$ & 0.578 & 1.422 & 0.250 & 19.00 & 27.00 & 33.85 & 28.10 & 1.86 & 3.57 & 0.30\\%26.99586, 33.85014826681723, 28.095396728979402
 & $H(z)$ + BAO + H$\beta$ mono & 0.0326 & 0.0930 & 0.282 & $\cdots$ & $-0.725$ & 66.93 & 0.566 & 1.379 & 0.243 & 46.42 & 60.42 & 77.51 & 61.64 & $-2.25$ & 0.19 & $-2.89$\\%60.4154, 77.51395879543222, 61.637049496676475
[6pt]
 & $H(z)$ + BAO & 0.0312 & 0.0990 & 0.293 & $-0.085$ & $-0.693$ & 66.84 & $\cdots$ & $\cdots$ & $\cdots$ & 26.00 & 36.00 & 44.92 & 36.17 & $-0.56$ & 3.01 & $-1.15$\\%35.996, 44.91694816959131, 36.17275325321086
 & H$\alpha$ mono & $\cdots$ & $\cdots$ & 0.994 & 1.936 & $-4.998$ & $\cdots$ & 0.593 & 1.757 & 0.243 & 15.04 & 25.04 & 33.60 & 26.40 & 1.39 & 4.82 & $-0.47$\\%25.03528, 33.60314033352154, 26.40044206169256
Nonflat XCDM & $H(z)$ + BAO + H$\alpha$ mono & 0.0324 & 0.0968 & 0.290 & $-0.073$ & $-0.680$ & 66.85 & 0.551 & 1.586 & 0.269 & 45.16 & 61.16 & 80.70 & 61.70 & $-0.86$ & 4.03 & $-1.99$\\%61.1562, 80.69741005192253, 61.69978036175635
 & H$\beta$ mono & $\cdots$ & $\cdots$ & 0.946 & 1.837 & $-4.699$ & $\cdots$ & 0.630 & 1.547 & 0.241 & 18.39 & 28.39 & 36.96 & 27.22 & 3.25 & 6.68 & $-0.58$\\%28.39012, 36.95798033352154, 27.222783707366442
 & $H(z)$ + BAO + H$\beta$ mono & 0.0306 & 0.0999 & 0.291 & $-0.067$ & $-0.721$ & 67.13 & 0.571 & 1.387 & 0.252 & 45.92 & 61.92 & 81.46 & 62.87 & $-0.75$ & 4.14 & $-1.65$\\%61.9224, 81.46361005192253, 62.87082240596633
[6pt]
 & $H(z)$ + BAO & 0.0337 & 0.0864 & 0.271 & $\cdots$ & 1.169 & 66.78 & $\cdots$ & $\cdots$ & $\cdots$ & 26.50 & 34.50 & 41.64 & 34.01 & $-2.05$ & $-0.27$ & $-3.31$\\%34.5036, 41.640358535673045, 34.00595987367331
 & H$\alpha$ mono & $\cdots$ & $\cdots$ & 0.998 & $\cdots$ & 4.991 & $\cdots$ & 0.552 & 1.625 & 0.256 & 17.64 & 25.64 & 32.50 & 25.79 & 2.00 & 3.71 & $-1.09$\\%25.64074, 32.49502826681723, 25.785283380059397
Flat \pcdm & $H(z)$ + BAO + H$\alpha$ mono & 0.0357 & 0.0793 & 0.263 & $\cdots$ & 1.454 & 66.26 & 0.539 & 1.574 & 0.261 & 45.58 & 59.58 & 76.68 & 59.72 & $-2.43$ & 0.01 & $-3.97$\\%59.5828, 76.68135879543222, 59.719582673361494
 & H$\beta$ mono & $\cdots$ & $\cdots$ & 0.996 & $\cdots$ & 8.454 & $\cdots$ & 0.577 & 1.423 & 0.248 & 19.14 & 27.14 & 34.00 & 26.93 & 2.00 & 3.72 & $-0.87$\\%27.14094, 33.995228266817236, 26.93333544256562
 & $H(z)$ + BAO + H$\beta$ mono & 0.0323 & 0.0891 & 0.275 & $\cdots$ & 1.074 & 66.57 & 0.568 & 1.377 & 0.247 & 46.36 & 60.36 & 77.46 & 60.85 & $-2.31$ & 0.13 & $-3.68$\\%60.3582, 77.45675879543222, 60.84784566383769
[6pt]
 & $H(z)$ + BAO & 0.0338 & 0.0878 & 0.273 & $-0.077$ & 1.441 & 66.86 & $\cdots$ & $\cdots$ & $\cdots$ & 25.92 & 35.92 & 44.84 & 35.12 & $-0.64$ & 2.93 & $-2.20$\\%35.9212, 44.842148169591304, 35.12310308752016 D
 & H$\alpha$ mono & $\cdots$ & $\cdots$ & 0.991 & $-0.966$ & 8.736 & $\cdots$ & 0.553 & 1.639 & 0.248 & 17.19 & 27.19 & 35.76 & 26.06 & 3.55 & 6.97 & $-0.81$\\%25.03528, 33.60314033352154, 26.40044206169256 D
Nonflat \pcdm & $H(z)$ + BAO + H$\alpha$ mono & 0.0357 & 0.0832 & 0.266 & $-0.080$ & 1.647 & 66.98 & 0.546 & 1.574 & 0.263 & 45.04 & 61.04 & 80.58 & 60.87 & $-0.98$ & 3.91 & $-2.81$\\%61.036, 80.57721005192253, 60.87497911598555 D
 & H$\beta$ mono & $\cdots$ & $\cdots$ & 0.978 & $-0.972$ & 9.751 & $\cdots$ & 0.593 & 1.431 & 0.250 & 18.89 & 28.89 & 37.46 & 27.16 & 3.76 & 7.18 & $-0.64$\\%28.89486, 37.462720333521546, 27.159087029744327 D
 & $H(z)$ + BAO + H$\beta$ mono & 0.0340 & 0.0893 & 0.275 & $-0.072$ & 1.420 & 67.15 & 0.585 & 1.392 & 0.249 & 45.84 & 61.84 & 81.38 & 61.83 & $-0.83$ & 4.05 & $-2.70$\\%61.8384, 81.37961005192253, 61.8274840106757 D
\bottomrule\bottomrule
\end{tabular}
%}
\begin{tablenotes}
\item [a] \wx\ corresponds to flat/nonflat XCDM and $\alpha$ corresponds to flat/nonflat \pcdm.
\item [b] \hunit.
% \item [c] $\Omega_b=0.05$ and $H_0=70$ \hunit.
\end{tablenotes}
\end{threeparttable}%
}
\end{table*}
\end{turnpage}

\begin{turnpage}
\begin{table*}
\centering
% \footnotesize
% \resizebox*{2.7\columnwidth}{1.2\columnwidth}{%
\resizebox{2.7\columnwidth}{!}{%
\begin{threeparttable}
\caption{One-dimensional marginalized posterior mean values and uncertainties ($\pm 1\sigma$ error bars or $2\sigma$ limits) of the parameters for all models from various combinations of data, where H$\alpha$ and H$\beta$ stand for H$\alpha$ and H$\beta$ time delays, respectively, and ``mono'' stands for monochromatic luminosity at 5100 \AA.}\label{tab:1d_BFP2}
\begin{tabular}{lcccccccccc}
\toprule\toprule
Model & dataset & $\Omega_{b}h^2$ & $\Omega_{c}h^2$ & $\Omega_{m0}$ & $\Omega_{k0}$ & $w_{\mathrm{X}}$/$\alpha$\tnote{a} & $H_0$\tnote{b} & $\gamma$ & $\beta$ & $\sigma_{\mathrm{int}}$\\
\midrule
 & $H(z)$ + BAO & $0.0260\pm0.0040$ & $0.1213^{+0.0091}_{-0.0103}$ & $0.298^{+0.015}_{-0.018}$ & $\cdots$ & $\cdots$ & $70.51\pm2.72$ & $\cdots$ & $\cdots$ & $\cdots$ \\
 & H$\alpha$ mono & $\cdots$ & $\cdots$ & $>0.487$\tnote{c} & $\cdots$ & $\cdots$ & $\cdots$ & $0.543\pm0.058$ & $1.601\pm0.058$ & $0.285^{+0.036}_{-0.051}$ \\
Flat \lcdm & $H(z)$ + BAO + H$\alpha$ mono & $0.0259^{+0.0036}_{-0.0040}$ & $0.1213^{+0.0087}_{-0.0099}$ & $0.298^{+0.015}_{-0.017}$ & $\cdots$ & $\cdots$ & $70.44\pm2.61$ & $0.536\pm0.057$ & $1.588\pm0.058$ & $0.287^{+0.035}_{-0.050}$ \\
 & H$\beta$ mono & $\cdots$ & $\cdots$ & $>0.437$\tnote{c} & $\cdots$ & $\cdots$ & $\cdots$ & $0.584\pm0.059$ & $1.408^{+0.058}_{-0.054}$ & $0.271^{+0.034}_{-0.049}$ \\
 & $H(z)$ + BAO + H$\beta$ mono & $0.0259^{+0.0036}_{-0.0040}$ & $0.1212^{+0.0088}_{-0.0099}$ & $0.298^{+0.015}_{-0.017}$ & $\cdots$ & $\cdots$ & $70.46\pm2.61$ & $0.579^{+0.058}_{-0.057}$ & $1.399^{+0.059}_{-0.055}$ & $0.272^{+0.034}_{-0.048}$ \\
[6pt]
 & $H(z)$ + BAO & $0.0275^{+0.0047}_{-0.0053}$ & $0.1132\pm0.0183$ & $0.289\pm0.023$ & $0.047^{+0.083}_{-0.091}$ & $\cdots$ & $69.81\pm2.87$ & $\cdots$ & $\cdots$ & $\cdots$ \\
 & H$\alpha$ mono & $\cdots$ & $\cdots$ & $>0.472$\tnote{c} & $>-1.047$ & $\cdots$ & $\cdots$ & $0.549\pm0.058$ & $1.617\pm0.060$ & $0.282^{+0.035}_{-0.050}$ \\
Nonflat \lcdm & $H(z)$ + BAO + H$\alpha$ mono & $0.0275^{+0.0046}_{-0.0052}$ & $0.1126^{+0.0178}_{-0.0179}$ & $0.289\pm0.023$ & $0.050^{+0.081}_{-0.089}$ & $\cdots$ & $69.72\pm2.80$ & $0.536^{+0.057}_{-0.056}$ & $1.584\pm0.058$ & $0.287^{+0.035}_{-0.049}$ \\
 & H$\beta$ mono & $\cdots$ & $\cdots$ & $>0.450$\tnote{c} & $>-1.256$ & $\cdots$ & $\cdots$ & $0.589\pm0.059$ & $1.420\pm0.060$ & $0.270^{+0.034}_{-0.049}$ \\
 & $H(z)$ + BAO + H$\beta$ mono & $0.0275^{+0.0046}_{-0.0052}$ & $0.1128\pm0.0179$ & $0.289\pm0.023$ & $0.048^{+0.081}_{-0.089}$ & $\cdots$ & $69.75\pm2.79$ & $0.579\pm0.057$ & $1.394\pm0.057$ & $0.271^{+0.033}_{-0.048}$ \\
[6pt]
 & $H(z)$ + BAO & $0.0308^{+0.0053}_{-0.0046}$ & $0.0980^{+0.0182}_{-0.0161}$ & $0.286\pm0.019$ & $\cdots$ & $-0.778^{+0.132}_{-0.104}$ & $67.20^{+3.05}_{-3.06}$ & $\cdots$ & $\cdots$ & $\cdots$ \\
 & H$\alpha$ mono & $\cdots$ & $\cdots$ & $>0.551$\tnote{c} & $\cdots$ & $-1.995^{+2.142}_{-0.939}$ & $\cdots$ & $0.540\pm0.058$ & $1.597\pm0.059$ & $0.285^{+0.035}_{-0.050}$ \\
Flat XCDM & $H(z)$ + BAO + H$\alpha$ mono & $0.0312^{+0.0054}_{-0.0045}$ & $0.0961^{+0.0183}_{-0.0161}$ & $0.285\pm0.019$ & $\cdots$ & $-0.762^{+0.130}_{-0.099}$ & $66.95\pm2.97$ & $0.539\pm0.057$ & $1.571\pm0.058$ & $0.285^{+0.035}_{-0.049}$ \\
 & H$\beta$ mono & $\cdots$ & $\cdots$ & $>0.507$\tnote{c} & $\cdots$ & $-2.105^{+2.241}_{-0.858}$ & $\cdots$ & $0.581\pm0.059$ & $1.401^{+0.062}_{-0.056}$ & $0.272^{+0.034}_{-0.049}$ \\
 & $H(z)$ + BAO + H$\beta$ mono & $0.0311^{+0.0054}_{-0.0046}$ & $0.0966^{+0.0182}_{-0.0161}$ & $0.285\pm0.019$ & $\cdots$ & $-0.767^{+0.131}_{-0.100}$ & $67.03^{+2.97}_{-2.98}$ & $0.582\pm0.057$ & $1.380\pm0.057$ & $0.271^{+0.033}_{-0.048}$ \\
[6pt]
 & $H(z)$ + BAO & $0.0305^{+0.0055}_{-0.0047}$ & $0.1011\pm0.0196$ & $0.292\pm0.024$ & $-0.059\pm0.106$ & $-0.746^{+0.135}_{-0.090}$ & $67.17^{+2.96}_{-2.97}$ & $\cdots$ & $\cdots$ & $\cdots$ \\
 & H$\alpha$ mono & $\cdots$ & $\cdots$ & $>0.465$\tnote{c} & $>-1.100$ & $-2.415^{+1.779}_{-1.729}$ & $\cdots$ & $0.558\pm0.059$ & $1.642\pm0.068$ & $0.278^{+0.035}_{-0.049}$ \\
Nonflat XCDM & $H(z)$ + BAO + H$\alpha$ mono & $0.0308^{+0.0058}_{-0.0047}$ & $0.0995\pm0.0197$ & $0.291\pm0.024$ & $-0.063\pm0.107$ & $-0.733^{+0.135}_{-0.087}$ & $66.96\pm2.93$ & $0.539^{+0.057}_{-0.056}$ & $1.571\pm0.058$ & $0.285^{+0.035}_{-0.049}$ \\
 & H$\beta$ mono & $\cdots$ & $\cdots$ & $>0.439$\tnote{c} & $>-1.178$ & $-2.278^{+2.331}_{-1.127}$ & $\cdots$ & $0.595\pm0.060$ & $1.438\pm0.066$ & $0.269^{+0.034}_{-0.048}$ \\
 & $H(z)$ + BAO + H$\beta$ mono & $0.0307^{+0.0056}_{-0.0048}$ & $0.1001^{+0.0196}_{-0.0197}$ & $0.291\pm0.024$ & $-0.063\pm0.107$ & $-0.736^{+0.135}_{-0.088}$ & $67.03^{+2.94}_{-2.93}$ & $0.582\pm0.057$ & $1.380^{+0.059}_{-0.055}$ & $0.270^{+0.033}_{-0.048}$ \\
[6pt]
 & $H(z)$ + BAO & $0.0327^{+0.0060}_{-0.0031}$ & $0.0866^{+0.0192}_{-0.0176}$ & $0.272\pm0.022$ & $\cdots$ & $1.261^{+0.494}_{-0.810}$ & $66.23^{+2.87}_{-2.86}$ & $\cdots$ & $\cdots$ & $\cdots$ \\%$1.261^{+1.276}_{-1.206}$ 2$\sigma$
 & H$\alpha$ mono & $\cdots$ & $\cdots$ & $>0.427$\tnote{c} & $\cdots$ & $\cdots$ & $\cdots$ & $0.546\pm0.057$ & $1.609\pm0.056$ & $0.282^{+0.035}_{-0.049}$ \\
Flat \pcdm & $H(z)$ + BAO + H$\alpha$ mono & $0.0330^{+0.0068}_{-0.0021}$ & $0.0856^{+0.0186}_{-0.0187}$ & $0.271\pm0.022$ & $\cdots$ & $1.307^{+0.498}_{-0.830}$ & $66.16\pm2.84$ & $0.539\pm0.057$ & $1.566^{+0.058}_{-0.057}$ & $0.286^{+0.035}_{-0.049}$ \\
 & H$\beta$ mono & $\cdots$ & $\cdots$ & $\cdots$ & $\cdots$ & $\cdots$ & $\cdots$ & $0.587\pm0.058$ & $1.416^{+0.057}_{-0.053}$ & $0.270^{+0.034}_{-0.048}$ \\
 & $H(z)$ + BAO + H$\beta$ mono & $0.0329^{+0.0068}_{-0.0022}$ & $0.0859\pm0.0187$ & $0.271\pm0.022$ & $\cdots$ & $1.292^{+0.496}_{-0.831}$ & $66.21\pm2.85$ & $0.582\pm0.057$ & $1.374\pm0.057$ & $0.271^{+0.033}_{-0.048}$ \\
[6pt]
 & $H(z)$ + BAO & $0.0324^{+0.0062}_{-0.0031}$ & $0.0900\pm0.0200$ & $0.277\pm0.025$ & $-0.072^{+0.093}_{-0.107}$ & $1.435^{+0.579}_{-0.788}$ & $66.50\pm2.88$ & $\cdots$ & $\cdots$ & $\cdots$ \\%$1.435^{+1.319}_{-1.290}$
 & H$\alpha$ mono & $\cdots$ & $\cdots$ & $>0.424$\tnote{c} & $-0.067^{+0.410}_{-0.414}$ & $\cdots$ & $\cdots$ & $0.546\pm0.057$ & $1.610\pm0.056$ & $0.282^{+0.035}_{-0.049}$ \\
Nonflat \pcdm & $H(z)$ + BAO + H$\alpha$ mono & $0.0327^{+0.0070}_{-0.0022}$ & $0.0889^{+0.0203}_{-0.0204}$ & $0.276\pm0.025$ & $-0.073^{+0.092}_{-0.107}$ & $1.484^{+0.595}_{-0.819}$ & $66.41\pm2.87$ & $0.539\pm0.056$ & $1.567^{+0.058}_{-0.057}$ & $0.285^{+0.035}_{-0.049}$ \\
 & H$\beta$ mono & $\cdots$ & $\cdots$ & $\cdots$ & $-0.049^{+0.411}_{-0.415}$ & $\cdots$ & $\cdots$ & $0.588\pm0.058$ & $1.417^{+0.057}_{-0.053}$ & $0.269^{+0.034}_{-0.048}$ \\
 & $H(z)$ + BAO + H$\beta$ mono & $0.0326^{+0.0070}_{-0.0023}$ & $0.0893\pm0.0203$ & $0.276\pm0.025$ & $-0.073^{+0.092}_{-0.107}$ & $1.465^{+0.582}_{-0.816}$ & $66.47\pm2.87$ & $0.582\pm0.057$ & $1.376^{+0.056}_{-0.057}$ & $0.271^{+0.032}_{-0.048}$ \\
\bottomrule\bottomrule
\end{tabular}
\begin{tablenotes}
\item [a] \wx\ corresponds to flat/nonflat XCDM and $\alpha$ corresponds to flat/nonflat \pcdm.
\item [b] \hunit. For the RM AGN alone cases, $\Omega_b=0.05$ and $H_0=70$ \hunit.
\item [c] This is the 1$\sigma$ limit. The 2$\sigma$ limit is set by the prior and not shown here.
\end{tablenotes}
\end{threeparttable}%
}
\end{table*}
\end{turnpage}

\section{Results}
\label{sec:results}

Plots of posterior one-dimensional probability distributions and two-dimensional confidence regions for the parameters of six cosmological models, along with the $R-L$ relation (and intrinsic scatter) parameters for H$\alpha$ broad ($R_{\text{H}\alpha} - L_{\text{H}\alpha}$) is shown in Figs.~\ref{fig1} and \ref{fig2}. The plots for the remaining relations, H$\alpha$ mono ($R_{\text{H}\alpha} - L_{5100}$), H$\beta$ broad ($R_{\text{H}\beta} - L_{\text{H}\alpha}$), and H$\beta$ mono ($R_{\text{H}\beta} - L_{5100}$), can be found in Appendix \ref{A1}. Note that the plotted 1D marginalized density distributions range from a minimum of zero to a maximum of 1. The unmarginalized best-fitting parameter values, maximum likelihood values ($\mathcal{L}_{\rm max}$), and model selection criteria including AIC, BIC, DIC, as well as their differences ($\Delta \mathrm{AIC}$, $\Delta \mathrm{BIC}$, $\Delta \mathrm{DIC}$), are summarized in Tables \ref{tab:BFP1} and \ref{tab:BFP2} for H$\alpha$/H$\beta$ broad and mono results, respectively. Marginalized posterior mean parameters and their uncertainties (including $1\sigma$ error bars and 1 or $2\sigma$ limits) are reported in Tables \ref{tab:1d_BFP1} and \ref{tab:1d_BFP2} for the same datasets.

Table~\ref{tab:diff1} lists the largest differences in the $R-L$ relation parameters and intrinsic scatter parameters across the cosmological models studied. These results indicate that all four RM AGN datasets can be standardized using their respective $R-L$ relations. Comparisons between the broad and mono cases for H$\alpha$ and H$\beta$ within each cosmological model reveal that the constraints for H$\alpha$ and H$\beta$ broad are consistent, with differences within $1.2\sigma$, while significant differences are observed for the intercept parameter $\beta$ in the mono cases, with $\Delta\beta > 2\sigma$, see Table~\ref{tab:diff2}.

The slope parameter $\gamma$ for H$\alpha$ broad ranges from $0.569\pm0.055$ in flat XCDM to $0.579\pm0.055$ in nonflat XCDM, while for H$\beta$ broad it ranges from $0.611\pm0.053$ in flat XCDM to $0.620\pm0.054$ in nonflat XCDM. For H$\alpha$ mono $\gamma$ ranges from $0.540\pm0.058$ in flat XCDM to $0.558\pm0.059$ in nonflat XCDM, and for H$\beta$ mono it ranges from $0.581\pm0.059$ in flat XCDM to $0.595\pm0.060$ in nonflat XCDM.

The intercept parameter $\beta$ for H$\alpha$ broad ranges from $2.325^{+0.101}_{-0.088}$ in flat XCDM to $2.373\pm0.103$ in nonflat XCDM, while for H$\beta$ broad it ranges from $2.166^{+0.100}_{-0.092}$ in flat XCDM to $2.208\pm0.106$ in nonflat XCDM. For H$\alpha$ mono $\beta$ ranges from $1.597\pm0.059$ in flat XCDM to $1.642\pm0.068$ in nonflat XCDM, and for H$\beta$ mono it ranges from $1.401^{+0.062}_{-0.056}$ in flat XCDM to $1.438\pm0.066$ in nonflat XCDM.

\begin{table}
\centering
% \resizebox{2\columnwidth}{!}{%
\setlength\tabcolsep{17pt}
\begin{threeparttable}
\caption{The largest differences between considered cosmological models (flat and nonflat \lcdm, XCDM, and \pcdm) from various combinations of data with $1\sigma$ being the quadrature sum of the two corresponding $1\sigma$ error bars.}\label{tab:diff1}
\begin{tabular}{lccc}
\toprule\toprule
 Data set & $\Delta\gamma$ & $\Delta\beta$ & $\Delta\sigma_{\mathrm{int}}$ \\
\midrule
H$\alpha$ broad & $0.13\sigma$ & $0.33\sigma$ & $0.05\sigma$ \\
H$\alpha$ mono & $0.22\sigma$ & $0.50\sigma$ & $0.11\sigma$\\
H$\beta$ broad & $0.12\sigma$ & $0.29\sigma$ & $0.02\sigma$ \\
H$\beta$ mono & $0.17\sigma$ & $0.41\sigma$ & $0.05\sigma$\\
\bottomrule\bottomrule
\end{tabular}
%}
% \begin{tablenotes}[flushleft]
% \item [a] Only for flat/Nonflat \lcdm.

% \end{tablenotes}
\end{threeparttable}%
% }
\end{table}

\begin{table}
\centering
% \resizebox{2\columnwidth}{!}{%
\setlength\tabcolsep{4pt}
\begin{threeparttable}
\caption{The differences between H$\alpha$ and H$\beta$ of broad/mono for a given cosmological model with $1\sigma$ being the quadrature sum of the two corresponding $1\sigma$ error bars.}\label{tab:diff2}
\begin{tabular}{lccc}
\toprule\toprule
 Model & $\Delta\gamma$ & $\Delta\beta$ & $\Delta\sigma_{\mathrm{int}}$ \\
\midrule
Flat \lcdm & $0.59\sigma$/$0.50\sigma$ & $1.19\sigma$/$2.35\sigma$ & $0.27\sigma$/$0.23\sigma$ \\
Nonflat \lcdm & $0.56\sigma$/$0.48\sigma$ & $1.19\sigma$/$2.32\sigma$ & $0.27\sigma$/$0.20\sigma$ \\
Flat XCDM & $0.55\sigma$/$0.50\sigma$ & $1.19\sigma$/$2.29\sigma$ & $0.29\sigma$/$0.22\sigma$ \\
Nonflat XCDM & $0.53\sigma$/$0.44\sigma$ & $1.12\sigma$/$2.15\sigma$ & $0.25\sigma$/$0.15\sigma$ \\
Flat \pcdm & $0.61\sigma$/$0.50\sigma$ & $1.20\sigma$/$2.42\sigma$ & $0.27\sigma$/$0.20\sigma$ \\
Nonflat \pcdm & $0.61\sigma$/$0.52\sigma$ & $1.20\sigma$/$2.42\sigma$ & $0.26\sigma$/$0.22\sigma$ \\
\bottomrule\bottomrule
\end{tabular}
%}
% \begin{tablenotes}[flushleft]
% \item [a] Only for flat/Nonflat \lcdm.

% \end{tablenotes}
\end{threeparttable}%
% }
\end{table}

The intrinsic scatter parameter $\sigma_{\rm int}$ for H$\alpha$ broad lies between $0.261^{+0.032}_{-0.045}$ in nonflat \pcdm\ and $0.264^{+0.033}_{-0.046}$ in flat XCDM. For H$\beta$ broad $\sigma_{\rm int}$ lies between $0.247^{+0.031}_{-0.046}$ in nonflat \pcdm\ and $0.248^{+0.032}_{-0.046}$ in flat \lcdm, nonflat \lcdm, flat XCDM, and nonflat XCDM. For H$\alpha$ mono $\sigma_{\rm int}$ ranges from $0.278^{+0.035}_{-0.049}$ in nonflat XCDM to $0.285^{+0.036}_{-0.051}$ in flat \lcdm, while for H$\beta$ mono it ranges from $0.269^{+0.034}_{-0.048}$ in nonflat XCDM and nonflat \pcdm\ to $0.272^{+0.034}_{-0.049}$ in flat XCDM. Among these datasets, the H$\beta$ broad data have the lowest intrinsic scatter, making them the most reliable for constraining the $R-L$ relation. The differences however are not significant. For example, in the flat \lcdm\ model $\sigma_{\rm int}$ is largest for H$\alpha$ mono, $0.285^{+0.036}_{-0.051}$, only $\sim0.6\sigma$ larger than the smallest, H$\beta$ broad, value of $0.248^{+0.032}_{-0.046}$.

Among the six cosmological models analyzed, in the flat \lcdm\ model, the flat XCDM parametrization, and in the $\wX-\Om$ plane of the nonflat XCDM parametrization these RM AGN data favor currently accelerating cosmological expansion more. In contrast, in the remaining cases they more favor currently decelerating expansion, but with accelerating expansion within $2\sigma$. 

We summarize the cosmological parameter constraints next. Unlike the resulting constraints on the $R-L$ relation parameters, these small, 41 source, RM AGN datasets provide only weak cosmological parameter constraints. This can be seen from the gray contours and likelihoods in Figs.~\ref{fig1} and \ref{fig2}, as well as Figs.~\ref{fig3}--\ref{fig8}, and the corresponding parameter values listed in Tables \ref{tab:1d_BFP1} and \ref{tab:1d_BFP2}. 

The \om\ $1\sigma$ lower limits range from a low of 0.406 (H$\beta$ broad nonflat XCDM) and a high of 0.551 (H$\alpha$ mono flat XCDM) with limits not provided in some cases from these AGN data. All $2\sigma$ limits on \om\ are set by the prior and so are not shown in the tables.

The \ok\ constraints are all consistent with flat spatial hypersurfaces. In nonflat \lcdm\ and nonflat XCDM, the H$\beta$ broad cases yield constraints of $0.323^{+1.496}_{-0.696}$ and $0.480^{+1.219}_{-0.657}$, respectively, mildly favoring open hypersurfaces, with other cases providing only $2\sigma$ lower limits. For nonflat \pcdm, the \ok\ constraints are $-0.041\pm0.411$ (H$\alpha$ broad), $-0.027\pm0.410$ (H$\beta$ broad), $-0.067^{+0.410}_{-0.414}$ (H$\alpha$ mono), and $-0.049^{+0.411}_{-0.415}$ (H$\beta$ mono), all mildly favoring closed hypersurfaces.

XCDM parameterization \wx\ constraints are similarly weak whereas these AGN data do not provide \pcdm\ model $\alpha$ constraints. In flat XCDM \wx\ constraints of $-2.103^{+2.238}_{-0.891}$, $-2.231^{+2.281}_{-0.939}$, $-1.995^{+2.142}_{-0.939}$, and $-2.415^{+1.779}_{-1.729}$ are derived in the H$\alpha$ broad, H$\beta$ broad, H$\alpha$ mono, and H$\beta$ mono cases, respectively. For nonflat XCDM the \wx\ constraints are $-2.251^{+2.362}_{-1.032}$ (H$\alpha$ broad), $-2.135^{+2.163}_{-0.885}$ (H$\beta$ broad), $-2.105^{+2.241}_{-0.858}$ (H$\alpha$ mono), and $-2.278^{+2.331}_{-1.127}$ (H$\beta$ mono).

Cosmological parameter constraints derived from these RM AGN datasets are consistent with those obtained from $H(z)$ + BAO data, indicating that these datasets can be reliably combined for joint analyses. However, the inclusion of RM AGN data results in only minor changes to the cosmological constraints compared to those from $H(z)$ + BAO data alone. The most notable differences are a $0.097\sigma$ shift in \wx\ for nonflat XCDM and a $0.059\sigma$ shift in $H_0$ for flat XCDM when using the $H(z)$ + BAO + H$\alpha$ mono dataset combination. Additionally, the $\alpha$ parameter shows slightly higher values when RM AGN data are included, with the largest difference being $0.049\sigma$ in the nonflat \pcdm\ model from the $H(z)$ + BAO + H$\alpha$ mono dataset combination.

The $R-L$ relation and intrinsic scatter parameters remain consistent when combining RM AGN datasets with $H(z)$ + BAO data, with the largest differences arising between standalone RM AGN datasets and their corresponding joint combinations detailed next.

For the slope parameter $\gamma$, the largest differences are $0.079\sigma$ and $0.068\sigma$ for H$\alpha$ broad and H$\beta$ broad, respectively, observed in the nonflat \lcdm\ model. For H$\alpha$ mono and H$\beta$ mono, the largest differences are $0.23\sigma$ and $0.16\sigma$, respectively, observed in the nonflat XCDM parametrization.

For the intercept parameter $\beta$, the largest differences are $0.47\sigma$ and $0.38\sigma$ for H$\alpha$ broad and H$\beta$ broad, respectively, observed in the nonflat \lcdm\ model. For H$\alpha$ mono and H$\beta$ mono, the largest differences are $0.79\sigma$ and $0.66\sigma$, respectively, observed in the nonflat XCDM model.

For the intrinsic scatter parameter $\sigma_{\rm int}$, the differences are generally very small, with one exception being $0.12\sigma$ for H$\alpha$ mono in nonflat XCDM, which remains modest.

Based on the more reliable DIC, H$\alpha$ broad and H$\beta$ broad datasets favor nonflat \pcdm\ the most, while H$\alpha$ mono and H$\beta$ mono datasets favor flat \pcdm\ the most, all with very weak evidence against the remaining models and parametrizations. In contrast, the joint datasets, dominated by $H(z)$ + BAO data, favor flat \pcdm\ the most, with weak or positive evidence against other models and parametrizations.

\begin{figure*}[htbp]
\centering
 \subfloat[Flat \lcdm]{%
    \includegraphics[width=0.45\textwidth,height=0.35\textwidth]{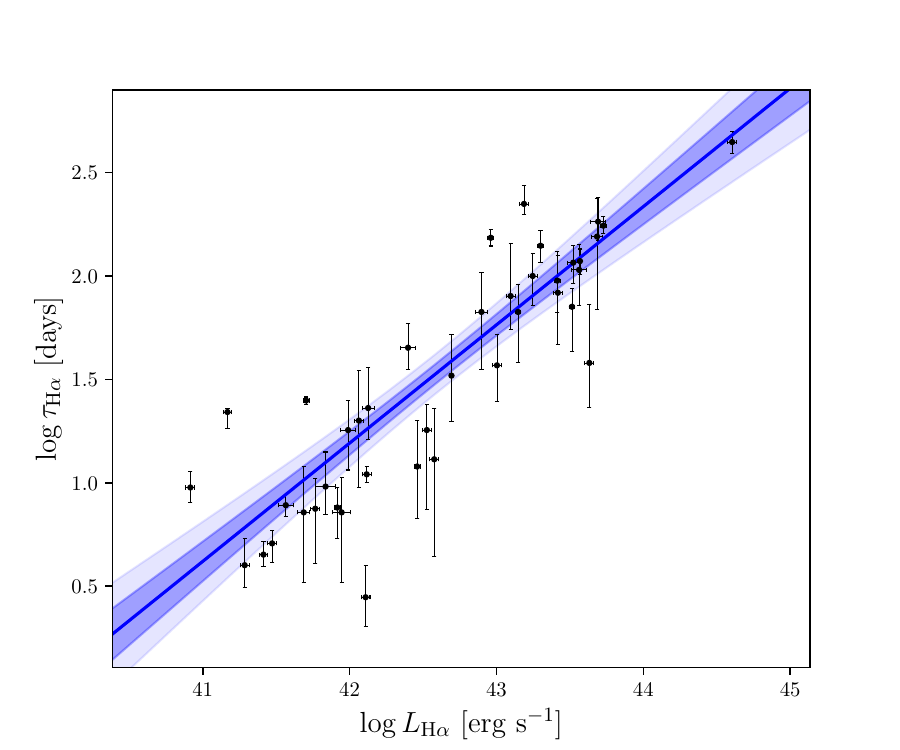}}
 \subfloat[Nonflat \lcdm]{%
    \includegraphics[width=0.45\textwidth,height=0.35\textwidth]{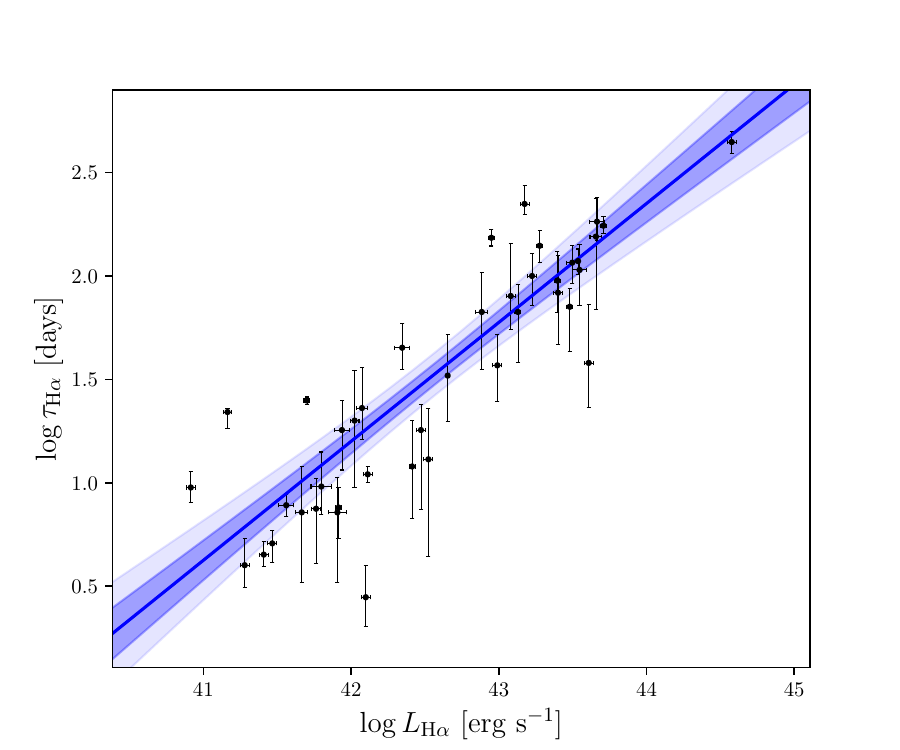}}\\
 \subfloat[Flat XCDM]{%
    \includegraphics[width=0.45\textwidth,height=0.35\textwidth]{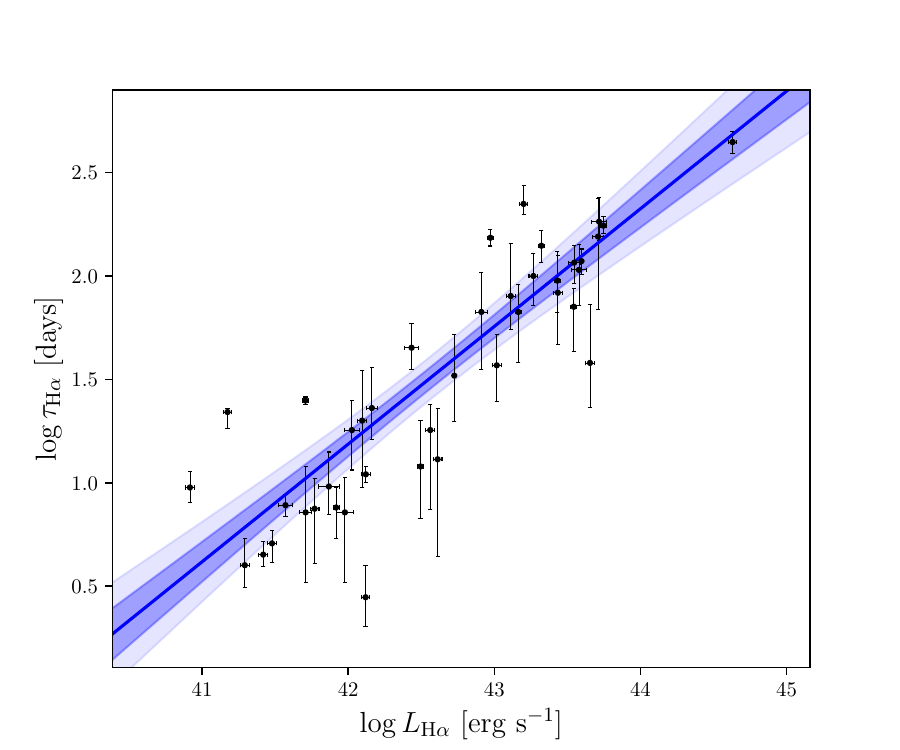}}
 \subfloat[Nonflat XCDM]{%
    \includegraphics[width=0.45\textwidth,height=0.35\textwidth]{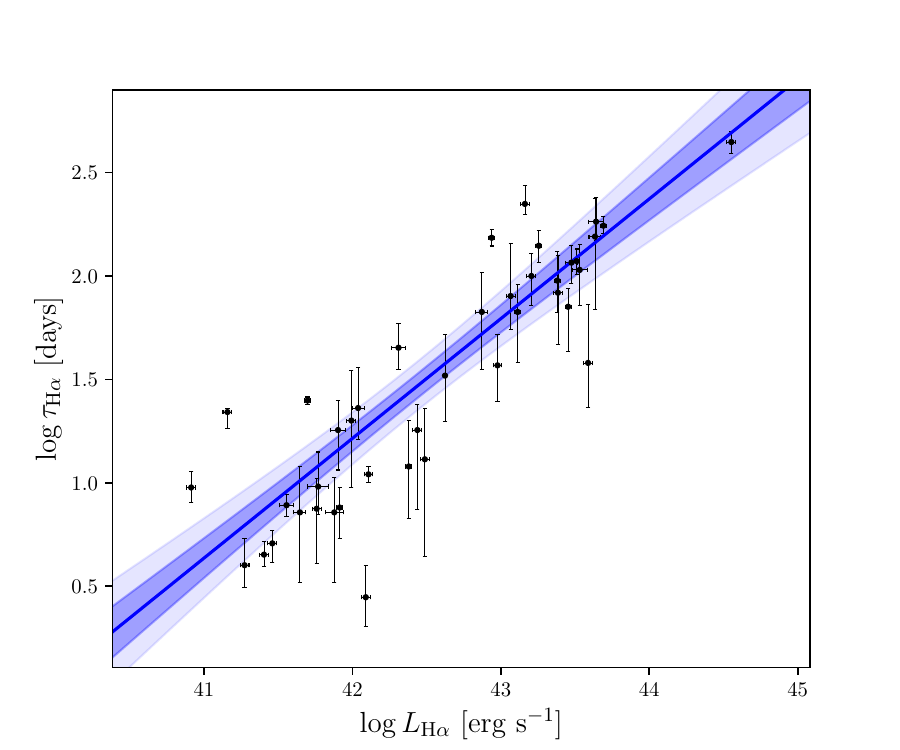}}\\
 \subfloat[Flat \pcdm]{%
    \includegraphics[width=0.45\textwidth,height=0.35\textwidth]{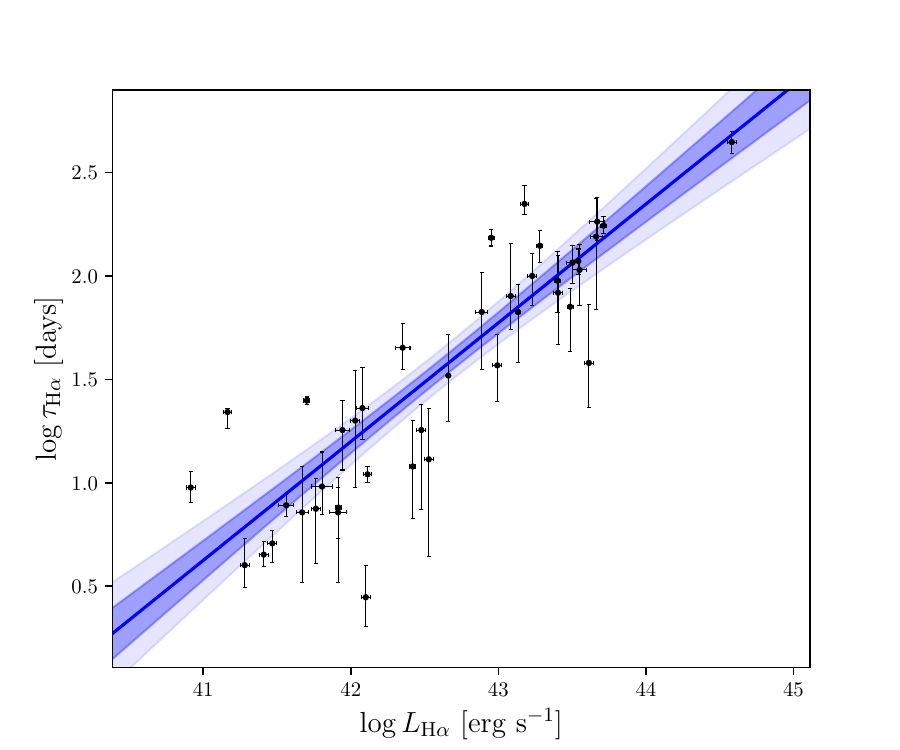}}
 \subfloat[Nonflat \pcdm]{%
    \includegraphics[width=0.45\textwidth,height=0.35\textwidth]{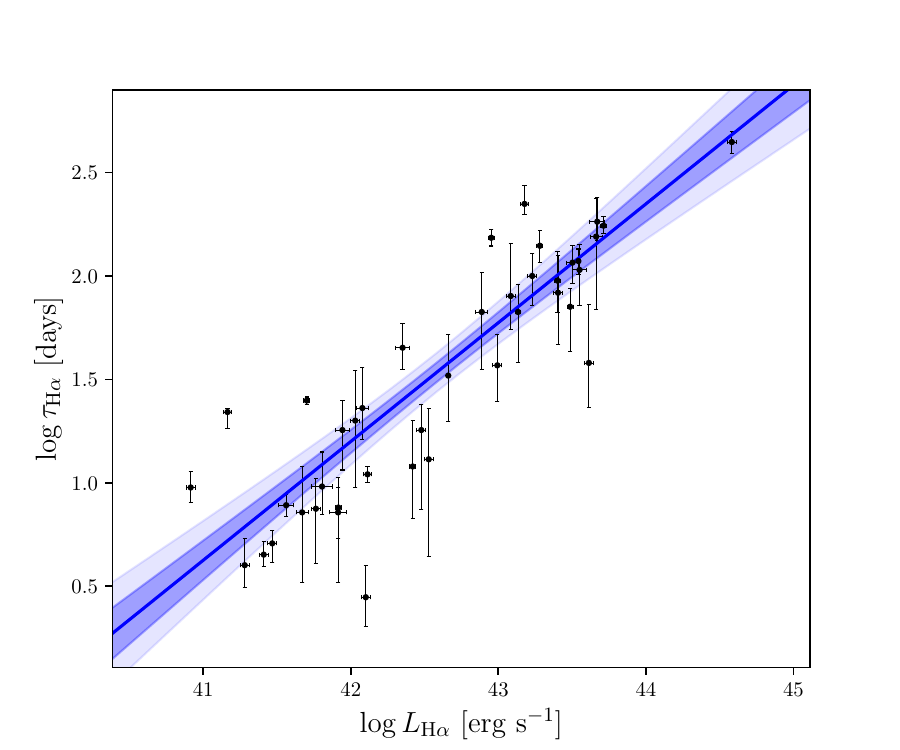}}\\
\caption{$R-L$ relations of H$\alpha$ broad for six different models, with \lcdm, XCDM, and \pcdm\ in the top, middle, and bottom rows, and flat (nonflat) models in the left (right) column. Here data point $D_L$ values are computed using the corresponding posterior mean values listed in Table \ref{tab:1d_BFP1} and the $R-L$ relations are derived from Monte Carlo simulations with given posterior mean values in Table \ref{tab:1d_BFP1} and covariance matrices of the intercept and slope parameters. Errors of $\log \tau$ are from error propagation of $\tau$.}
\label{R-L1}
\vspace{-50pt}
\end{figure*}

Figure \ref{R-L1} shows the $R-L$ relation of the H$\alpha$ broad AGN dataset for the six different cosmological models, while the corresponding relations for the H$\alpha$ mono, H$\beta$ broad, and H$\beta$ mono AGN datasets, demonstrating consistent patterns, are shown in Appendix \ref{A1}. Data point $D_L$ values are computed using the corresponding posterior mean values listed in Tables \ref{tab:1d_BFP1} and \ref{tab:1d_BFP2}, and the $R-L$ relations are derived from Monte Carlo simulations with given posterior mean values in Tables \ref{tab:1d_BFP1} and \ref{tab:1d_BFP1} and covariance matrices of the corresponding intercept and slope parameters, where errors of $\log \tau$ are from error propagation of $\tau$. There are six, six, five, and four outliers outside the 2$\sigma$ ranges of $R-L$ relation predictions for H$\alpha$ broad, H$\alpha$ mono, H$\beta$ broad, and H$\beta$ mono, respectively. There are common outliers among the four different datasets, with the same outliers consistently appearing in each of the six cosmological models for a given dataset. However, on average the data points are consistent with the predicted $R-L$ relations.

\section{Discussion}
\label{sec:discussion}

We showed that the samples of 41 RM AGNs with measured H$\alpha$ and H$\beta$ time lags are standardizable considering $R-L$ relations with both monochromatic luminosity at 5100\,\AA\, and broad H$\alpha$ luminosity. In our analysis here we have corrected source redshifts for peculiar velocity contamination. We also note that, unlike in almost all earlier analyses, our $R-L$ relation parameter values and error bars fully account for cosmological parameters uncertainties and so typically are a little less constraining.  While the $R-L$ relation parameters are well constrained by these data, the inferred cosmological parameter constraints are very weak. Therefore, in the following discussion we focus on the parameters of the $R-L$ relations in Subsection~\ref{subsec_RL}, a comparison between the $L_{\mathrm{H\alpha}}$ and $L_{5100}$ luminosities in Subsection~\ref{subsec_luminosity-luminosity}, the scatter in the $R-L$ relations in Subsection~\ref{subsec_scatter}, and comment, in Subsection~\ref{subsec_differences}, on the qualitative differences between the cosmological parameter constraints here and in our previous studies.

\subsection{BLR radius--luminosity relation}
\label{subsec_RL}

In this section we investigate the $R-L$ relations by examining H$\alpha$ and H$\beta$ BLR radii, relative to both $L_{\mathrm{H\alpha}}$ and $L_{5100}$ luminosity, across different data samples and cosmological models, using Eq.~\eqref{eq:R-L}.

Our results demonstrate that, within uncertainties, the $R-L$ relations exhibit similar intrinsic scatter and remain consistent across different cosmological models when similar luminosity measures, either $L_{\mathrm{H\alpha}}$ or $L_{5100}$, are used (see Section \ref{sec:results}). For illustrative purposes, in the flat $\Lambda$CDM model, which is currently the standard model of cosmology, the $R-L$ relations are
\begin{widetext}
\begin{equation}
    \label{eq:Ralp-Lalp}
    \log{\left(\frac{\tau_{H\alpha}}{\rm day}\right)}=2.336_{-0.086}^{+0.096} + (0.571\pm0.055) \log{\left(\frac{L_{\mathrm{H\alpha}}}{10^{44}\,{\rm erg\ s^{-1}}}\right)},
\end{equation}
\begin{equation}
    \label{eq:Rbet-Lalp}
    \log{\left(\frac{\tau_{H\beta}}{\rm day}\right)}=2.184_{-0.087}^{+0.095} + (0.616\pm0.053) \log{\left(\frac{L_{\mathrm{H\alpha}}}{10^{44}\,{\rm erg\ s^{-1}}}\right)},
\end{equation}
\begin{equation}
    \label{eq:Ralp-Lmon}
    \log{\left(\frac{\tau_{H\alpha}}{\rm day}\right)}=1.601 \pm 0.058 + (0.543\pm0.058) \log{\left(\frac{L_{5100}}{10^{44}\,{\rm erg\ s^{-1}}}\right)},
\end{equation}
and,
\begin{equation}
    \label{eq:Rbet-Lmon}
    \log{\left(\frac{\tau_{H\beta}}{\rm day}\right)}=1.408_{-0.054}^{+0.058} + (0.584\pm0.059) \log{\left(\frac{L_{5100}}{10^{44}\,{\rm erg\ s^{-1}}}\right)},
\end{equation}
\end{widetext}
while the $R-L$ relations for H$\alpha$ and H$\beta$ time delays in other cosmological models can be found in Tables~\ref{tab:1d_BFP1} and \ref{tab:1d_BFP2}, respectively.

Using a very similar RM AGN sample, \citet{Choetal2023} report a slope of $0.61 \pm 0.04$  ($0.58 \pm 0.05$ when excluding NGC 4395 and PG 1226+023) for the $R_{\mathrm{H\alpha}} - L_{\mathrm{H\alpha}}$ relation, and $0.58 \pm 0.04$ ($0.54 \pm 0.06$ excluding NGC 4395 and PG 1226+023) for the $R_{\mathrm{H\alpha}} - L_{5100}$ relation. In our analysis, we excluded NGC 4395 due to the lack of H$\beta$ lag measurement, while PG 1226+023 was included in our sample. The small differences between our results and those reported by \citet{Choetal2023} are likely due to the slightly different RM AGN samples used in the two analyses and because \citet{Choetal2023} used a fixed cosmology for fitting their  $R_{\mathrm{H\alpha}} - L_{\mathrm{H\alpha}}$ relations, whereas we treat cosmological parameters as free variables and account for their uncertainties. Additionally, we have corected source redshifts for peculiar velocities. Nonetheless, given the uncertainties, the two results are consistent.

However, we find steeper slopes than the expected value of 0.5 from a simple photoionization model of BLR clouds \citep{1972ApJ...171..213D,2021bhns.confE...1K}, with $0.74\sigma$ (H$\alpha$ mono), $1.3\sigma$ (H$\alpha$ broad), $1.4\sigma$ (H$\beta$ mono), and $2.2\sigma$ (H$\beta$ broad) significance in the flat \lcdm\ model. Notably, the discrepancy becomes more pronounced when using $L_{H\alpha}$ instead of $L_{5100}$ in the $R-L$ relations. This difference may be explained by the Baldwin effect \citep{1977ApJ...214..679B} observed in the broad H$\alpha$ line (see Section \ref{subsec_luminosity-luminosity} for further details).

In addition, previous H$\beta$ RM campaigns, such as \citet{2024ApJ...962...67W} and \citet{2024ApJS..275...13W}, reported a much shallower slope of $0.402_{-0.022}^{+0.020}$ but a relatively similar intercept of $1.405_{-0.023}^{+0.018}$ in the $R_{\mathrm{H\beta}} - L_{5100}$ relation. Given that their and our intercepts are quite similar, it is not unreasonable to note that the two slopes differ at a significance of $2.9\sigma$ (a more correct estimate of the significance would account for the covariance between the slope and intercept parameters). It is important to note that $R_{\mathrm{H\beta}} - L_{5100}$ relations in the literature typically are based on a much larger AGN sample, in this case containing about 240 AGNs that cover a broader range of Eddington ratios ($-3 < \mathrm{log \, \lambda_{Edd}} < 1.5$)  \citep[see figure 17 in][]{2024ApJ...962...67W}, including many super-Eddington accreting massive black holes. In contrast, our $R_{\mathrm{H\beta}} - L_{\mathrm{H\alpha}}$ and $R_{\mathrm{H\beta}} - L_{5100}$ relations are based on a relatively smaller sample of 41 AGNs with Eddington ratio values ranging between $-2.4 < \mathrm{log \, \lambda_{Edd}} < 0.4$, most of which have $\mathrm{log \, \lambda_{Edd}} < 0.0$. Therefore, the steeper slope we observe, compared to the value predicted by photoionization model and those reported in previous studies, is likely attributable to differences in sample selection, i.e., higher-accreting massive black holes tend to exhibit shorter BLR time lags \citep{2015ApJ...806...22D}, leading to a shallower slope in the $R-L$ relation. 

We also note that the intercept values in the $R-L$ relations are larger when H$\alpha$ lags are used compared to when H$\beta$ lags are employed. This difference arises because H$\alpha$ time lags are generally larger than H$\beta$ time lags for the same AGN, reflecting the radial ionization stratification of the BLR \citep{Fengetal2025}. Furthermore, the significantly larger error bars we find, relative to \citet{2024ApJ...962...67W}, are likely a result of the smaller sample size of only 41 AGNs, which increases statistical uncertainties. Additionally, our inclusion of cosmological parameter uncertainties contributes to our overall error budget.

\citet{Khadkaetal2021c} use a sample of 118 H$\beta$ RM AGNs to determine a flat \lcdm\ model $R_{\mathrm{H\beta}} - L_{5100}$ slope $\gamma=0.415^{+0.030}_{-0.029}$ and intercept $\beta=1.350^{+0.026}_{-0.028}$. They accounted for cosmological parameter uncertainties in this analysis. Again, as their and our intercept values are not that dissimilar, we compare the two slopes and see that they differ at $2.6\sigma$ significance. This difference can be attributed to a larger fraction of higher-accreting sources in the larger, 118 AGN, sample, as argued above.

Similarly, when we look at \mii\ and \civ\ RM QSO $R-L$ relation parameters for the case of free cosmological model parameters, we obtain $\gamma=0.296 \pm  0.047$ and $\beta=1.030 \pm 0.089$ for 78 \mii\ QSOs and $\gamma=0.440 \pm 0.042$ and $\beta=1.703^{+0.064}_{-0.057}$ for 38 \civ\ QSOs in the flat $\Lambda$CDM model \citep{Cao:2022pdv}. Since these QSOs are higher-redshift and typically more luminous sources, it is quite likely that their relative accretion rates (Eddington ratios) are also larger on average than for the H$\alpha$ and H$\beta$ AGNs studied here. This can partially address the overall flatter slopes since a correlation was also found between \mii\ time-lag shortening and the source Eddington ratio \citep{2020ApJ...896..146Z}, though a more detailed analysis with independently determined SMBH masses is required to study the correlation between the departure from the best-fit $R-L$ relation and the Eddington ratio. As for the intercepts, they differ among different ionic species due to the BLR radial stratification.

\begin{figure*}[htbp]
\centering
 \subfloat[Flat \lcdm]{%
    \includegraphics[width=0.45\textwidth,height=0.35\textwidth]{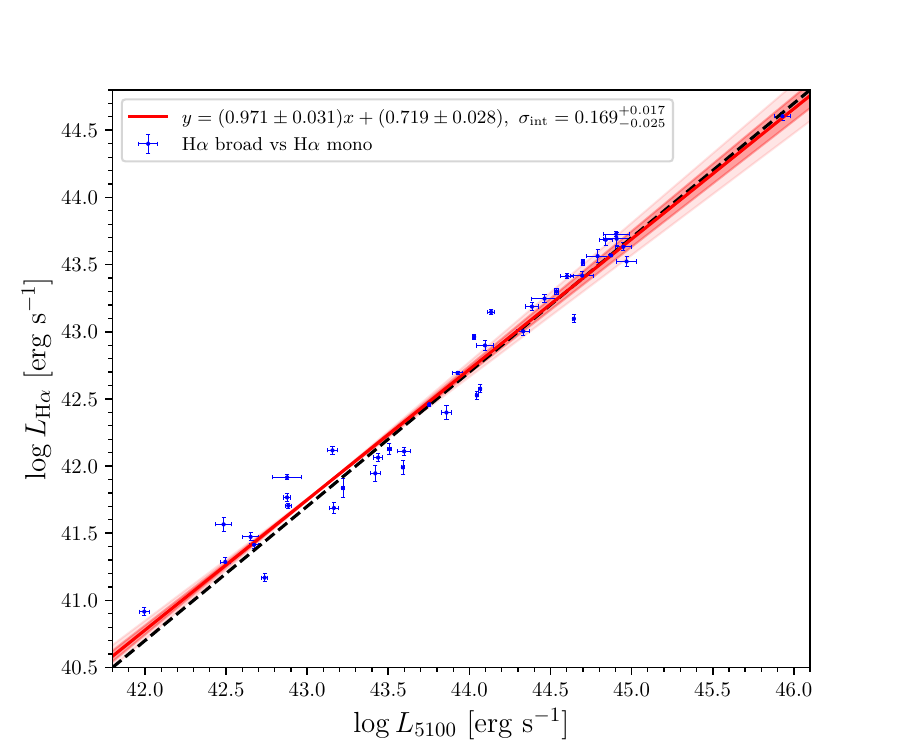}}
 \subfloat[Nonflat \lcdm]{%
    \includegraphics[width=0.45\textwidth,height=0.35\textwidth]{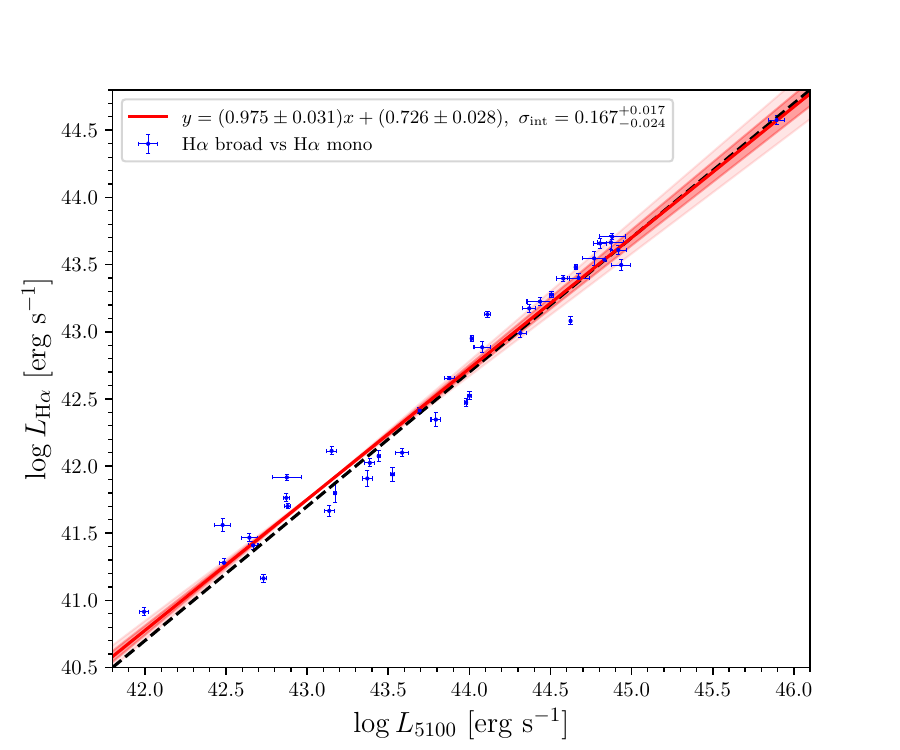}}\\
 \subfloat[Flat XCDM]{%
    \includegraphics[width=0.45\textwidth,height=0.35\textwidth]{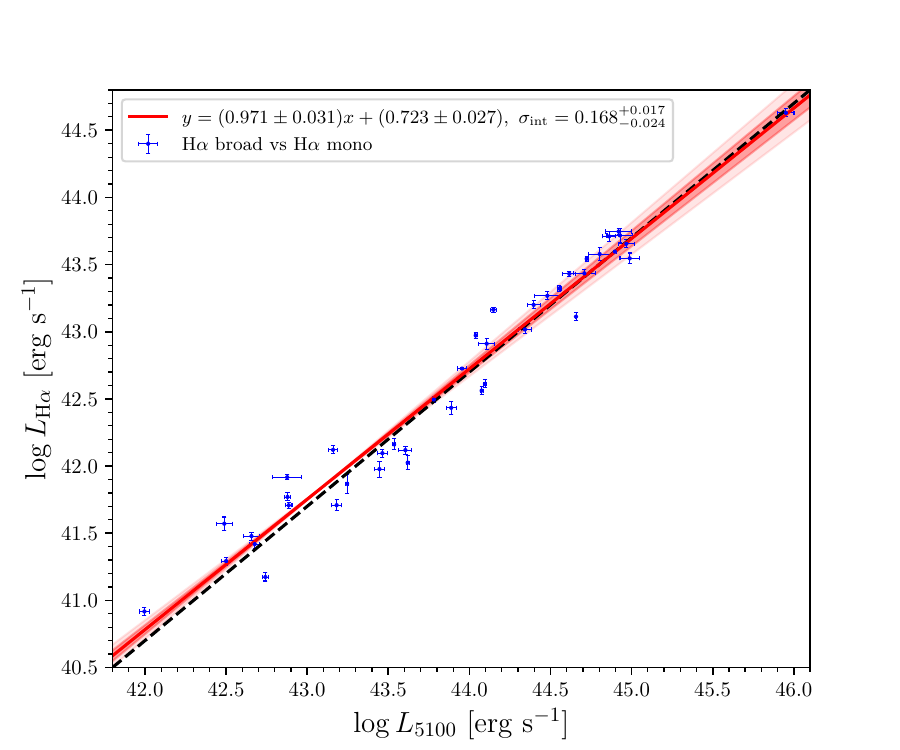}}
 \subfloat[Nonflat XCDM]{%
    \includegraphics[width=0.45\textwidth,height=0.35\textwidth]{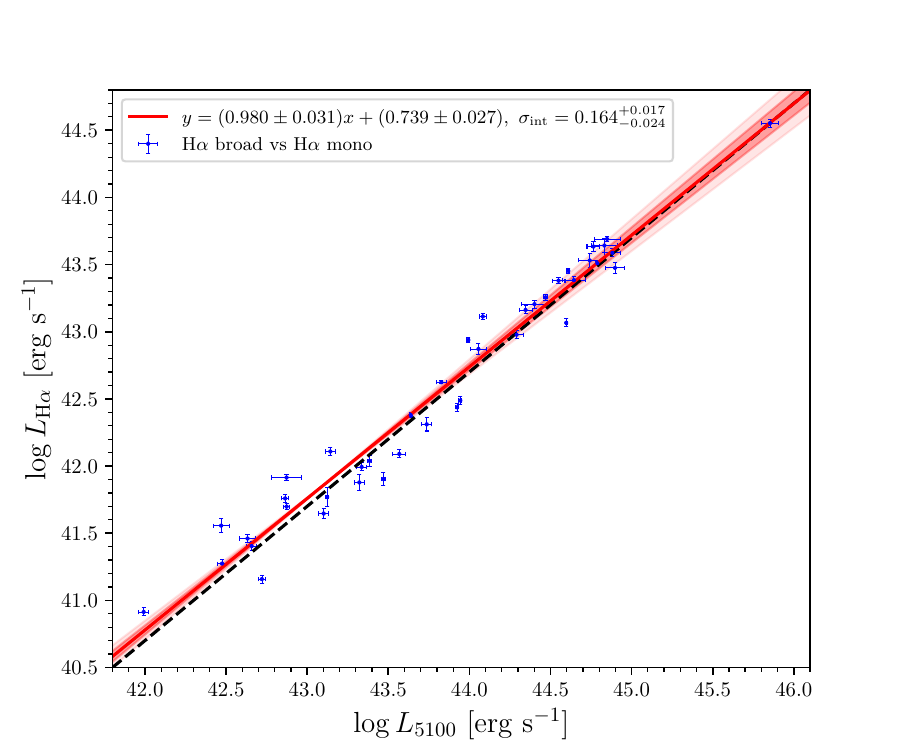}}\\
 \subfloat[Flat \pcdm]{%
    \includegraphics[width=0.45\textwidth,height=0.35\textwidth]{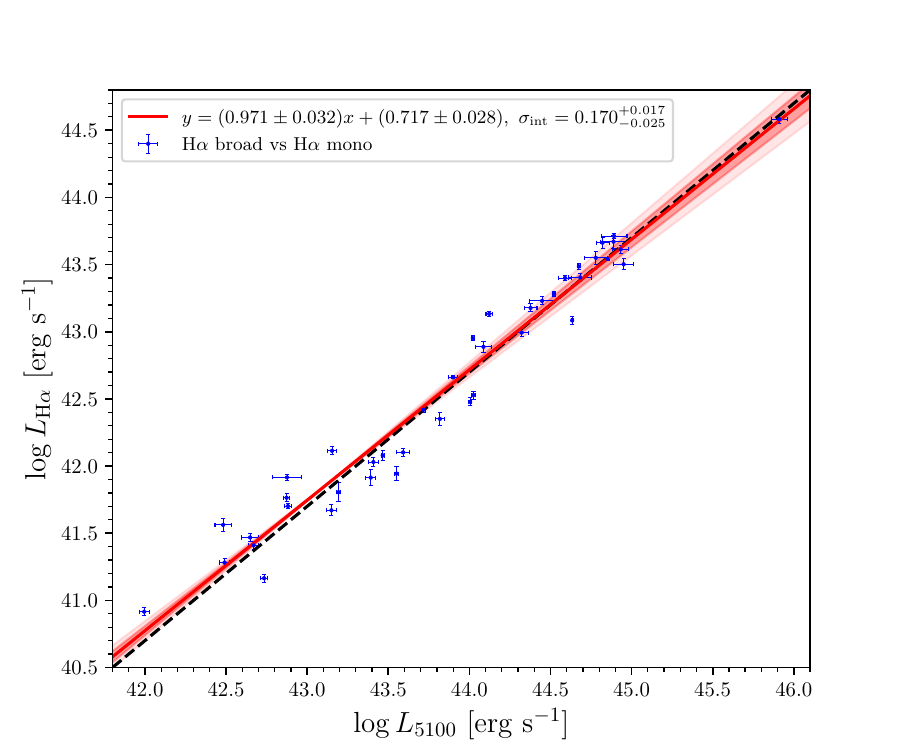}}
 \subfloat[Nonflat \pcdm]{%
    \includegraphics[width=0.45\textwidth,height=0.35\textwidth]{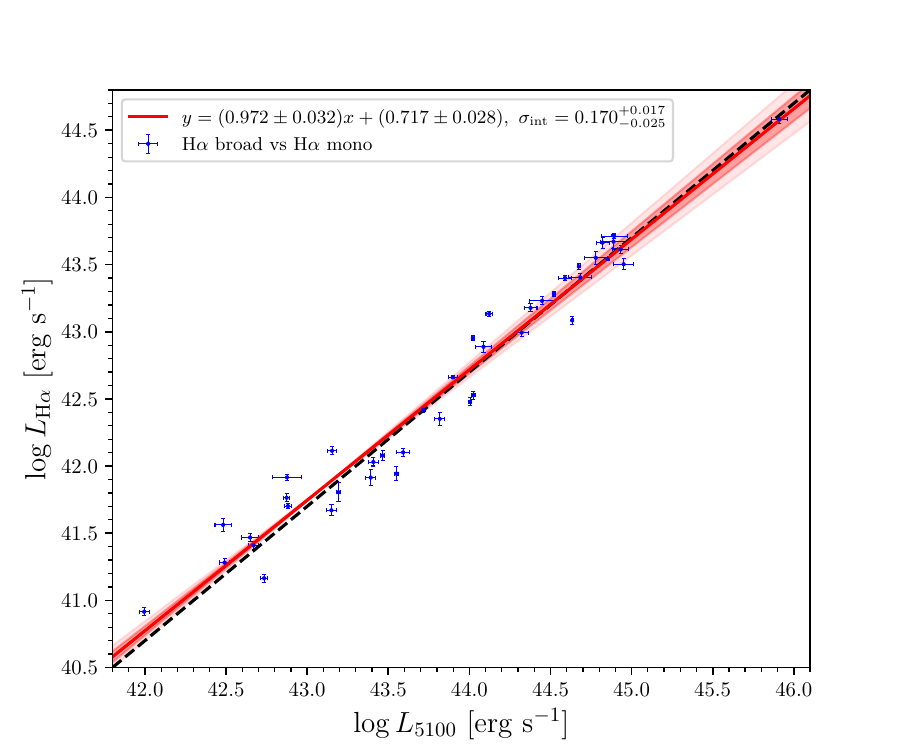}}\\
\caption{Comparisons between H$\alpha$ broad luminosity, $L_{\rm H\alpha}$, and monochromatic luminosity at 5100 \AA, $L_{5100}$, from posterior mean predictions of H$\alpha$ broad and H$\alpha$ mono for the corresponding cosmological models. In the legend, $y$ is $\log \left(L_{\rm broad}/10^{44}\, \rm erg\, s^{-1}\right)$ and $x$ is $\log \left(L_{5100}/10^{42}\, \rm erg\, s^{-1}\right)$. The black dashed line is $y=x-1.3$.}
\label{L-L1}
\end{figure*}

\begin{figure*}
\centering
\includegraphics[scale=0.5]{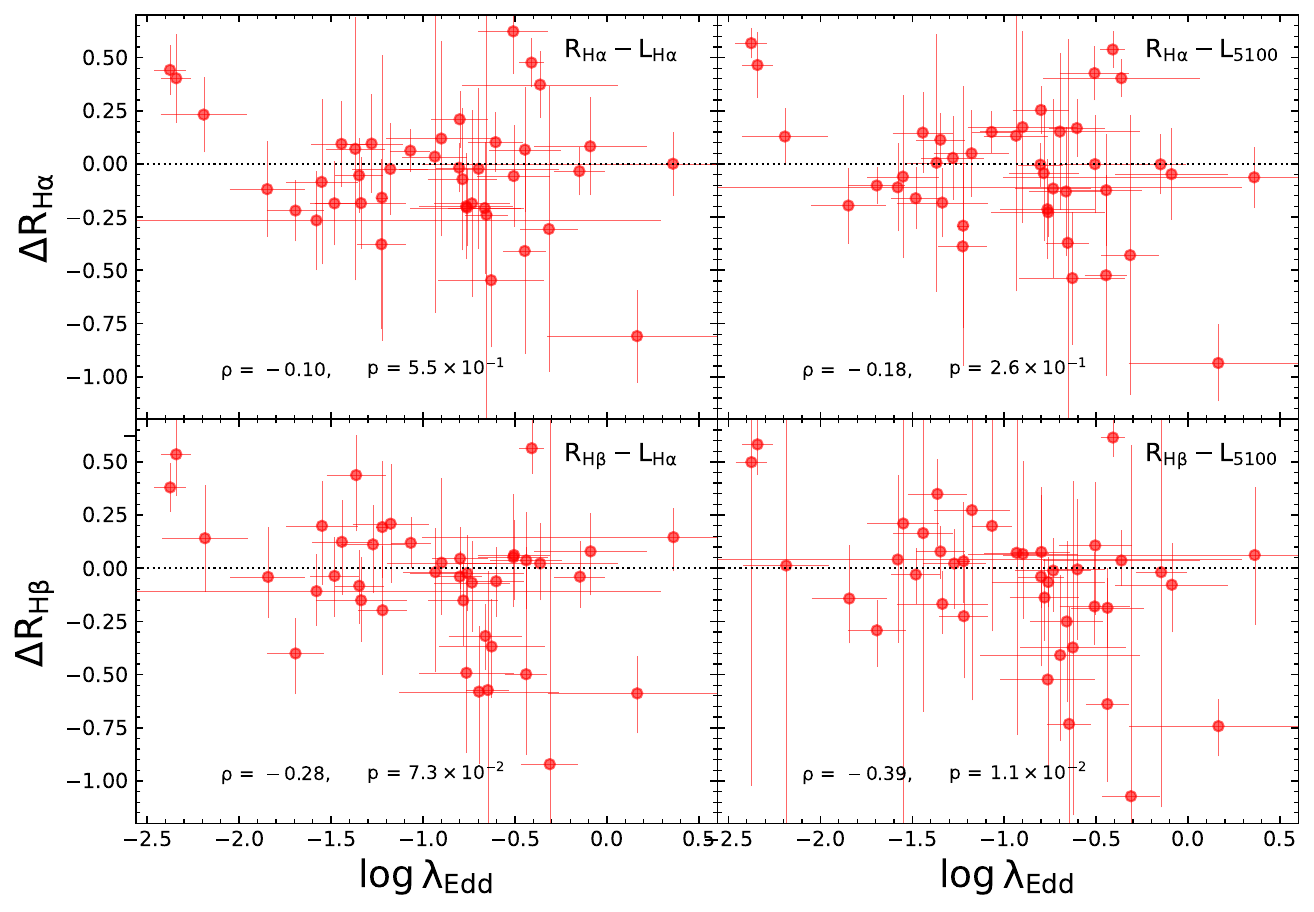}
%\resizebox{6cm}{5cm}{\includegraphics{grt_curve.pdf}}
\caption{The deviations in BLR radii, based on the H$\alpha$ and H$\beta$ lines from their respective best-fitting relations ($\Delta R_\mathrm{{H\alpha}}$, $\Delta R_\mathrm{{H\beta}}$), as a function of the Eddington ratio. Each panel also displays the Spearman's rank correlation coefficients along with the associated probabilities. The black dotted lines indicate $\Delta R$ = 0.}
\label{fig:scat}
\end{figure*}

\subsection{Comparison between $L_{\mathrm{H\alpha}}$ and $L_{5100}$ luminosities}
\label{subsec_luminosity-luminosity}

As noted above, the $R-L$ relations based on $L_{\mathrm{H\alpha}}$ exhibit slightly steeper slopes compared to those based on $L_{5100}$. To explore this difference we plot $L_{\mathrm{H\alpha}}$ as a function of $L_{5100}$ in Fig.~\ref{L-L1} for H$\alpha$ broad vs. H$\alpha$ mono combination, with the other combinations provided in Appendix \ref{A1}. We find similar patterns for the $L_{\text{H}\alpha} - L_{5100}$ relations in different data and cosmological model combinations. The best-fit slope of the $L_{\mathrm{H\alpha}} - L_{5100}$ relation is $\sim$ $0.97 \pm 0.03$, which is slightly less than unity, but consistent with unity given the uncertainties. Our results agree with those of \citet{Choetal2023}, demonstrating consistency even though we allow for varying cosmological parameters during our fitting process. This suggests that the observed trends are robust and not significantly influenced by changes in the cosmological framework.

The slightly sublinear relation observed in the $L_{\mathrm{H\alpha}} - L_{5100}$ relation suggests the presence of a weak Baldwin effect for H$\alpha$ \citep{1977ApJ...214..679B, 2024ApJS..275...13W}, i.e., an anti-correlation between the equivalent width of the BLR emission line and the central continuum luminosity. In the presence of a Baldwin effect, as the continuum luminosity increases the BLR H$\alpha$ emission line luminosity does not increase as rapidly. This results in a slightly sub-unity slope in the $L_{\mathrm{H\alpha}} - L_{5100}$ relation, consistent with the weak indication for a slope of 0.97 that we find here. Such a behavior potentially contributes to the steeper slopes observed in the $R-L$ relations based on $L_{\mathrm{H\alpha}}$ relative to those based on $L_{5100}$. However, considering the uncertainties, all of these $R-L$ relations are mutually consistent.

\subsection{Scatter in the radius-luminosity relation}
\label{subsec_scatter}

We now explore the potential correlation between the scatter around the $R-L$ relations and the Eddington ratio ($\lambda_{\text{Edd}}$). Since the various cosmological models used in this work yield consistent $R-L$ relations, we adopt the flat $\Lambda$CDM cosmological model $R-L$ relations for this analysis. To quantify the deviations of the measured BLR radii from the best-fit relations, we define $\Delta R = {\rm log}R - {\rm log}R_{\text{best-fit}}$ \citep[see e.g.][]{MartinezAldama2019,2020ApJ...896..146Z}, where $R$ is the measured BLR radius while $R_{\text{best-fit}}$ is the BLR radius expected from the best-fit $R-L$ relation. In order to compute the Eddington ratios ($L_{\text{bol}}/L_{\text{Edd}}$), we compile $M_{BH}$ values based on measurements of the H$\beta$ line from RM of our sample AGNs and use the relations $L_{\text{bol}} = 9.26 \times L_{5100}$ \citep{2006ApJS..166..470R} and $L_{\text{Edd}} = 1.26 \times 10^{38} M_{BH} / M_{\odot}$.

Figure \ref{fig:scat} presents the comparison between $\Delta R$ and $\lambda_{\text{Edd}}$. We detect a weak negative correlation between $\Delta R$ and $\lambda_{\text{Edd}}$, with Spearman’s rank correlation coefficient of $\rho$ = $-0.10$ ($-0.18$) and corresponding probability of $p = 5.5 \times 10^{-1}$ ($2.6 \times 10^{-1}$) for the $R_{\mathrm{H\alpha}} - L_{\mathrm{H\alpha}}$ ($R_{\mathrm{H\alpha}} - L_{5100}$) relation. Interestingly, this negative correlation becomes more pronounced for the $R_{\mathrm{H\beta}} - L_{\mathrm{H\alpha}}$ ($R_{\mathrm{H\beta}} - L_{5100}$) relation, where we find $\rho$ = $-0.28$ ($-0.39$) and $p = 7.3 \times 10^{-2}$ ($1.1 \times 10^{-2}$). These results suggest a trend in which the BLR radius decreases as $\lambda_{\text{Edd}}$ increases, reflecting a higher accretion rate.

However, it is crucial to note that $\lambda_{\text{Edd}}$ scales  with $\sim$ $L/R_{\text{BLR}}$, while the change in $\Delta R$ follows $\sim$ $R_{\text{BLR}}/L^{0.6}$, as indicated by our findings. Consequently, the observed negative correlation between $\Delta R$ and $\lambda_{\text{Edd}}$ could primarily arise from a self-correlation between $\lambda_{\text{Edd}}$ and the deviations from the best-fit relation \citep{2020ApJ...899...73F, 2024ApJ...962...67W}. Moreover, the stronger correlation observed when using $L_{5100}$ in the $R-L$ relations may also be attributable to this self-correlation effect.

\begin{table*}[htbp]
\centering
\caption{Summary of the AGN/QSO data constraints (marginalized 1$\sigma$ and 2$\sigma$ limits; unmarginalized best-fitting values)  for $\Omega_{m0}$ in the flat $\Lambda$CDM cosmological model from lower-redshift H$\alpha$ and H$\beta$ AGN and higher-redshift \mii\ and \civ\ QSO samples. The adopted radius-luminosity relation involves a corresponding monochromatic luminosity.}
\setlength\tabcolsep{11.5pt}
\begin{tabular}{l l c r r r c}
\toprule\toprule
$z_{\rm min}$ & $z_{\rm max}$ & BLR line &$\Omega_{m0}$ (1$\sigma$)  & $\Omega_{m0}$ (2$\sigma$) & $\Omega_{m0}$ (best) & Reference\\
\midrule
0.0041488 & 0.4741278  & H$\alpha$ (41) & $>0.487$ &  [0.05, 1]  &  0.999   &  this~work \\
0.0041488 & 0.4741278  & H$\beta$ (41) & $>0.437$ & [0.05, 1]  & 0.997   &  this~work \\
0.002 & 0.89   & H$\beta$ (118) & $>0.336$ &  [0,1]  &  0.998   &  \citet{Khadkaetal2021c} \\
%0.0033     &  1.89  & Mg II(78) &  0.060 &  0.270 &  0.670 & \citet{Khadkaetal_2021a} \\
0.0033 & 1.89 & \mii\ (78)  & $0.470^{+0.196}_{-0.430}$  & [0,0.862]  & 0.148  &  \citet{Cao:2022pdv}\\
0.001064 & 3.368 & \civ\ (38)  & $<0.503$  & [0,1] & $0.046$  &  \citet{Cao:2022pdv}\\
0.001064 & 3.368   & \mii\ + \civ\ (116) & $<0.444$  & [0,1]  & $0.068$  &  \citet{Cao:2022pdv}\\
\bottomrule\bottomrule
\end{tabular}
\label{tab:fit_trend}
\end{table*}

Prior studies have indicated that super-Eddington accreting massive AGNs generally exhibit smaller BLR radii for a given AGN luminosity, with this deviation becoming more pronounced at higher $\lambda_{\text{Edd}}$ or elevated accretion rates, \citep{2015ApJ...806...22D} and later confirmed by \citep{2024ApJS..275...13W}. The phenomenon was attributed to the self-shadowing effect of the slim disk model, where the thickness of the inner slim disk reduces the ionizing radiation reaching the BLR, leading to a shortening of H$\alpha$ and/or H$\beta$ lags as the accretion rate increases. However, \citet{2020ApJ...899...73F} suggested that the deviation from the radius--luminosity relation depends on the shape of the UV/optical SED, the relative contribution of ionizing radiation, or the influence of black hole spin \citep[see][for details]{2019ApJ...870...84C}. Furthermore, similar shortening of torus sizes with increasing accretion rates have been observed by \citet{2024ApJ...968...59M} and \citet{2023MNRAS.522.3439C}. All of these findings align with our results concerning the observed shortening of H$\alpha$ and H$\beta$ lags in relation to Eddington ratios.

\subsection{Comparison to our earlier analyses}
\label{subsec_differences}

Here we first discuss the difference between our previous cosmological model parameter results for $R_{\mathrm{H\beta}} - L_{5100}$ low-redshift sources \citep{Khadkaetal2021c} and the current results. We emphasize that in both cases the RM AGN datasets do not significantly constrain cosmological parameters so our discussion here is more qualitative than for the case of the $R-L$ relations, which these data constrain quite well. The previous H$\beta$ sample was larger (118 objects) than the current 41 AGN one by a factor of nearly three, but the current sample is a more uniform one compared to the earlier sample. The dispersions from the two datasets are comparable, being $0.236^{+0.020}_{-0.018}$ there and $0.271^{+0.034}_{-0.049}$ here, for the flat \lcdm\ cosmological model, and consistent with the sample sizes the error bars here are more than twice as large as those in the larger 118 AGN case.

In both cases there are only weak limits on \om, no 2$\sigma$ AGN data constraints on \om\ here (all 2$\sigma$ limits here are determined by the prior), and the same is true for the earlier 118 source dataset. However, from the 1D \om\ likelihoods in figure 3 of \citep{Khadkaetal2021c} we see that the 118 source sample favors large \om\ while from the corresponding panels in Fig.~\ref{fig2} for the H$\alpha$ broad AGN dataset, the 1D \om\ distributions from the current 41 AGN sample are relatively flat and do not as significantly favor large \om\ values, a conclusion that is supported by the other datasets shown in Appendix \ref{A1}.

These findings are consistent with two other results. \citet{Khadkaetal2021c} noted that their 118 AGN cosmological constraint contours were more consistent with currently decelerated cosmological expansion as well as being $\sim 2 \sigma$ away from the corresponding better-established $H(z)$ + BAO data cosmological constraint contours. This differs from what we find here, where the AGN data cosmological constraint contours are all within 2$\sigma$ of the currently accelerating cosmological expansion part of parameter space, and also are consistent with better-established $H(z)$ + BAO data cosmological constraint contours.   

These differences are presumably partly a consequence of the 41 AGN dataset being more homogeneous than the 118 AGN one, and partly a consequence of the smaller dataset resulting in correspondingly larger error bars. Determining which of these is the dominant effect is important. If a larger but equally homogeneous AGN compilation, with correspondingly smaller error bars, gives cosmological constraints that remain consistent with those from better established data, such as $H(z)$ + BAO data, then the RM AGN $R-L$ method we have developed could become a valuable cosmological test, provided that more and better RM AGN data also result in a smaller dispersion.    

Our previous results based on larger \mii\ and \civ\ RM QSO compilations, that span a larger redshift range, gave tighter constraints on the current matter density parameter, see Table~\ref{tab:fit_trend} for the comparison of the $\Omega_{\rm m0}$ constraints for the flat $\Lambda$CDM model between the lower-redshift H$\alpha$ and H$\beta$ AGNs and the higher-redshift \mii\ and \civ\ QSOs. The intermediate-redshift \mii\ QSOs gave both upper and lower limits on $\Omega_{m0}$ \citep{Khadkaetal_2021a,Cao:2022pdv}, while the highest-redshift \civ\ sample gives only an upper limit on $\Omega_{m0}$ \citep{Cao:2022pdv}. Assuming there are no undiscovered systematic errors, this may suggest that by combining samples and so covering a larger redshift range we can get more restrictive cosmological constraints. We did this from a joint analysis of the combined \mii\ and \civ\ samples \citep{Cao:2022pdv}. Our  current low-redshift H$\alpha$ and H$\beta$ samples are too small to justify doing a combined analysis with the \mii\ + \civ\ compilation, but this could be viable in the future with a significantly larger, homogeneous H$\alpha$ or H$\beta$ sample.

\section{Conclusion}
\label{sec:conclusion}

In this paper we investigate BLR radius--luminosity relations while simultaneously constraining cosmological model parameters. This study considers six cosmological models and uses the $R-L$ relations derived from a small but homogeneous sample of AGNs with a consistent time-lag determination based on both H$\alpha$ and H$\beta$ emission lines. Specifically, we derive $R-L$ relations for four distinct combinations of datasets: H$\alpha$ time lag -- H$\alpha$ luminosity, H$\alpha$ time lag -- monochromatic luminosity, H$\beta$ time lag -- H$\alpha$ luminosity, and H$\beta$ time lag -- monochromatic luminosity. We improve upon previous analyses by correcting for peculiar velocity contamination, incorporating cosmological parameter uncertainties, and examining six cosmological models to assess source standardizability.

Our results demonstrate consistent $R-L$ relations across various cosmological models for all four RM AGN datasets when using comparable luminosities ($L_{\text{H}\alpha}$ or $L_{5100}$). This consistency suggests that these $R-L$ relations can effectively standardize all four datasets. However, we observe a steeper slope than the theoretically expected value of 0.5 derived from a simple photoionization model of the BLR and observed values from prior H$\beta$ RM campaigns. This deviation is primarily due to differences in sample selection, as our sample is predominantly comprised of AGNs with low to moderate Eddington ratios that exhibit zero to minimal shortened time lags with respect to the best-fitting $R-L$ relation.

We also find that the $R-L$ relations have larger intercepts when derived using H$\alpha$ lags compared to those derived using H$\beta$ lags, likely due to ionization stratification within the BLR. Moreover, the presence of a mild Baldwin effect in H$\alpha$ could contribute to the steeper slopes seen in $R-L$ relations derived from $L_{H\alpha}$ compared to those based on $L_{5100}$.

Regarding cosmological parameters, our samples of 41 H$\alpha$ and H$\beta$ AGNs provide weak constraints, consistent within 2$\sigma$ (or better) with those from better-established cosmological probes, in particular the $H(z)$ + BAO dataset. Given this consistency, joint analysis of the RM AGN sample with the $H(z)$ + BAO data compilation is warranted. The resulting constraints differ only slightly (typically by $\lesssim 0.1\sigma$, depending on the parameter and model), from those derived using $H(z)$ + BAO. Based on the deviance information criterion, the joint analysis favors the flat $\phi$CDM model, with only weak to positive evidence against the other models.

These results are primarily due to the small size of our RM AGN samples. Determining the viability of RM H$\alpha$ and H$\beta$ sources as cosmological probes requires a larger, homogeneous sample. Future surveys, such as the SDSS-V Black Hole Mapper \citep{SDSS_BH_Mapper_2017arXiv171103234K} and the Vera C. Rubin Observatory Legacy Survey of Space and Time \citep{Kovacevic:2022msi,2023A&A...675A.163C}, are expected to provide thousands of such measurements, sufficient for this purpose.
\\

\begin{acknowledgments}
 The computations for this project were partially performed on the Beocat Research Cluster at Kansas State University, which is funded in part by NSF grants CNS-1006860, EPS-1006860, EPS-0919443, ACI-1440548, CHE-1726332, and NIH P20GM113109. This project has received funding from the European Research Council (ERC) under the European Union’s Horizon 2020 research and innovation program (grant agreement No.~[951549]). MZ acknowledges the support of the Czech Science Foundation Junior Star grant no.~GM24-10599M. BC and MZ ackowledge the OPUS-LAP/GA\v{C}R-LA bilateral project (2021/43/I/ST9/01352/OPUS22 and GF23-04053L).
\end{acknowledgments}

\appendix

\section{Figures for supporting data}
\label{A1}

\begin{figure*}
\centering
 \subfloat[Flat \lcdm]{%
    \includegraphics[width=0.4\textwidth,height=0.35\textwidth]{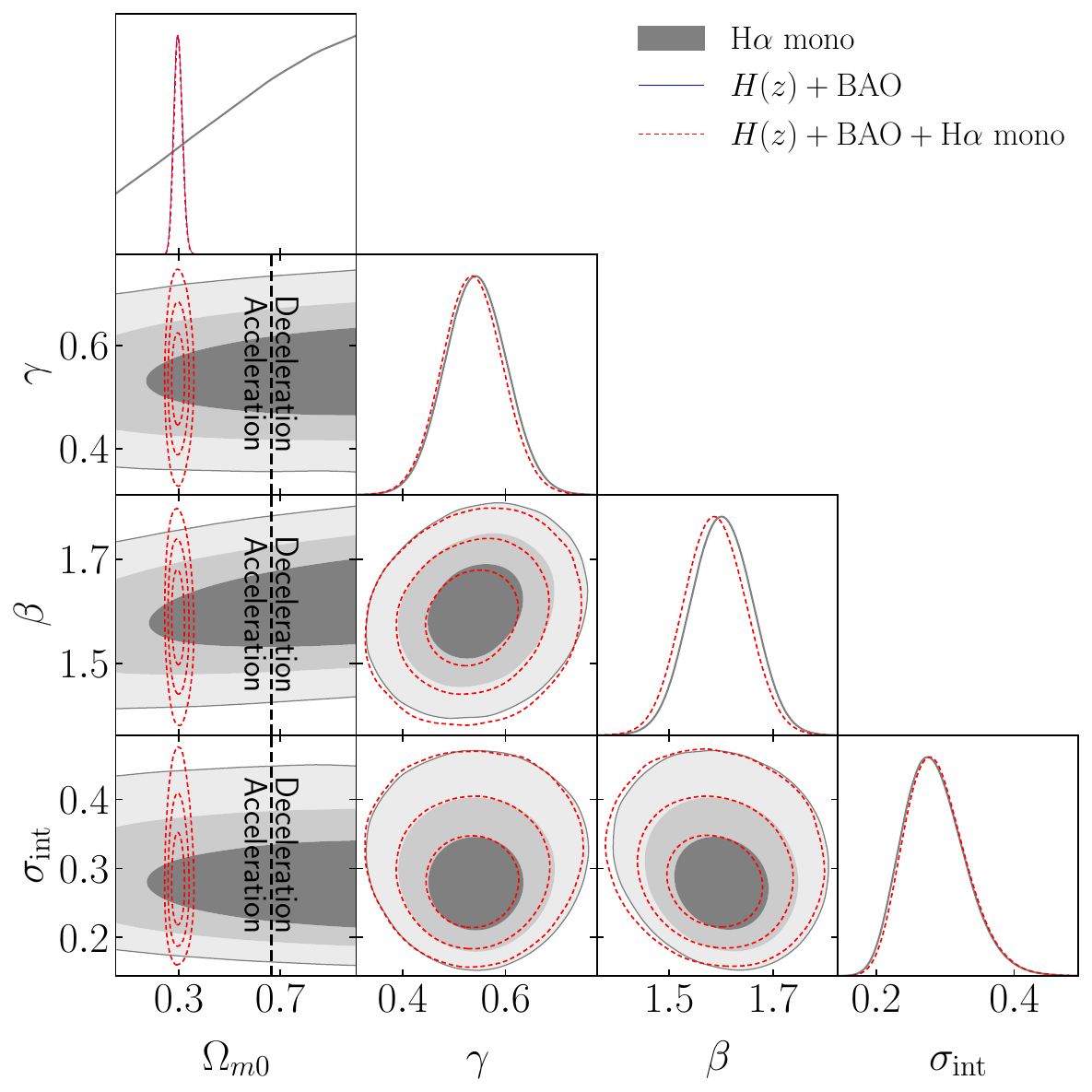}}
 \subfloat[Nonflat \lcdm]{%
    \includegraphics[width=0.4\textwidth,height=0.35\textwidth]{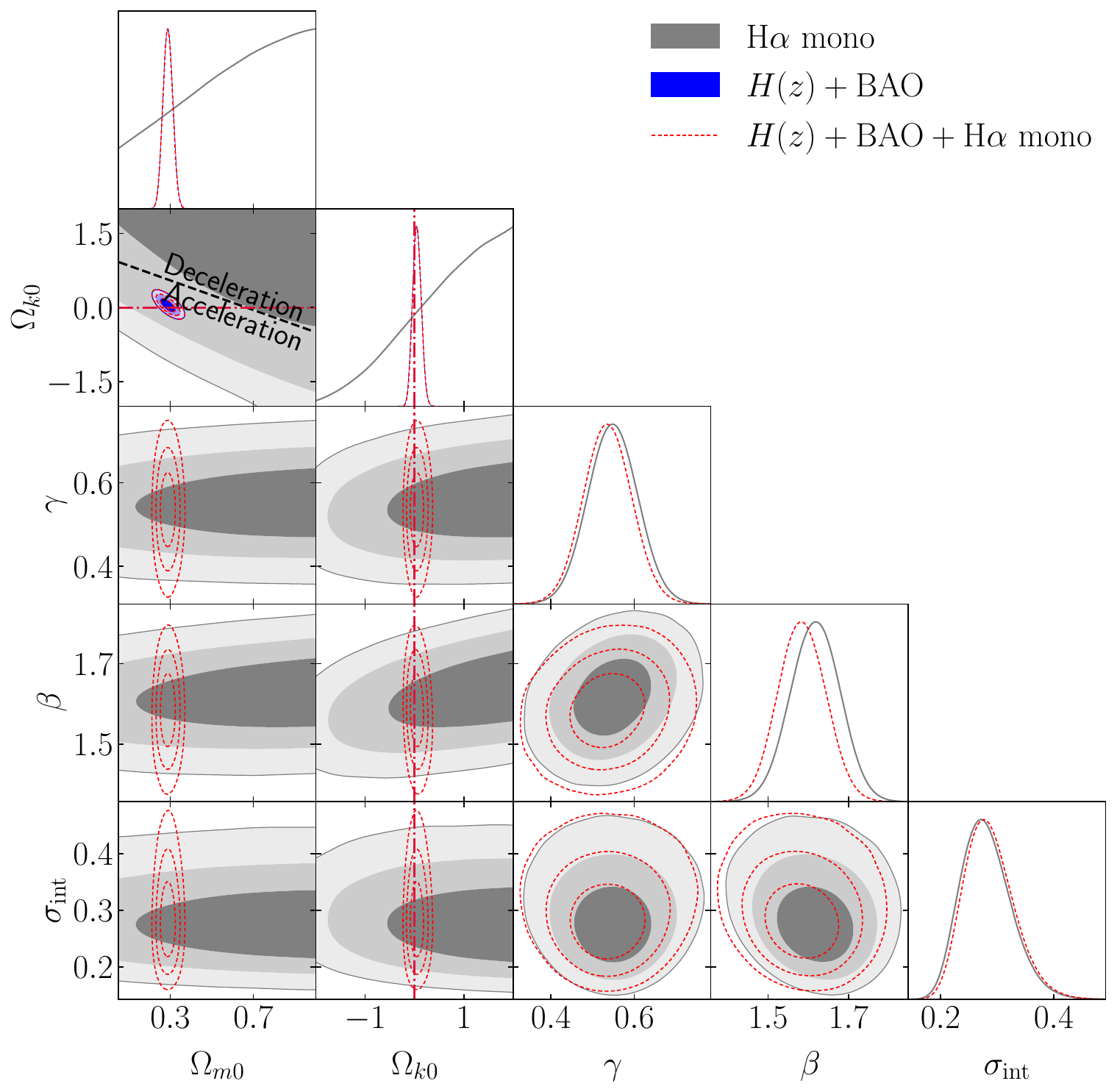}}\\
 \subfloat[Flat XCDM]{%
    \includegraphics[width=0.4\textwidth,height=0.35\textwidth]{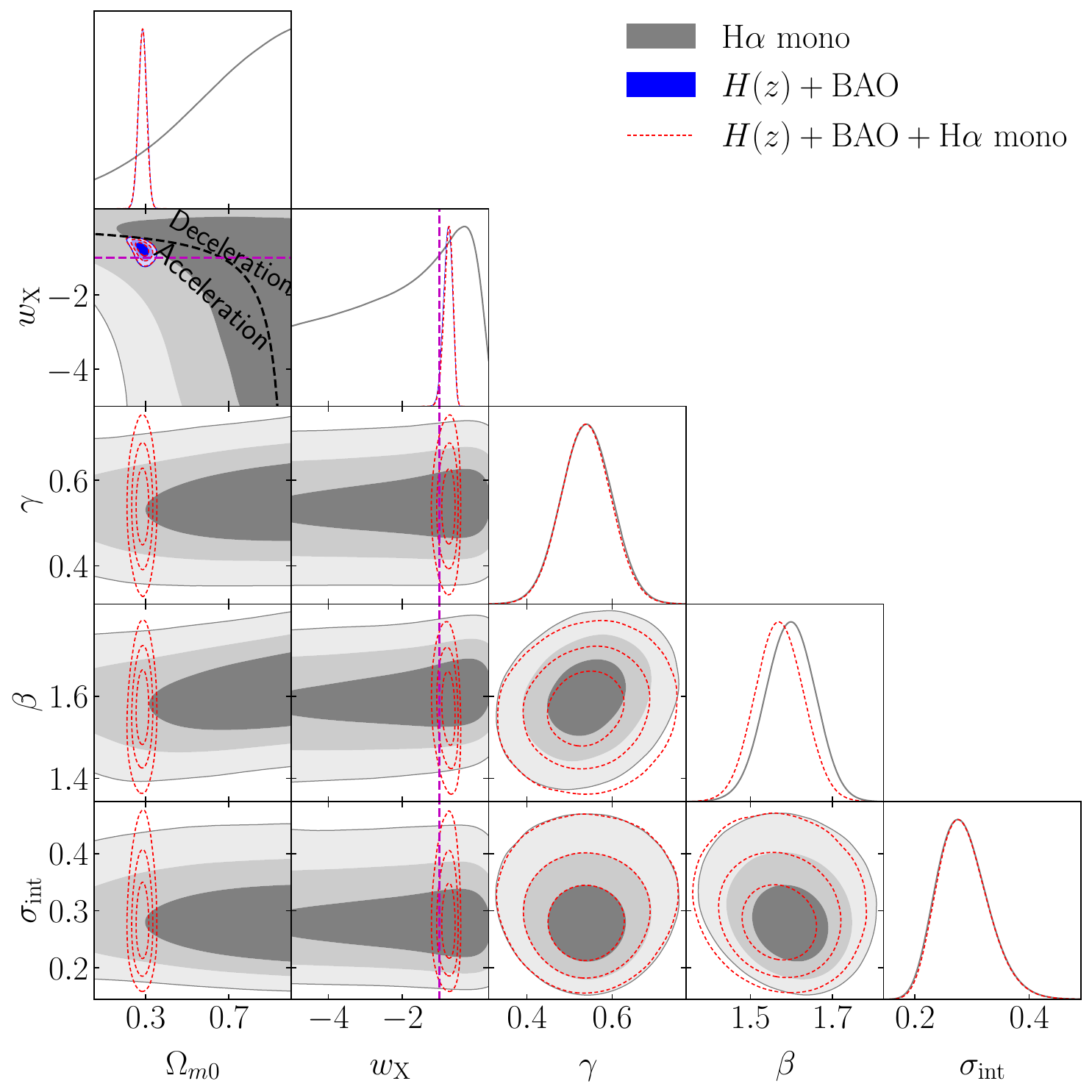}}
 \subfloat[Nonflat XCDM]{%
    \includegraphics[width=0.4\textwidth,height=0.35\textwidth]{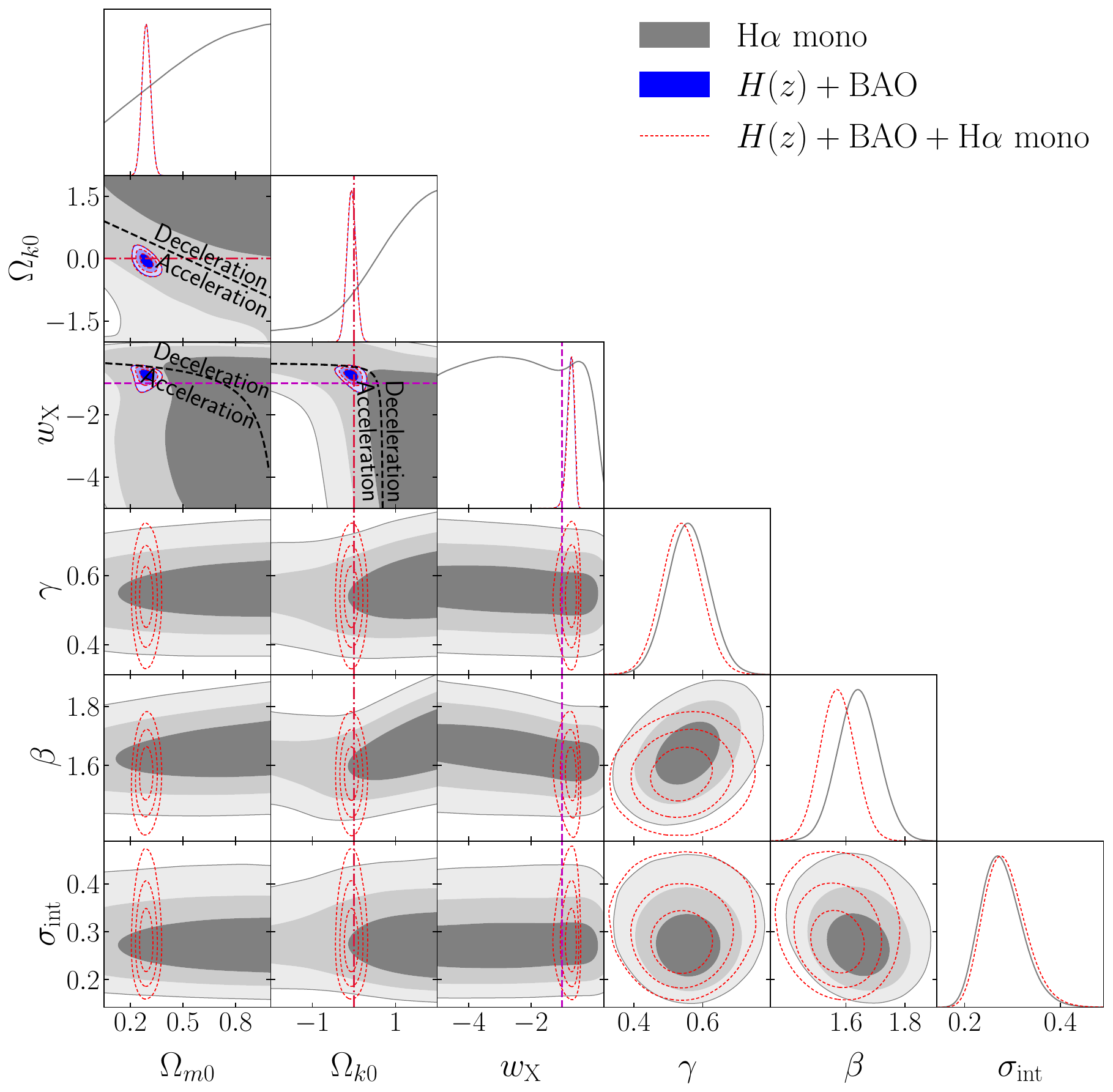}}\\
 \subfloat[Flat \pcdm]{%
    \includegraphics[width=0.4\textwidth,height=0.35\textwidth]{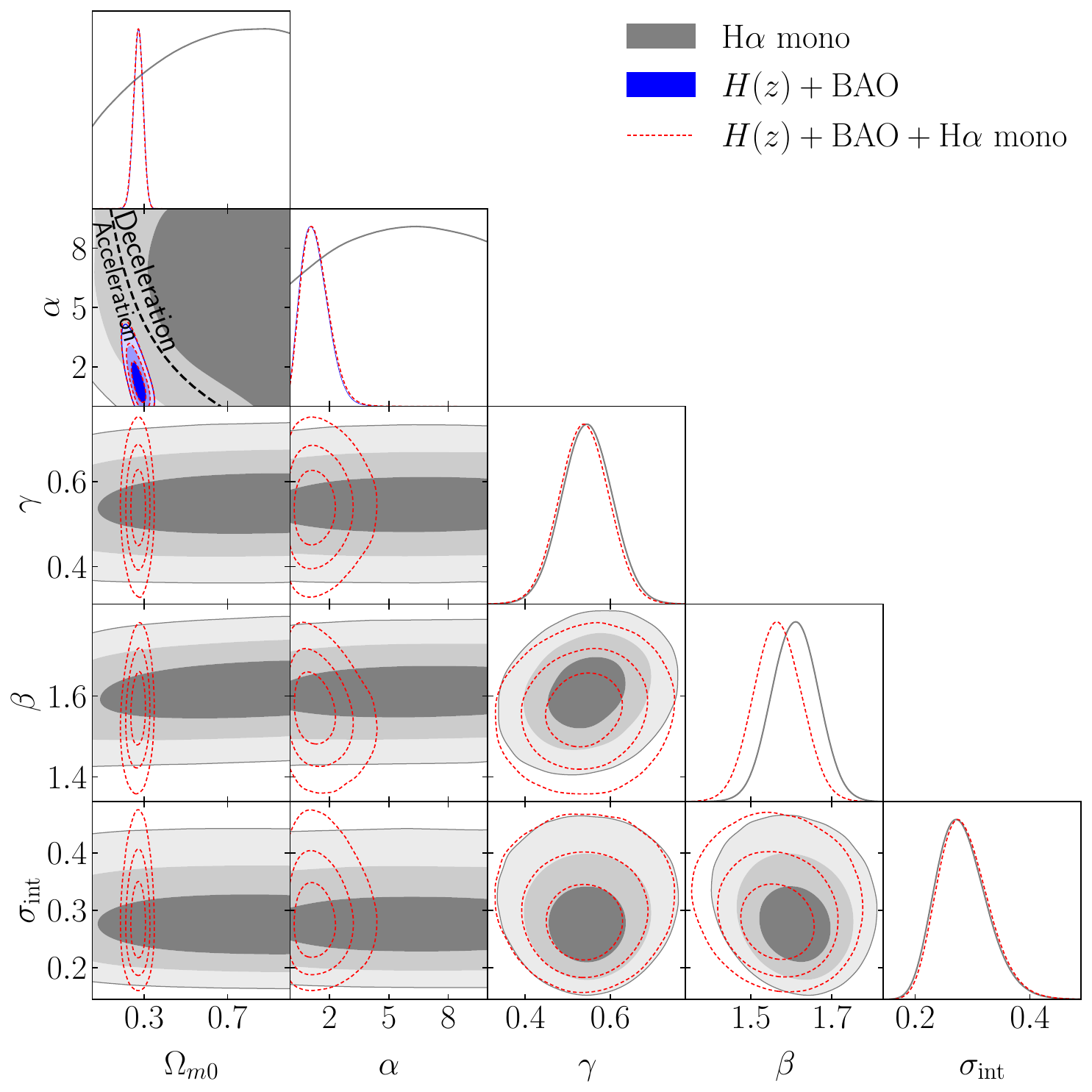}}
 \subfloat[Nonflat \pcdm]{%
    \includegraphics[width=0.4\textwidth,height=0.35\textwidth]{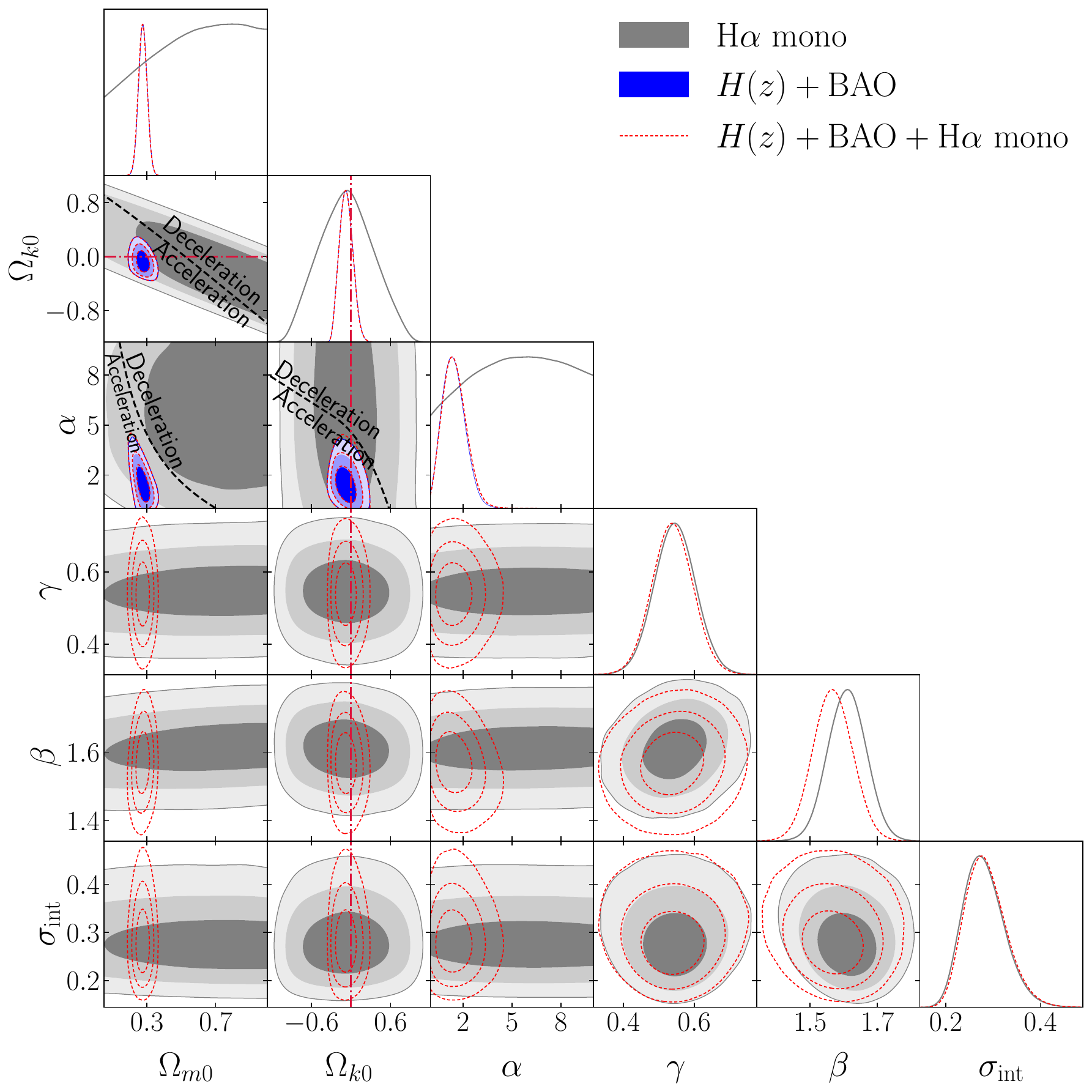}}\\
\caption{Same as Fig.\ \ref{fig1}, but for RM AGN of H$\alpha$ mono (gray), $H(z)$ + BAO (blue), and $H(z)$ + BAO + H$\alpha$ mono (dashed red) data.}
\label{fig3}
\end{figure*}

\begin{figure*}
\centering
 \subfloat[Flat \lcdm]{%
    \includegraphics[width=0.4\textwidth,height=0.35\textwidth]{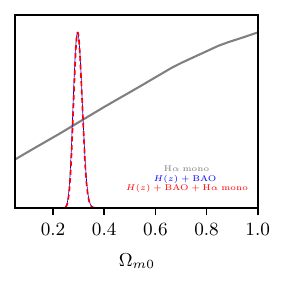}}
 \subfloat[Nonflat \lcdm]{%
    \includegraphics[width=0.4\textwidth,height=0.35\textwidth]{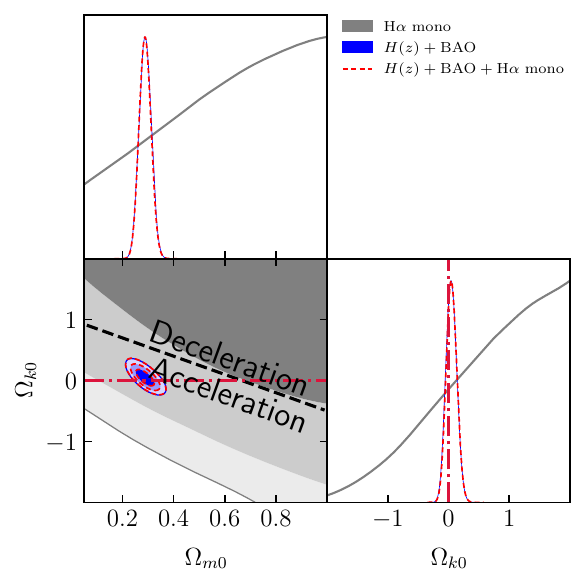}}\\
 \subfloat[Flat XCDM]{%
    \includegraphics[width=0.4\textwidth,height=0.35\textwidth]{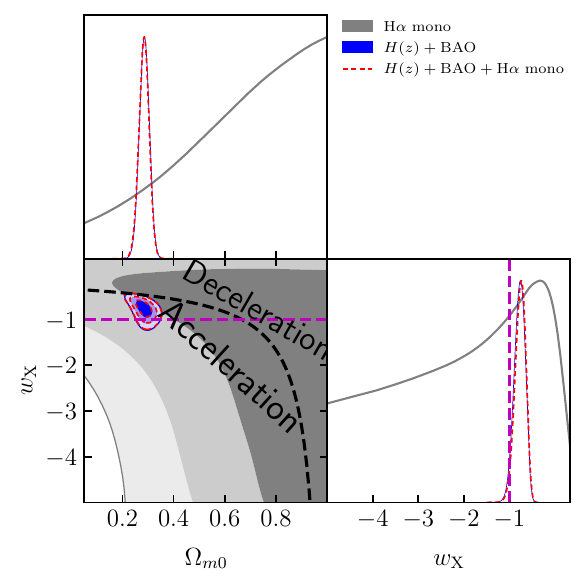}}
 \subfloat[Nonflat XCDM]{%
    \includegraphics[width=0.4\textwidth,height=0.35\textwidth]{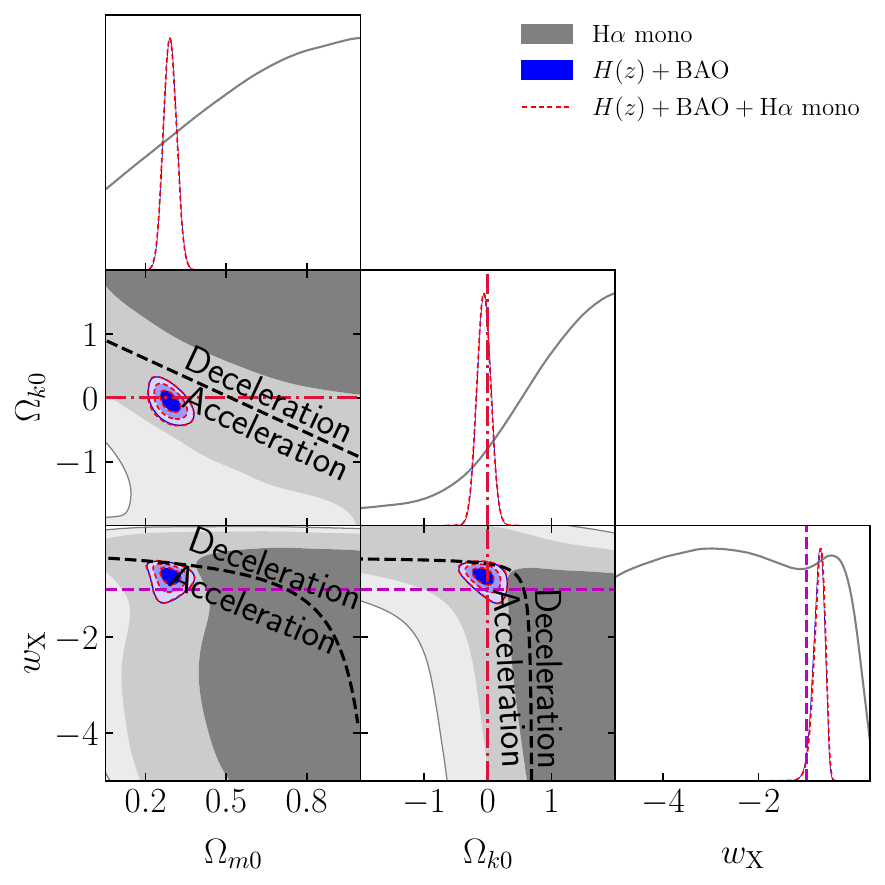}}\\
 \subfloat[Flat \pcdm]{%
    \includegraphics[width=0.4\textwidth,height=0.35\textwidth]{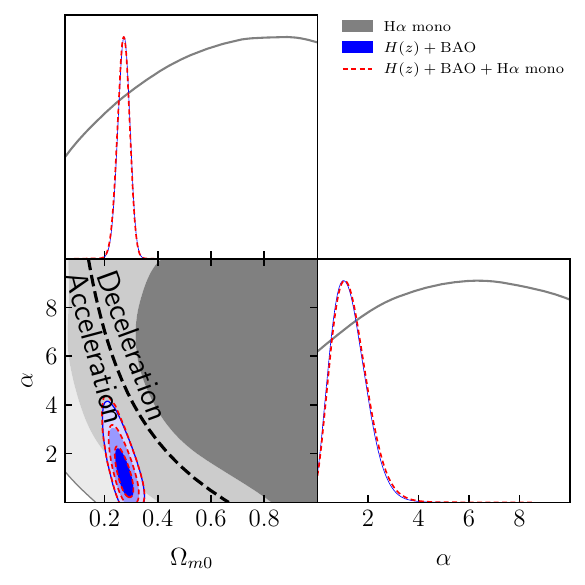}}
 \subfloat[Nonflat \pcdm]{%
    \includegraphics[width=0.4\textwidth,height=0.35\textwidth]{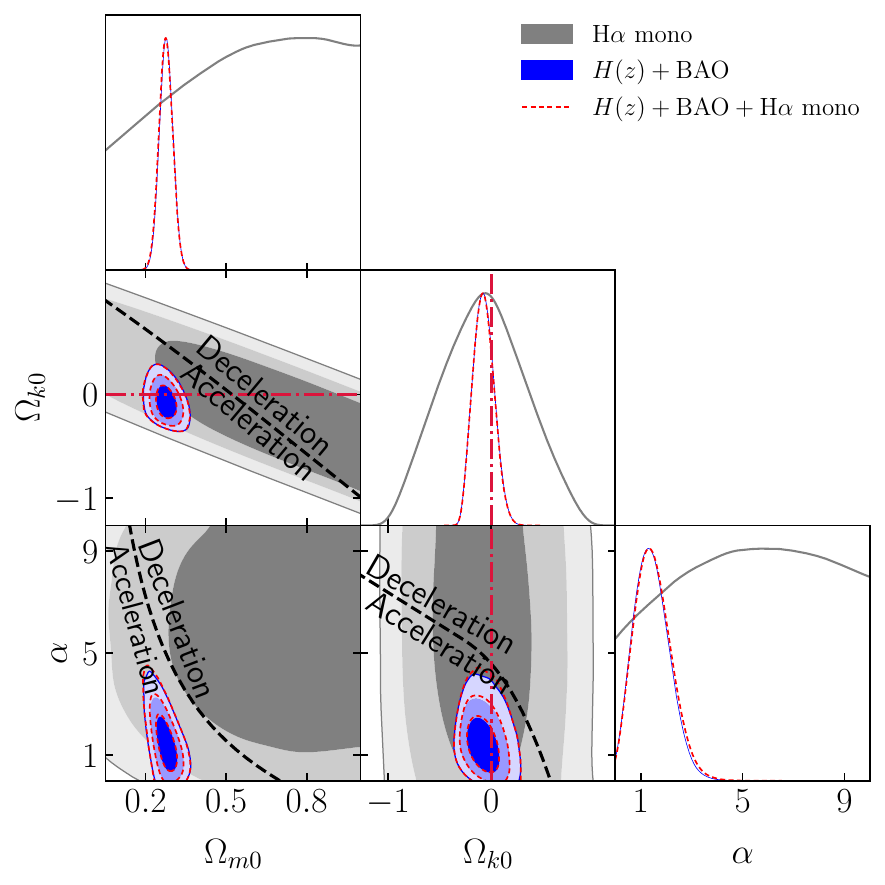}}\\
\caption{Same as Fig.\ \ref{fig3}, but for cosmological parameters only.}
\label{fig4}
\end{figure*}

\begin{figure*}
\centering
 \subfloat[Flat \lcdm]{%
    \includegraphics[width=0.4\textwidth,height=0.35\textwidth]{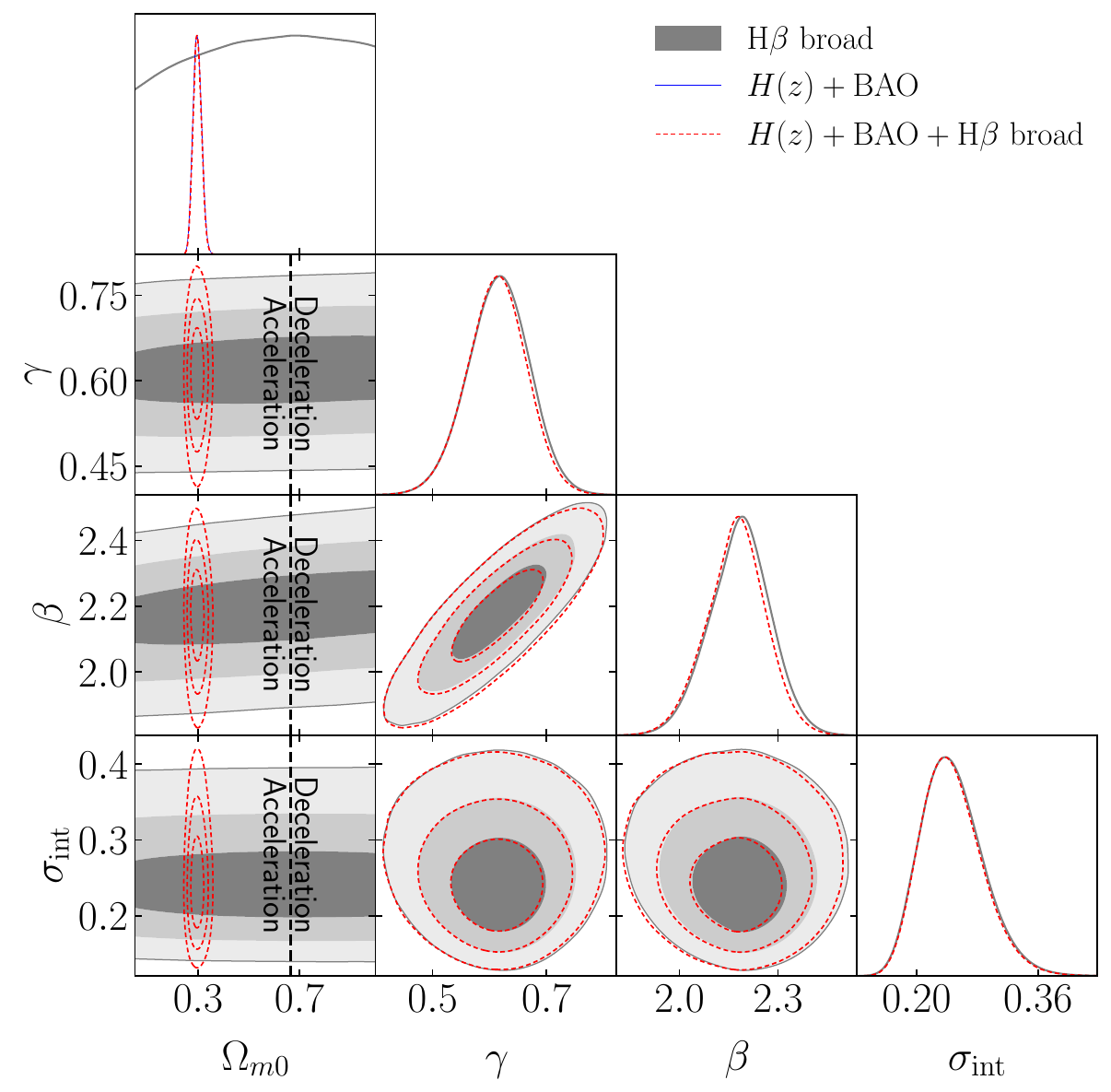}}
 \subfloat[Nonflat \lcdm]{%
    \includegraphics[width=0.4\textwidth,height=0.35\textwidth]{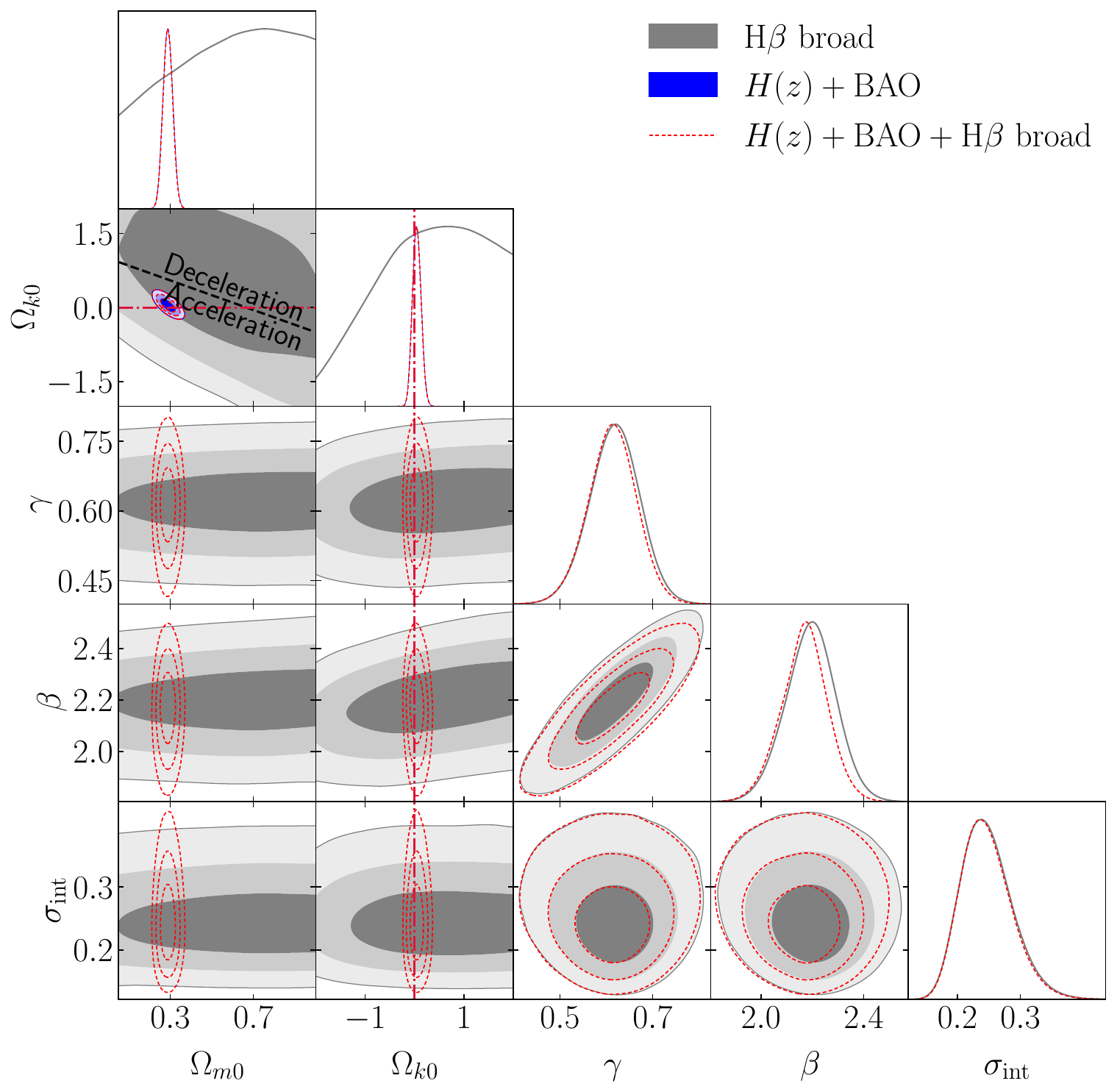}}\\
 \subfloat[Flat XCDM]{%
    \includegraphics[width=0.4\textwidth,height=0.35\textwidth]{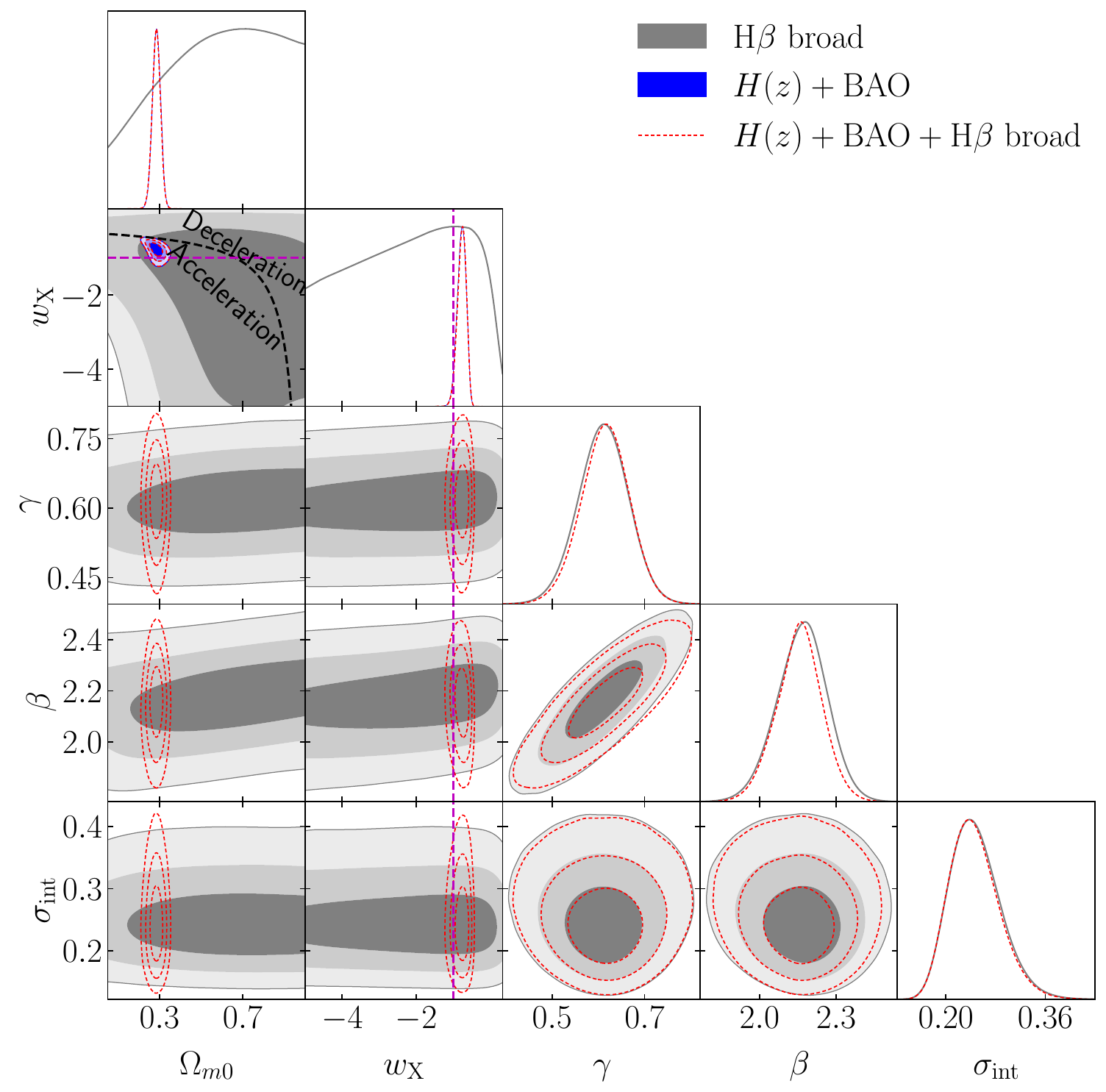}}
 \subfloat[Nonflat XCDM]{%
    \includegraphics[width=0.4\textwidth,height=0.35\textwidth]{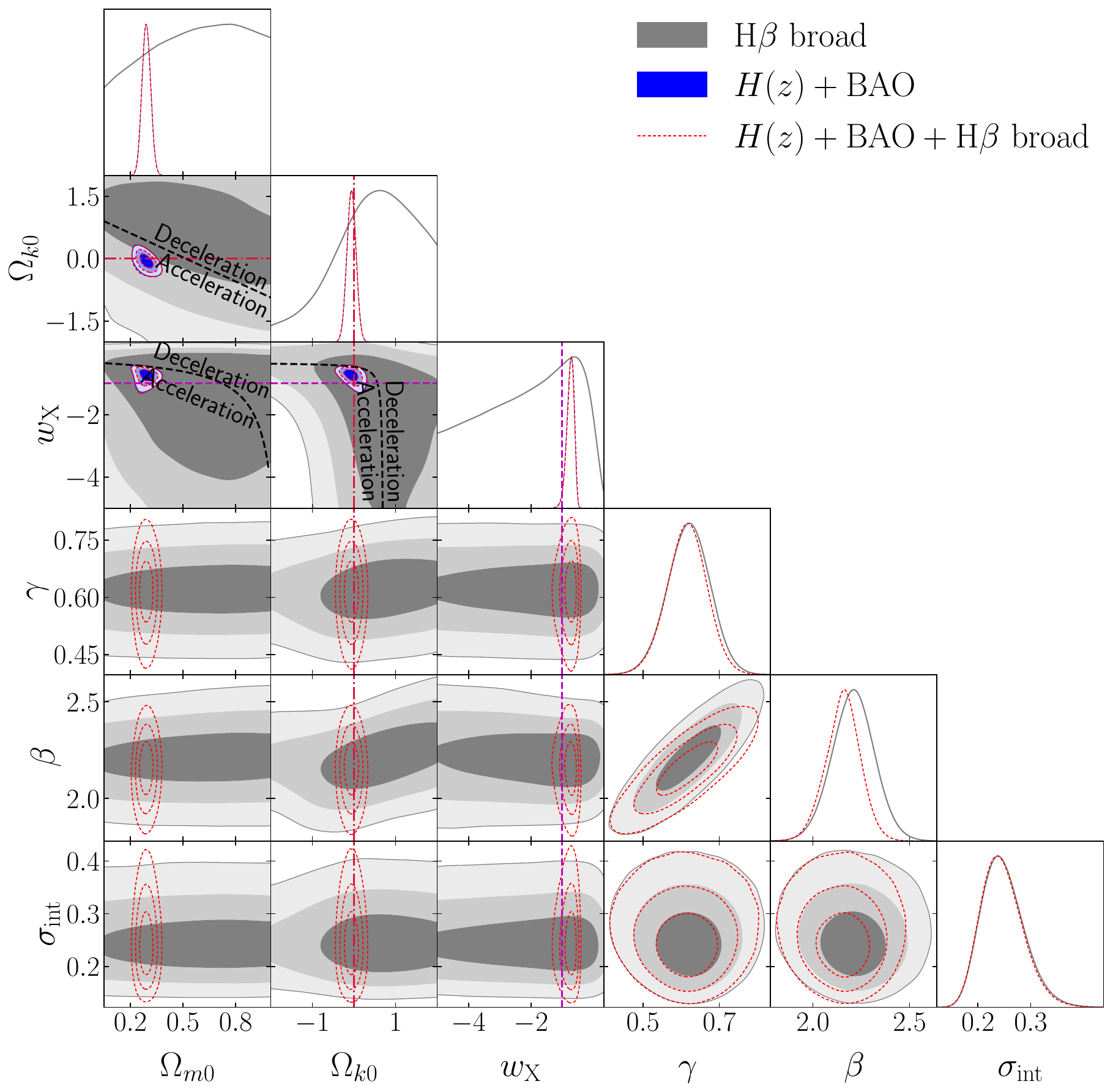}}\\
 \subfloat[Flat \pcdm]{%
    \includegraphics[width=0.4\textwidth,height=0.35\textwidth]{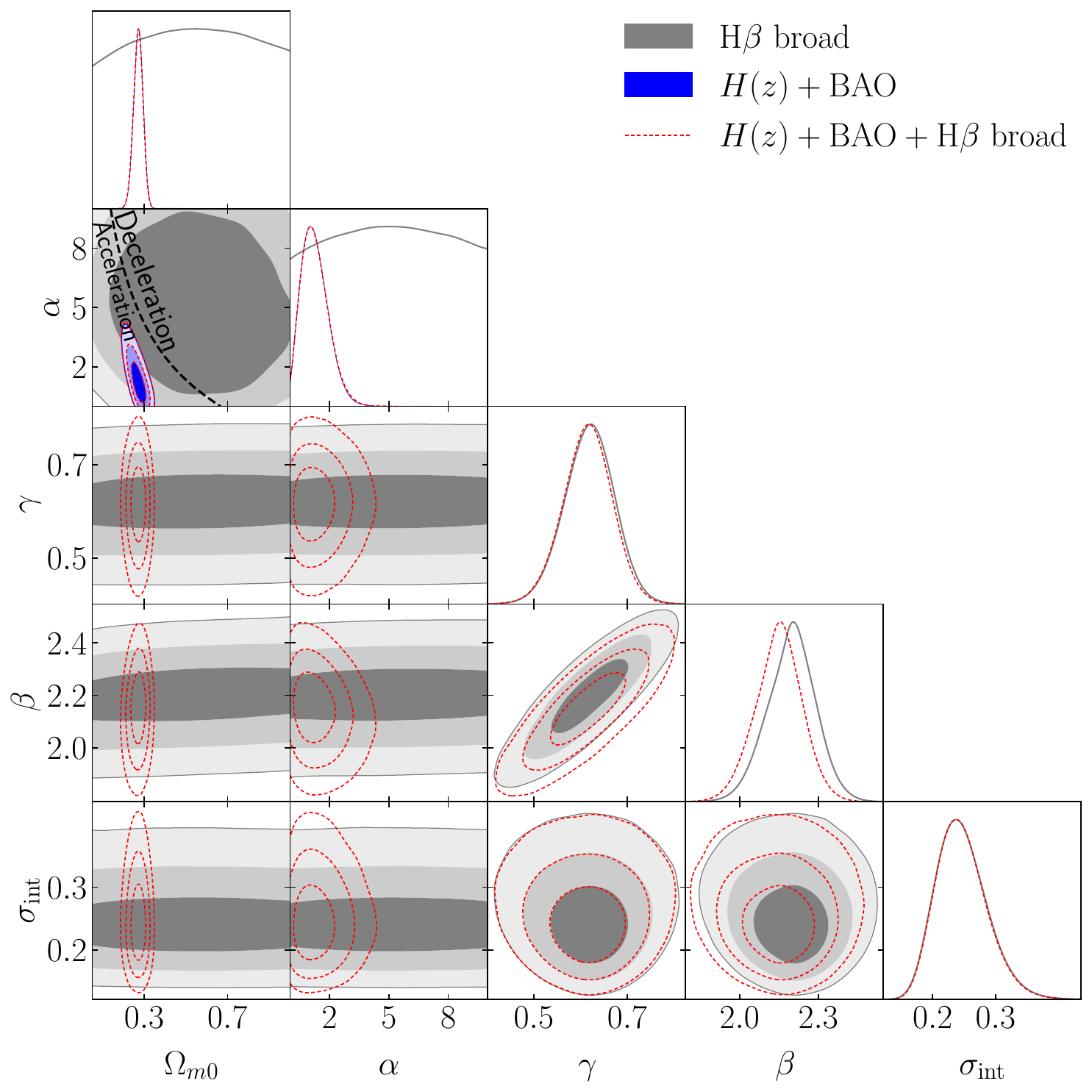}}
 \subfloat[Nonflat \pcdm]{%
    \includegraphics[width=0.4\textwidth,height=0.35\textwidth]{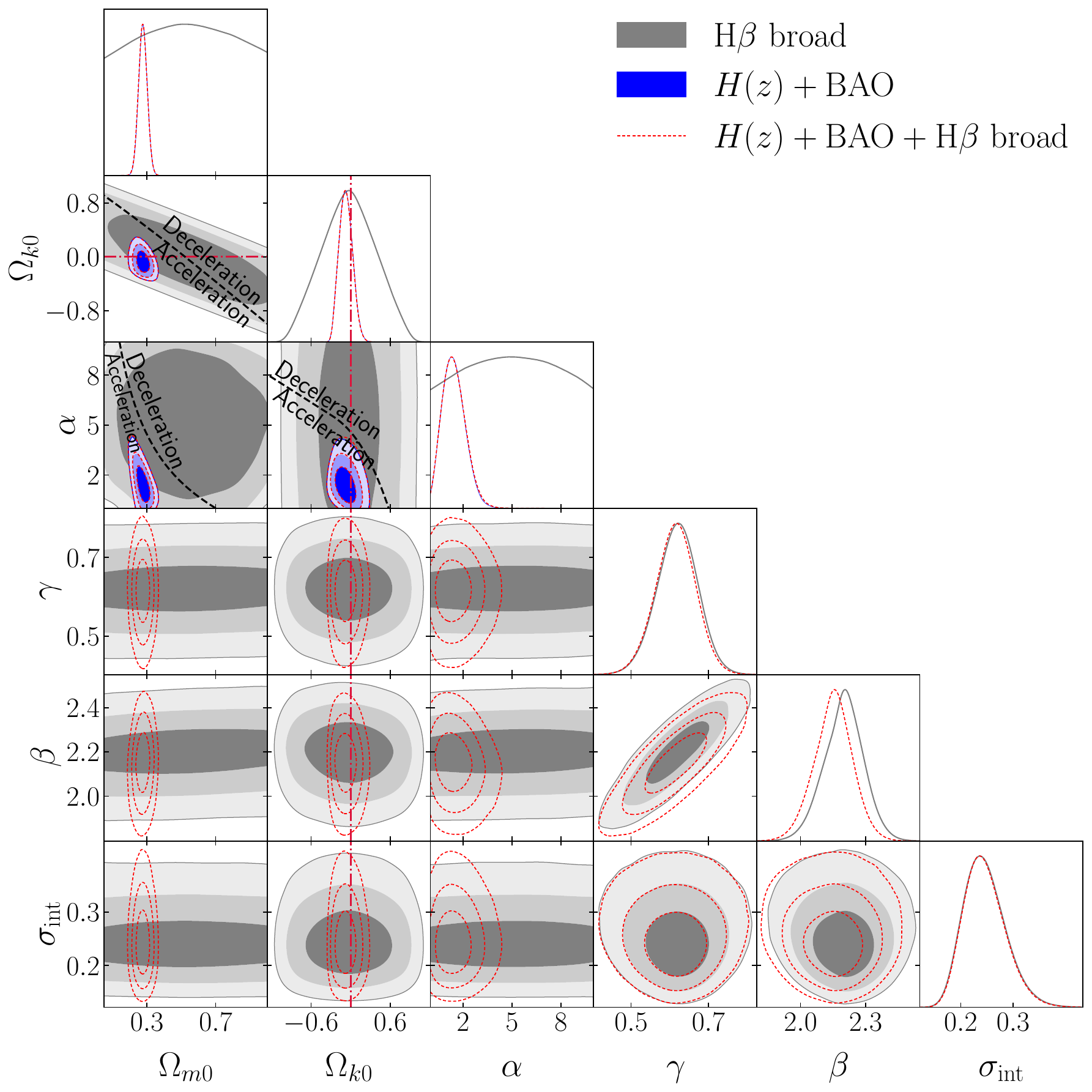}}\\
\caption{Same as Fig.\ \ref{fig1}, but for RM AGN of H$\beta$ broad (gray), $H(z)$ + BAO (blue), and $H(z)$ + BAO + H$\beta$ broad (dashed red) data.}
\label{fig5}
\end{figure*}

\begin{figure*}
\centering
 \subfloat[Flat \lcdm]{%
    \includegraphics[width=0.4\textwidth,height=0.35\textwidth]{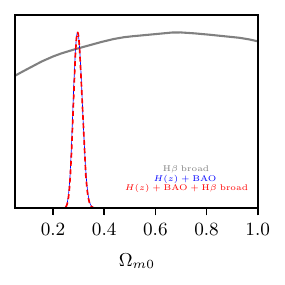}}
 \subfloat[Nonflat \lcdm]{%
    \includegraphics[width=0.4\textwidth,height=0.35\textwidth]{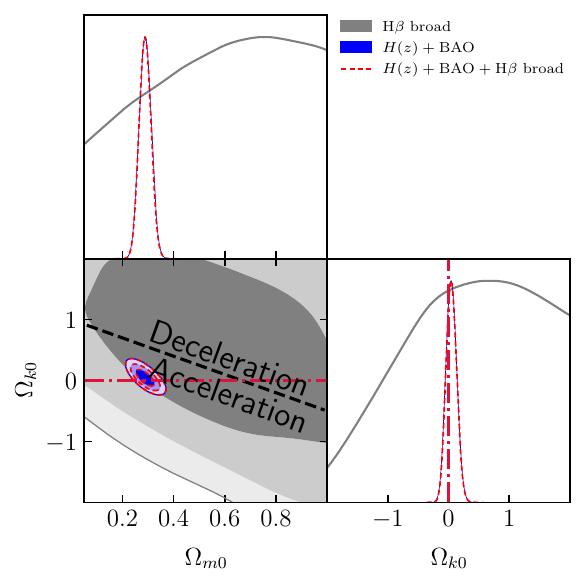}}\\
 \subfloat[Flat XCDM]{%
    \includegraphics[width=0.4\textwidth,height=0.35\textwidth]{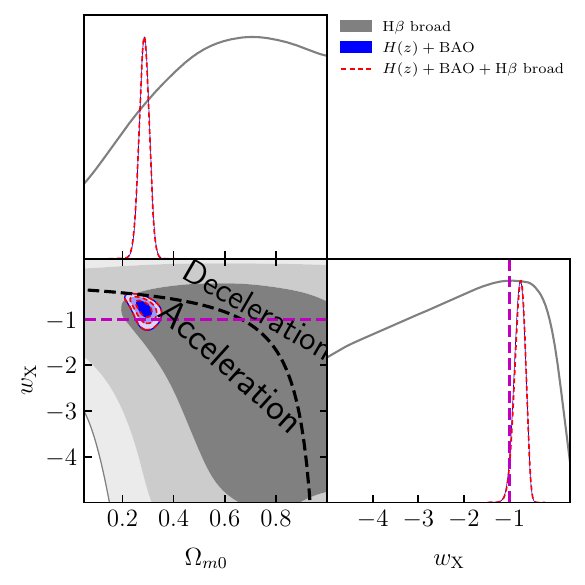}}
 \subfloat[Nonflat XCDM]{%
    \includegraphics[width=0.4\textwidth,height=0.35\textwidth]{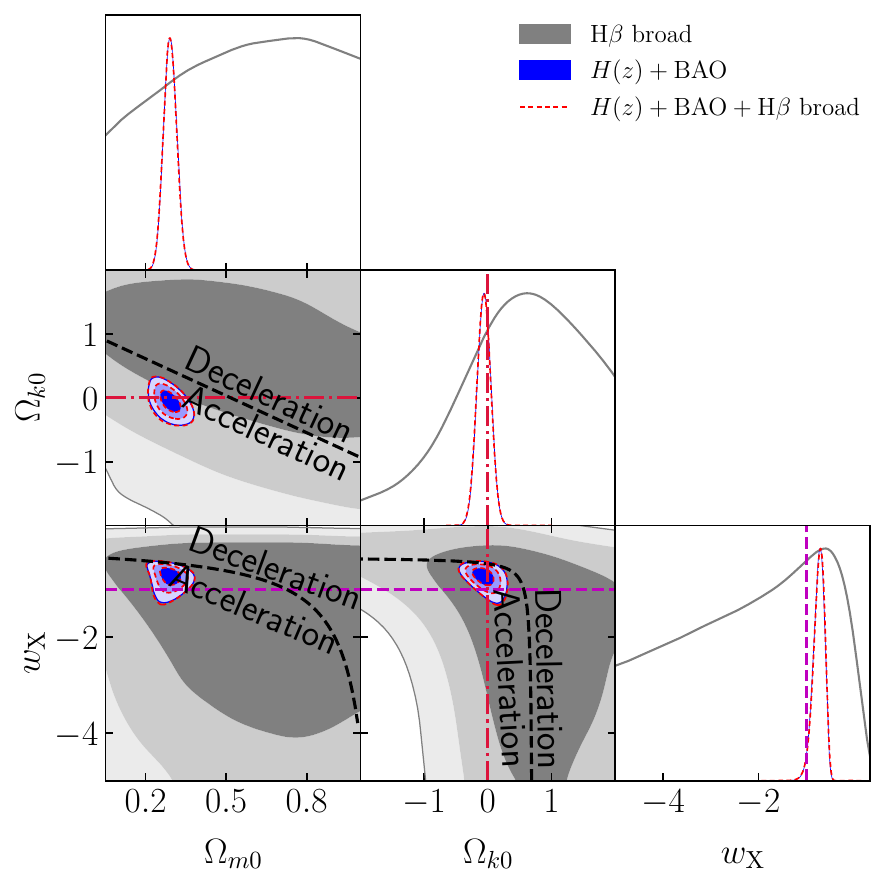}}\\
 \subfloat[Flat \pcdm]{%
    \includegraphics[width=0.4\textwidth,height=0.35\textwidth]{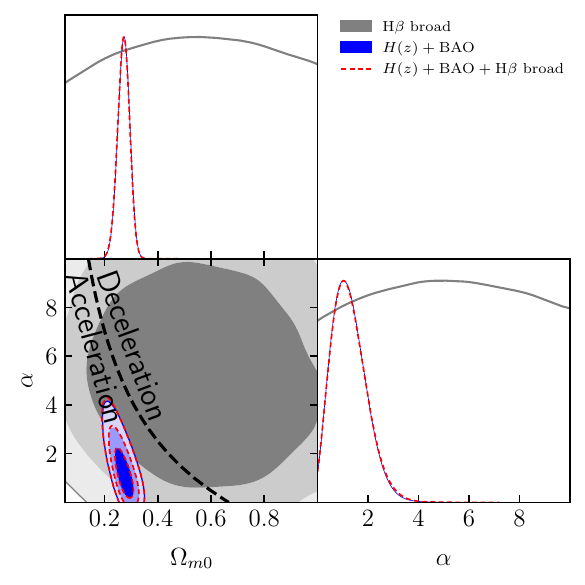}}
 \subfloat[Nonflat \pcdm]{%
    \includegraphics[width=0.4\textwidth,height=0.35\textwidth]{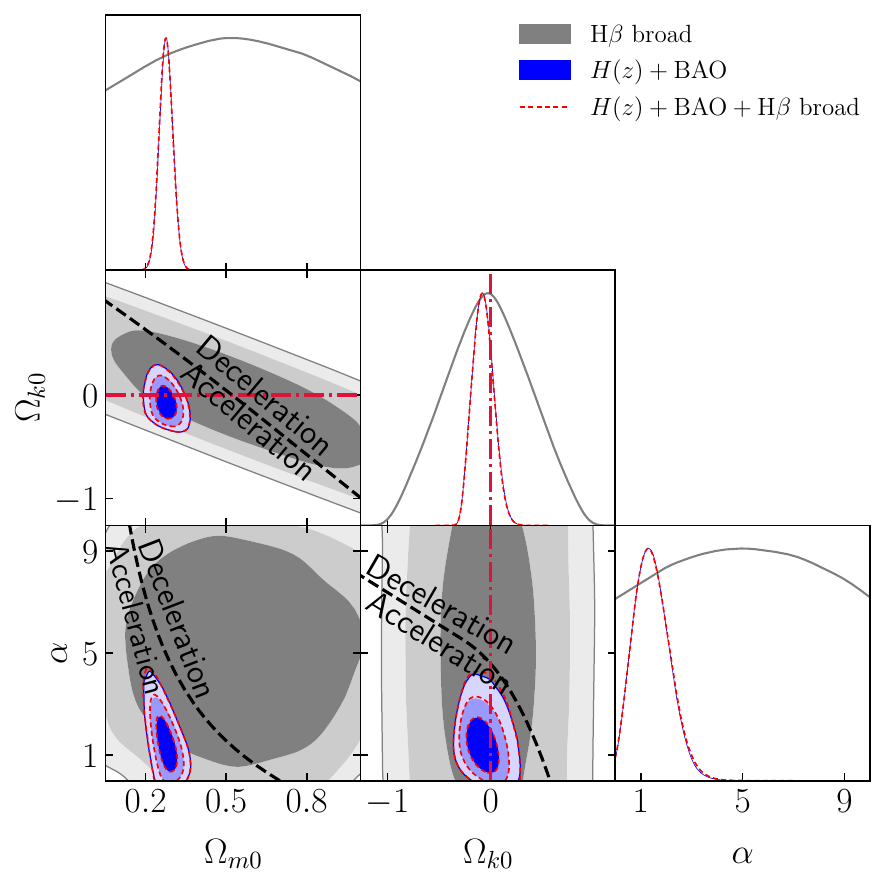}}\\
\caption{Same as Fig.\ \ref{fig5}, but for cosmological parameters only.}
\label{fig6}
\end{figure*}

\begin{figure*}
\centering
 \subfloat[Flat \lcdm]{%
    \includegraphics[width=0.4\textwidth,height=0.35\textwidth]{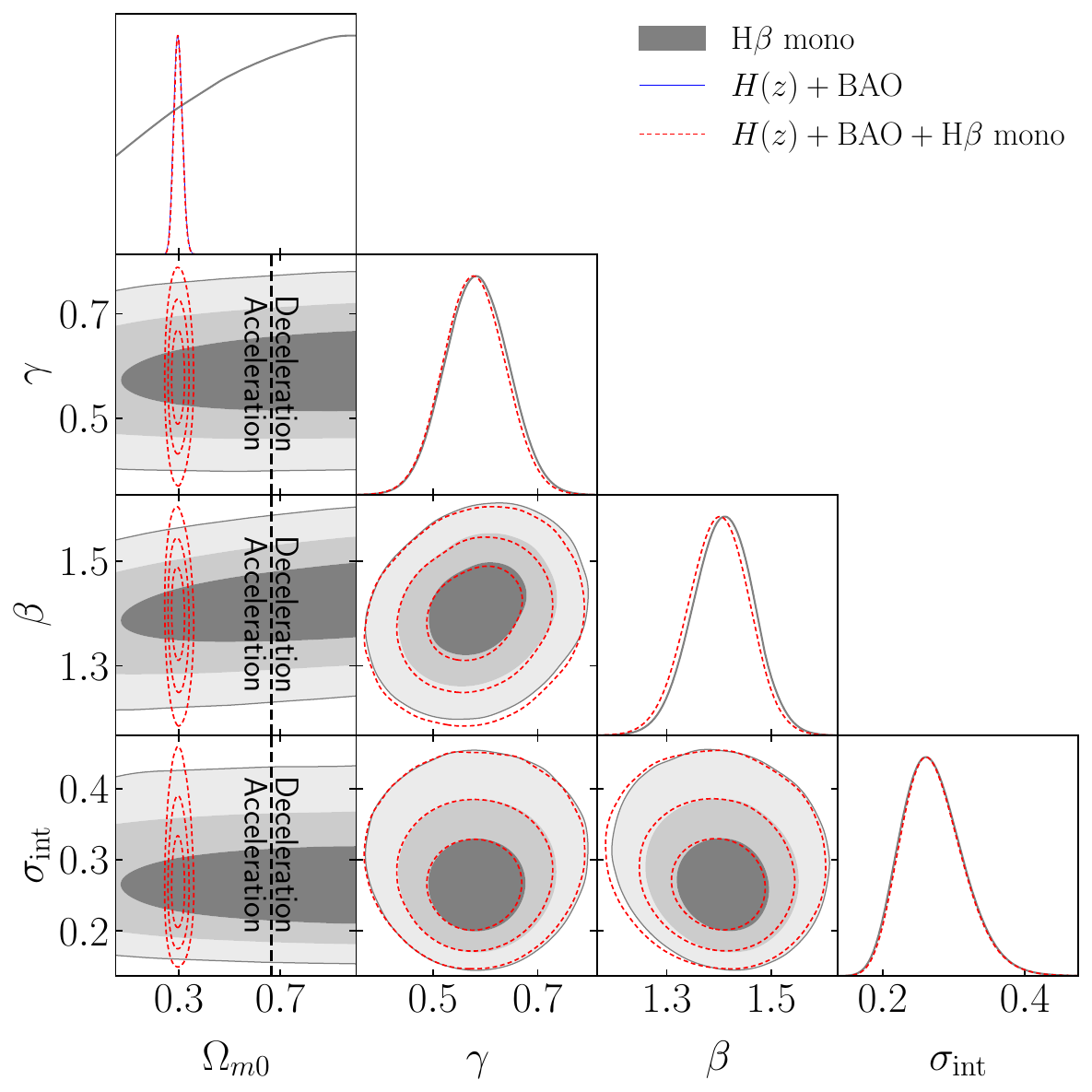}}
 \subfloat[Nonflat \lcdm]{%
    \includegraphics[width=0.4\textwidth,height=0.35\textwidth]{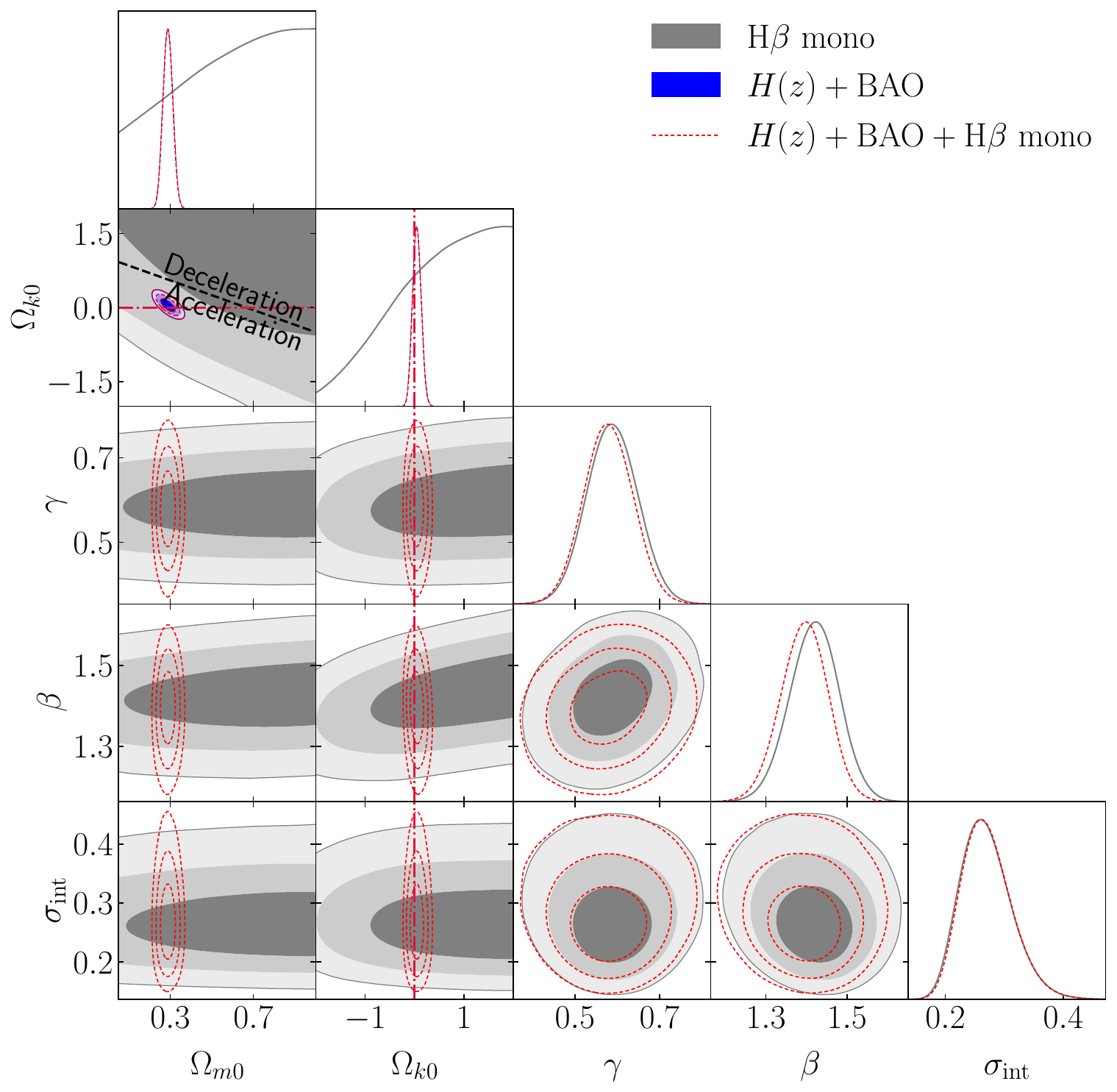}}\\
 \subfloat[Flat XCDM]{%
    \includegraphics[width=0.4\textwidth,height=0.35\textwidth]{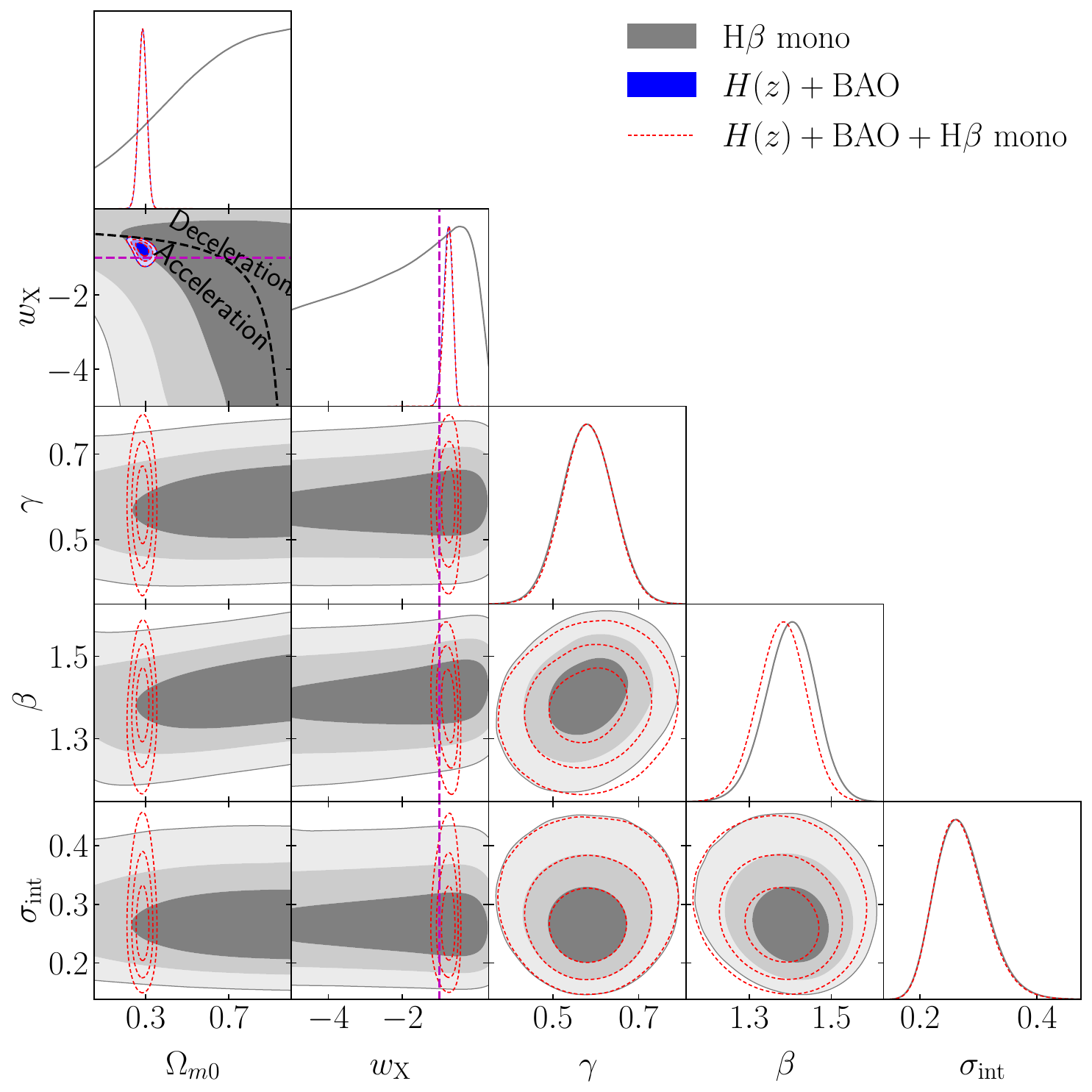}}
 \subfloat[Nonflat XCDM]{%
    \includegraphics[width=0.4\textwidth,height=0.35\textwidth]{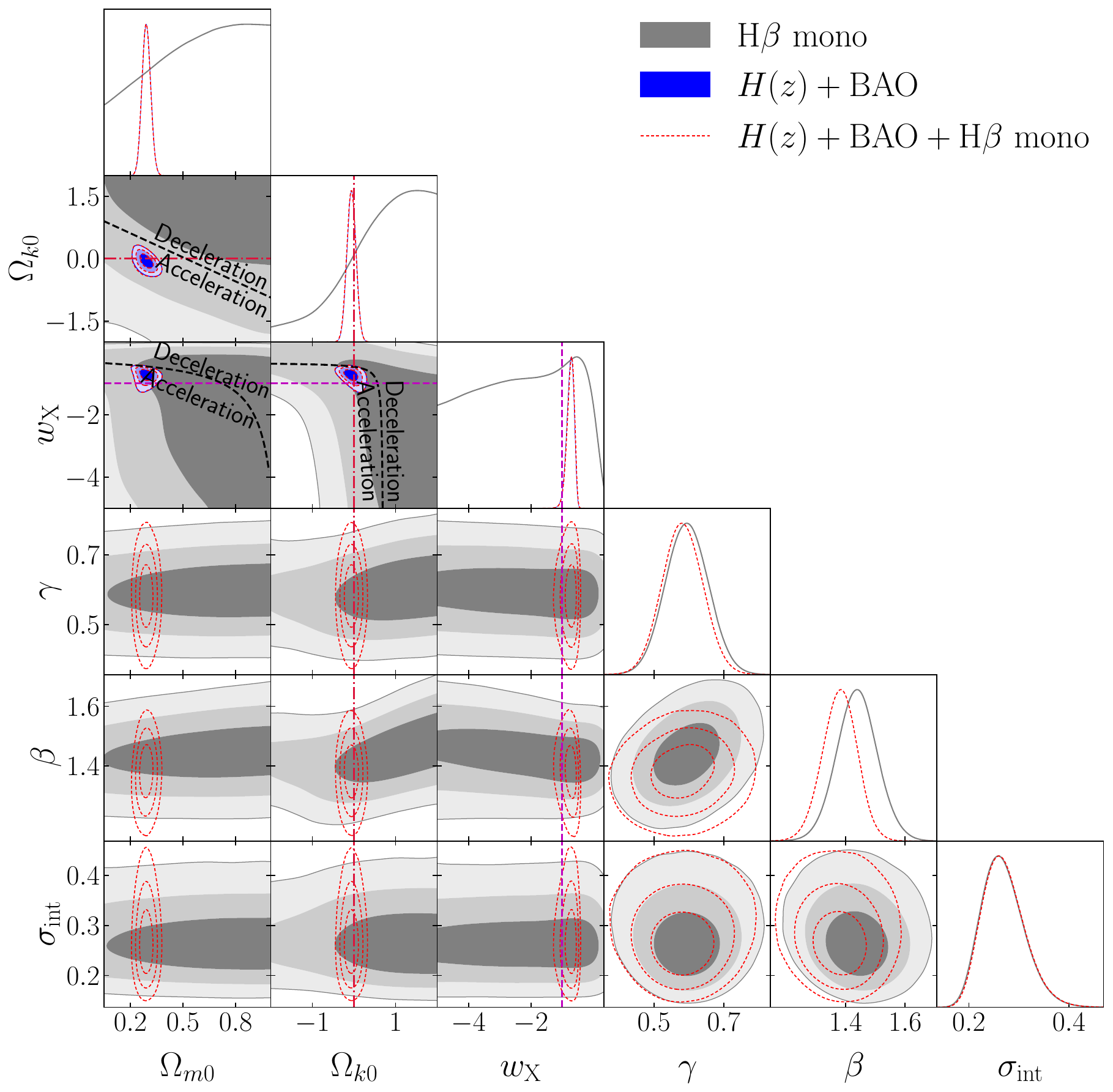}}\\
 \subfloat[Flat \pcdm]{%
    \includegraphics[width=0.4\textwidth,height=0.35\textwidth]{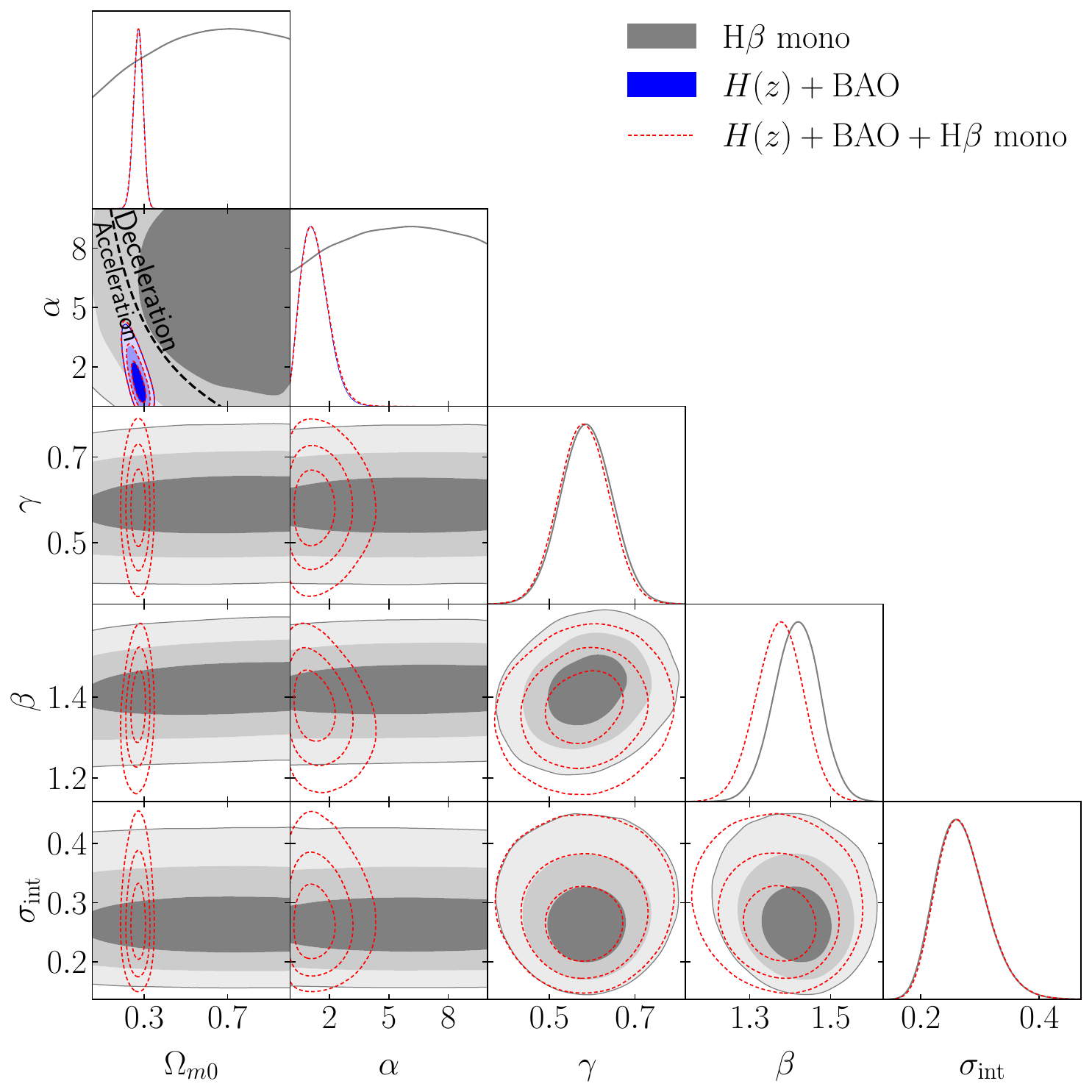}}
 \subfloat[Nonflat \pcdm]{%
    \includegraphics[width=0.4\textwidth,height=0.35\textwidth]{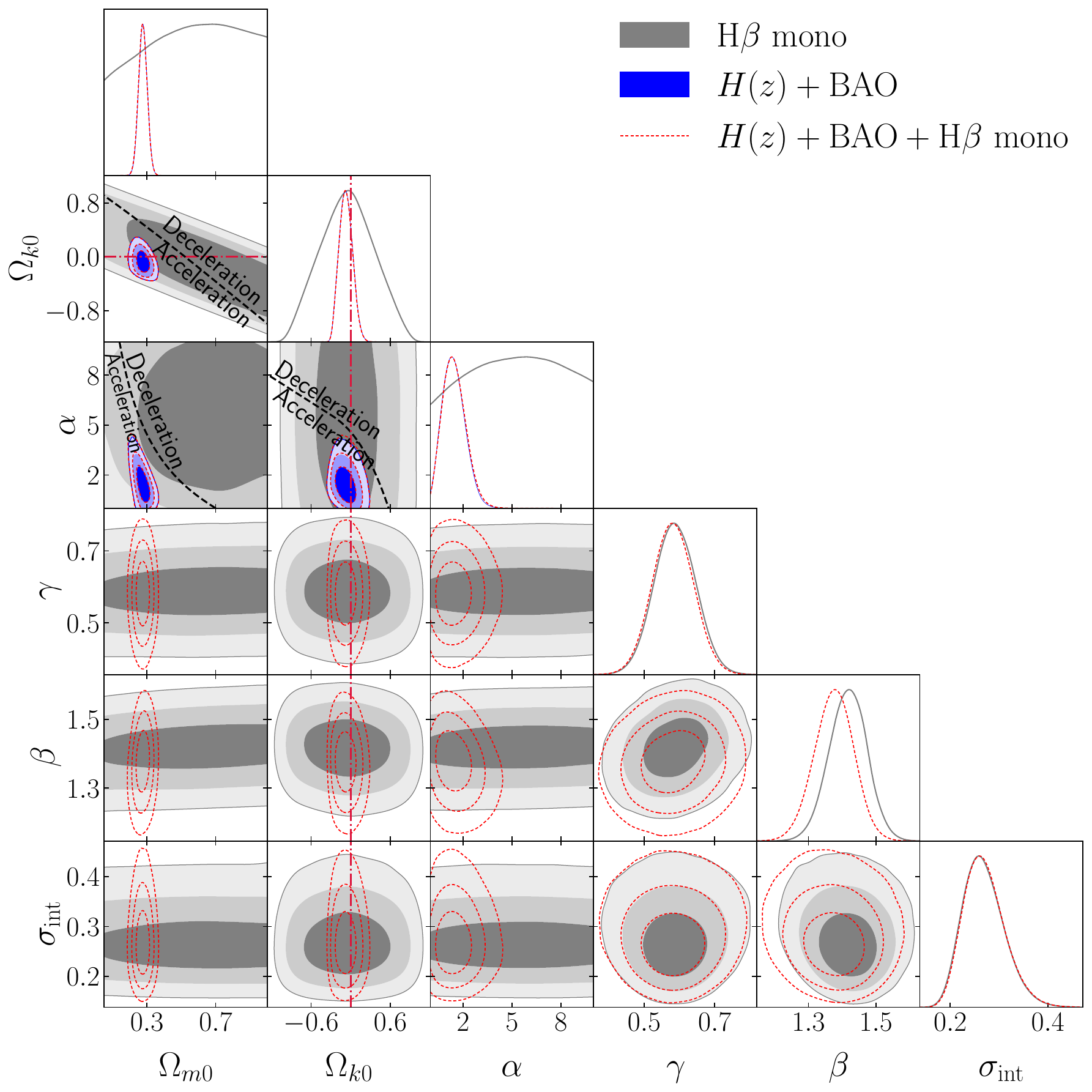}}\\
\caption{Same as Fig.\ \ref{fig1}, but for RM AGN of H$\beta$ mono (gray), $H(z)$ + BAO (blue), and $H(z)$ + BAO + H$\beta$ mono (dashed red) data.}
\label{fig7}
\end{figure*}

\begin{figure*}
\centering
 \subfloat[Flat \lcdm]{%
    \includegraphics[width=0.4\textwidth,height=0.35\textwidth]{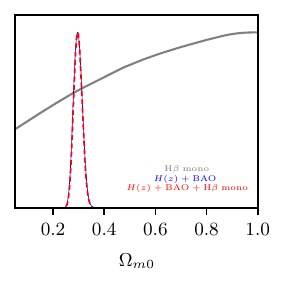}}
 \subfloat[Nonflat \lcdm]{%
    \includegraphics[width=0.4\textwidth,height=0.35\textwidth]{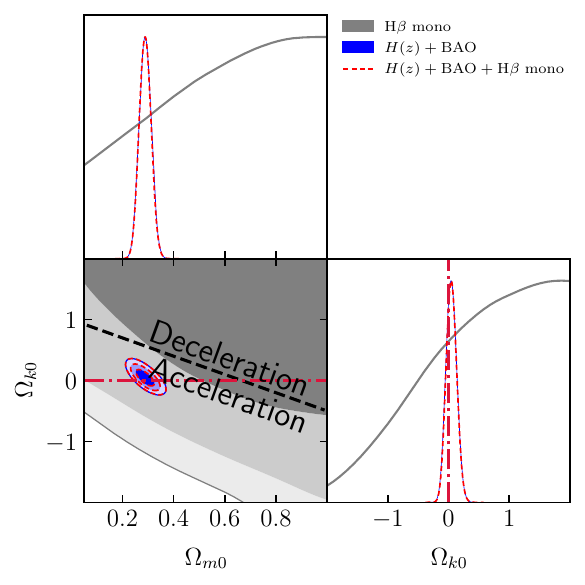}}\\
 \subfloat[Flat XCDM]{%
    \includegraphics[width=0.4\textwidth,height=0.35\textwidth]{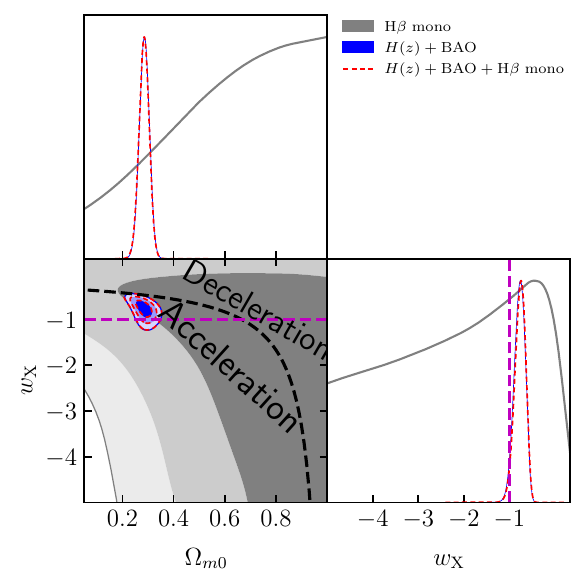}}
 \subfloat[Nonflat XCDM]{%
    \includegraphics[width=0.4\textwidth,height=0.35\textwidth]{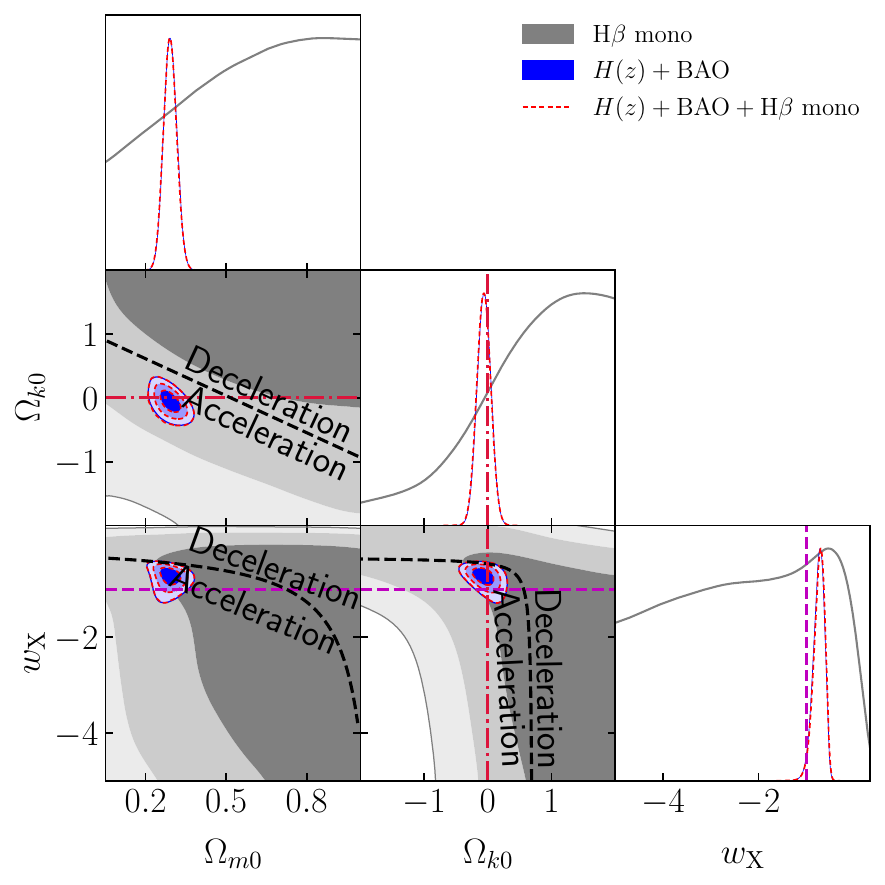}}\\
 \subfloat[Flat \pcdm]{%
    \includegraphics[width=0.4\textwidth,height=0.35\textwidth]{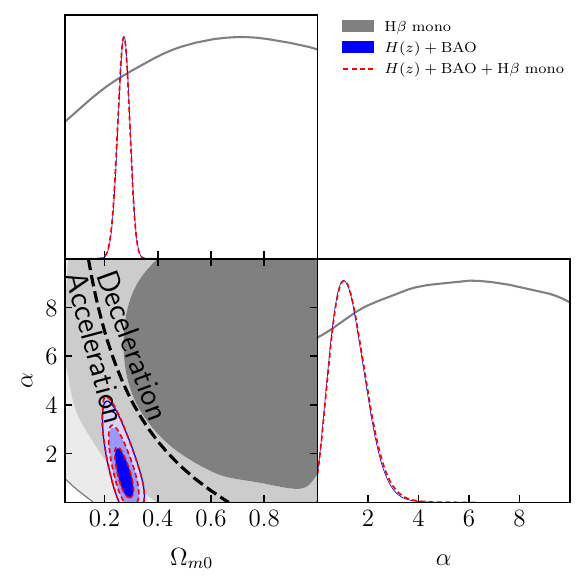}}
 \subfloat[Nonflat \pcdm]{%
    \includegraphics[width=0.4\textwidth,height=0.35\textwidth]{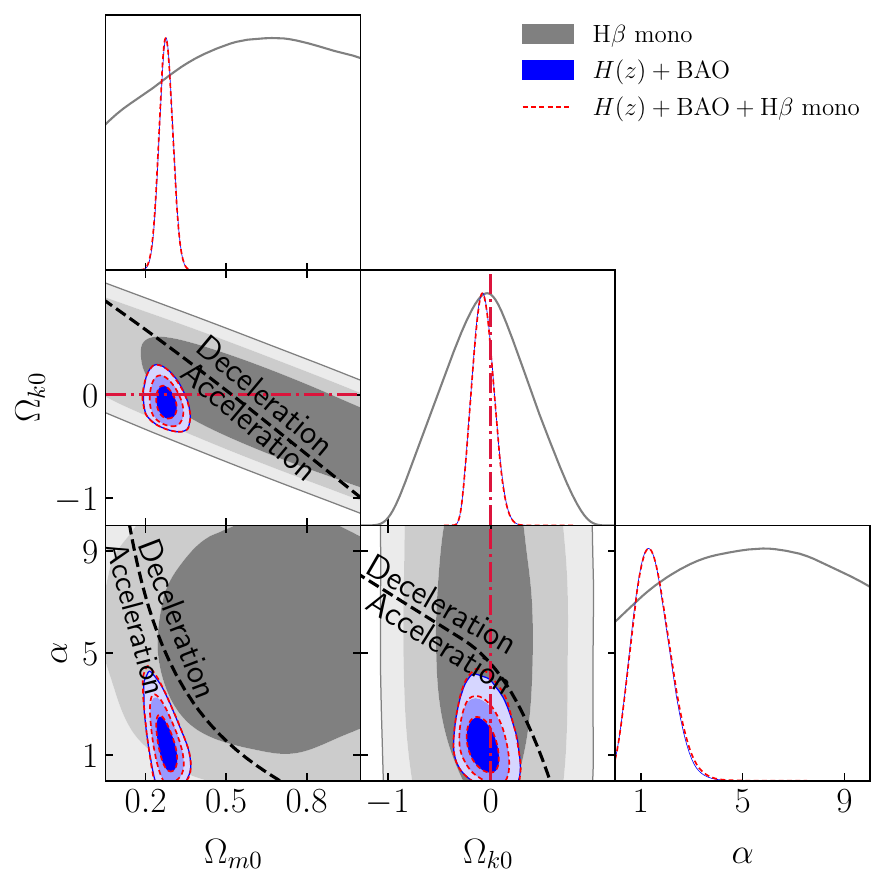}}\\
\caption{Same as Fig.\ \ref{fig7}, but for cosmological parameters only.}
\label{fig8}
\end{figure*}

\begin{figure*}[htbp]
\centering
 \subfloat[Flat \lcdm]{%
    \includegraphics[width=0.45\textwidth,height=0.35\textwidth]{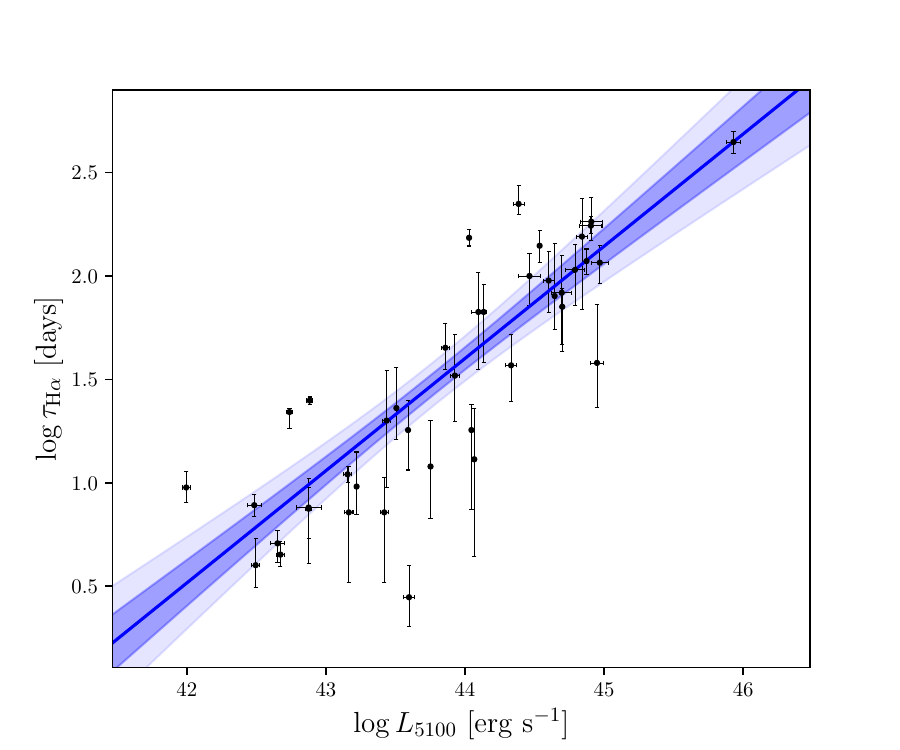}}
 \subfloat[Nonflat \lcdm]{%
    \includegraphics[width=0.45\textwidth,height=0.35\textwidth]{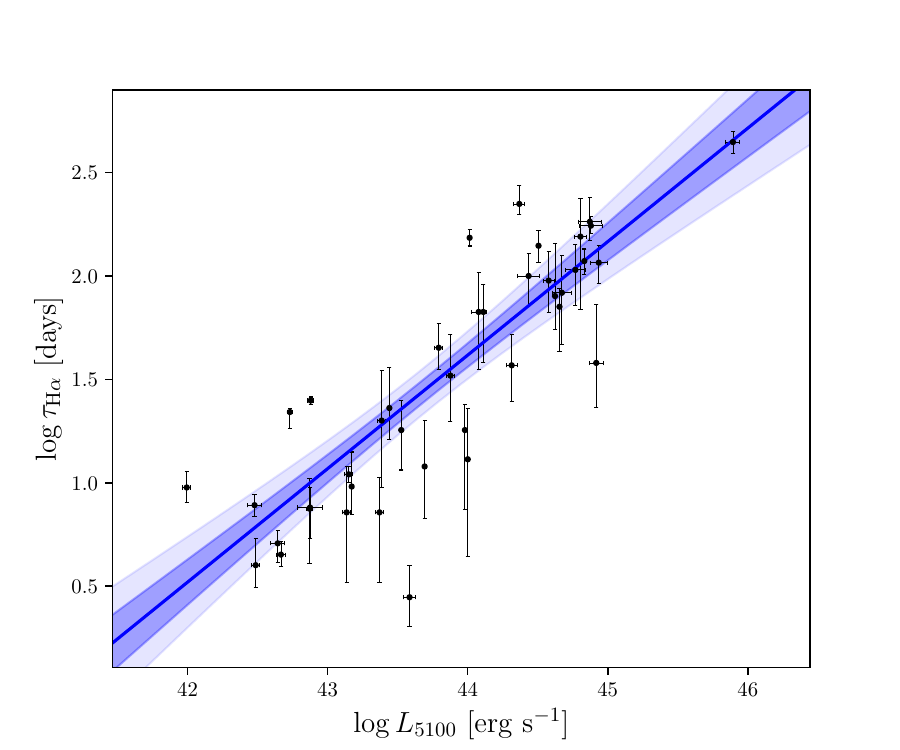}}\\
 \subfloat[Flat XCDM]{%
    \includegraphics[width=0.45\textwidth,height=0.35\textwidth]{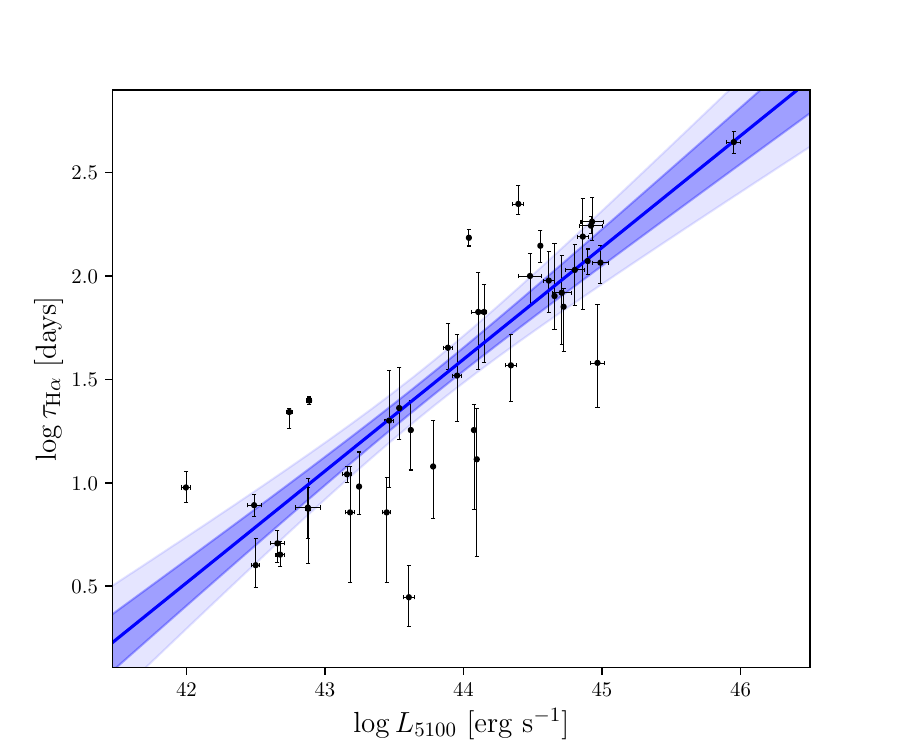}}
 \subfloat[Nonflat XCDM]{%
    \includegraphics[width=0.45\textwidth,height=0.35\textwidth]{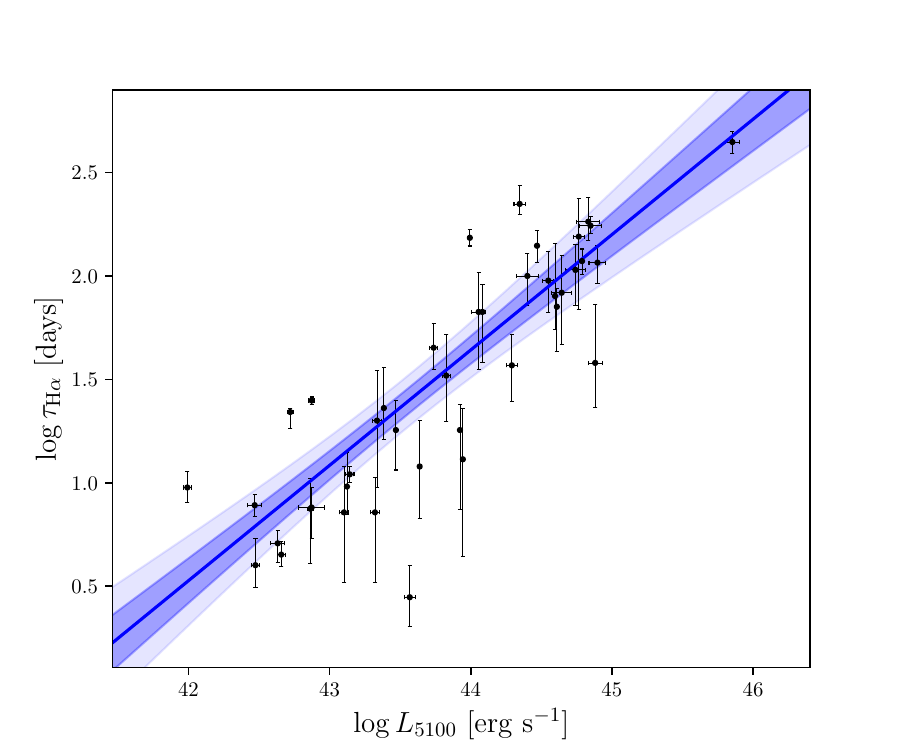}}\\
 \subfloat[Flat \pcdm]{%
    \includegraphics[width=0.45\textwidth,height=0.35\textwidth]{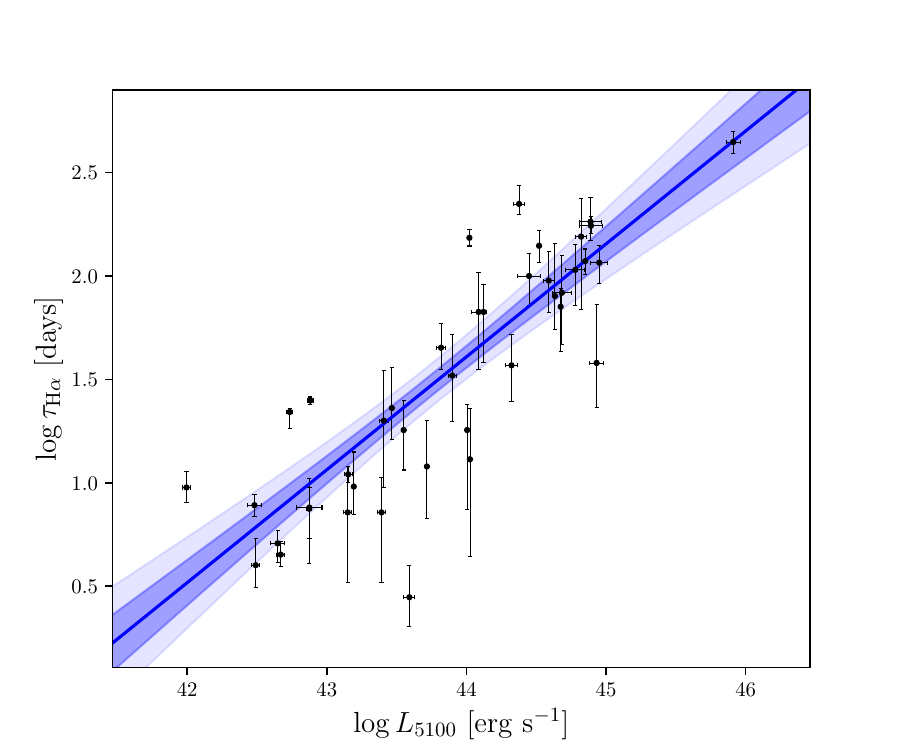}}
 \subfloat[Nonflat \pcdm]{%
    \includegraphics[width=0.45\textwidth,height=0.35\textwidth]{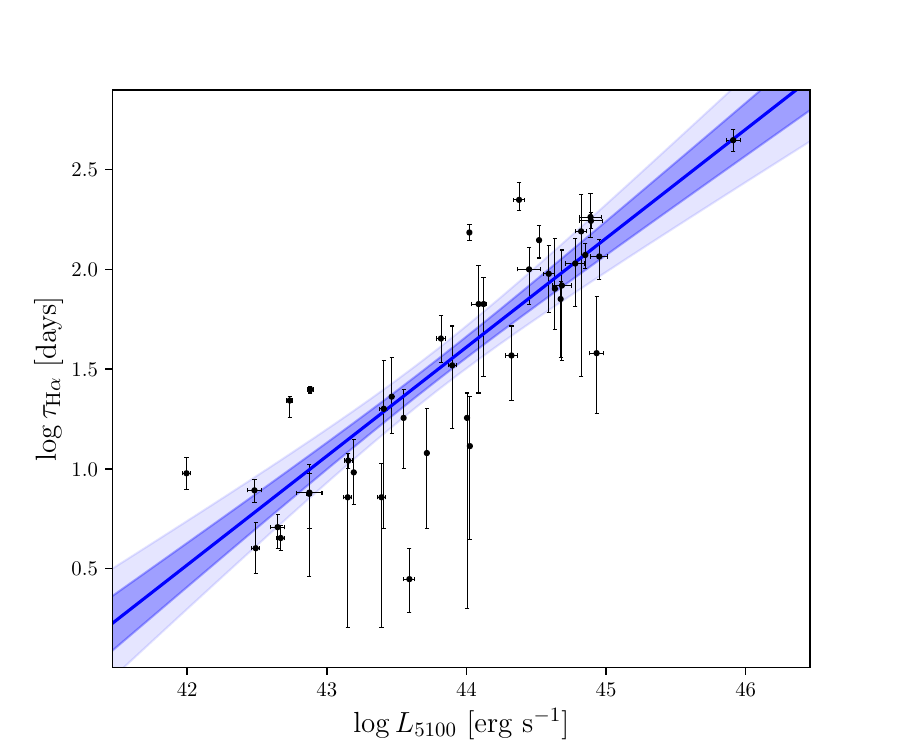}}\\
\caption{Same as Fig.\ \ref{R-L1}, but for H$\alpha$ mono.}
\label{R-L2}
\end{figure*}

\begin{figure*}[htbp]
\centering
 \subfloat[Flat \lcdm]{%
    \includegraphics[width=0.45\textwidth,height=0.35\textwidth]{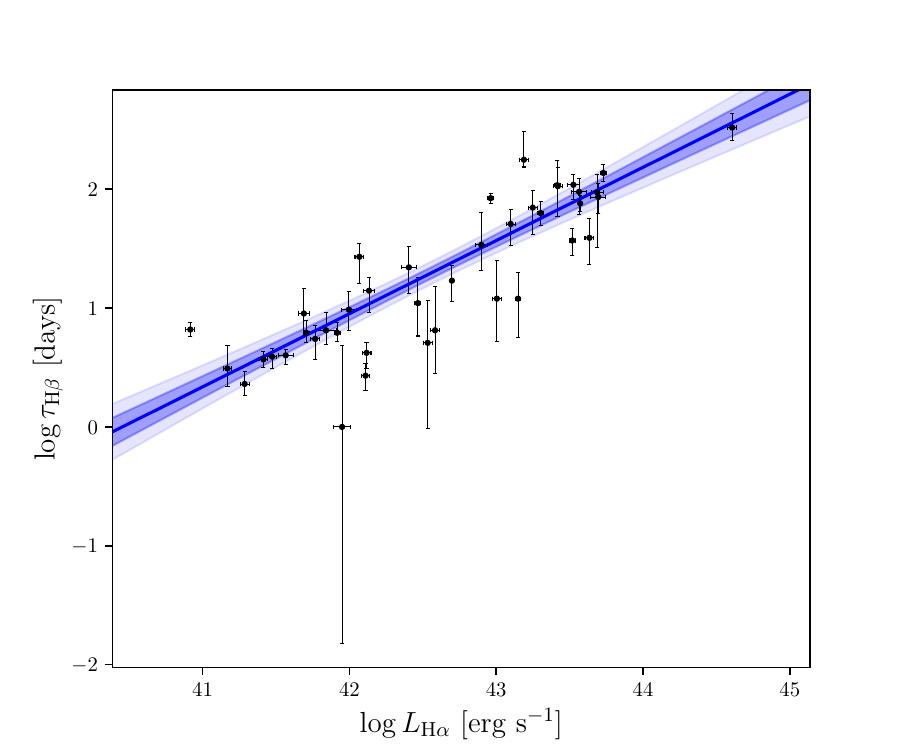}}
 \subfloat[Nonflat \lcdm]{%
    \includegraphics[width=0.45\textwidth,height=0.35\textwidth]{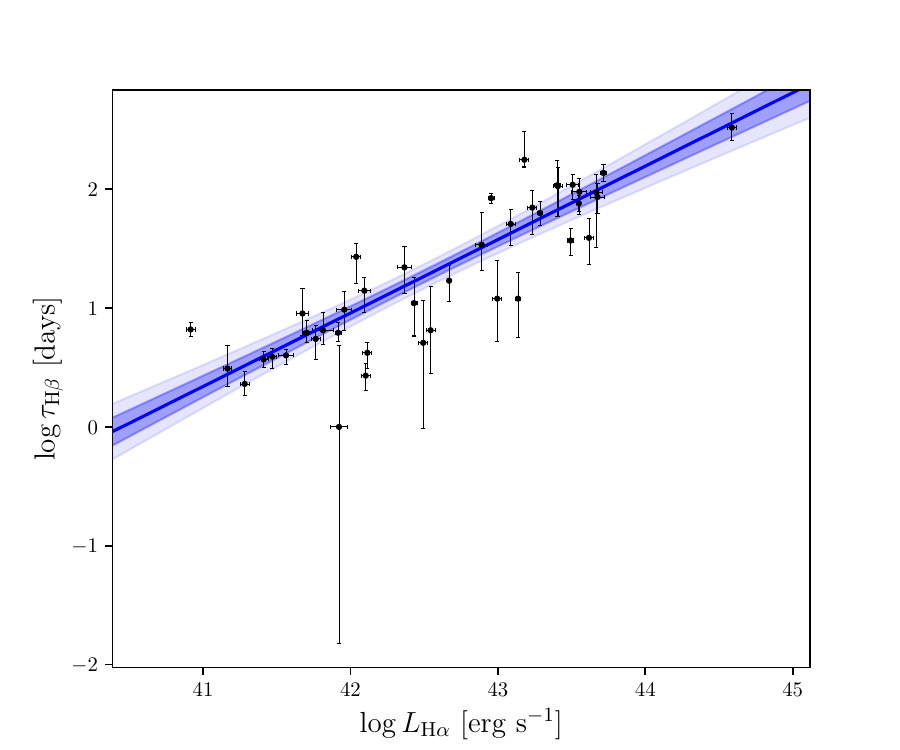}}\\
 \subfloat[Flat XCDM]{%
    \includegraphics[width=0.45\textwidth,height=0.35\textwidth]{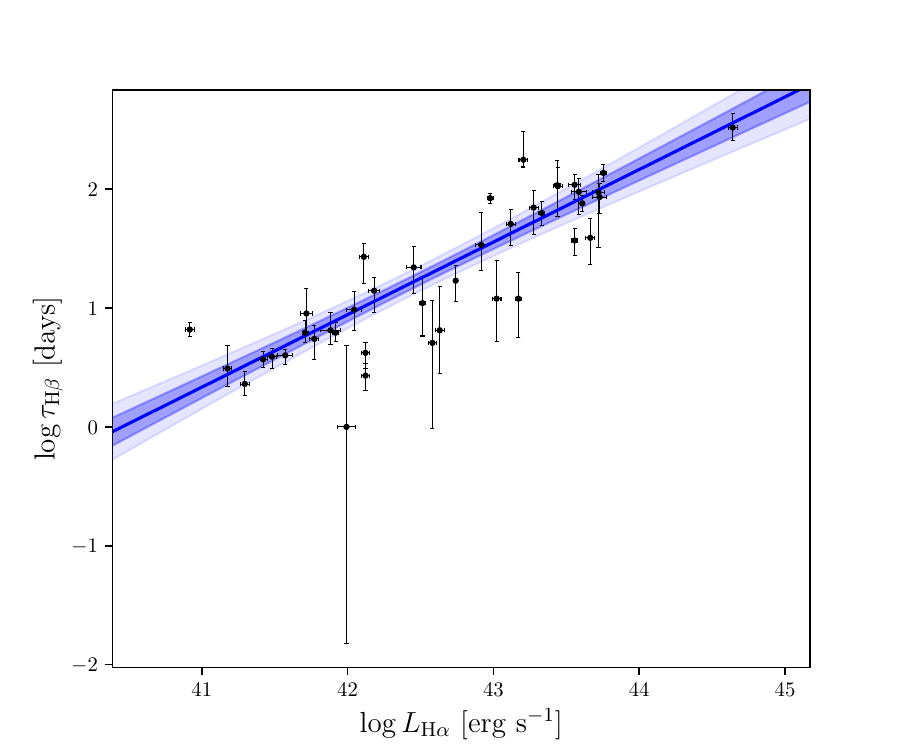}}
 \subfloat[Nonflat XCDM]{%
    \includegraphics[width=0.45\textwidth,height=0.35\textwidth]{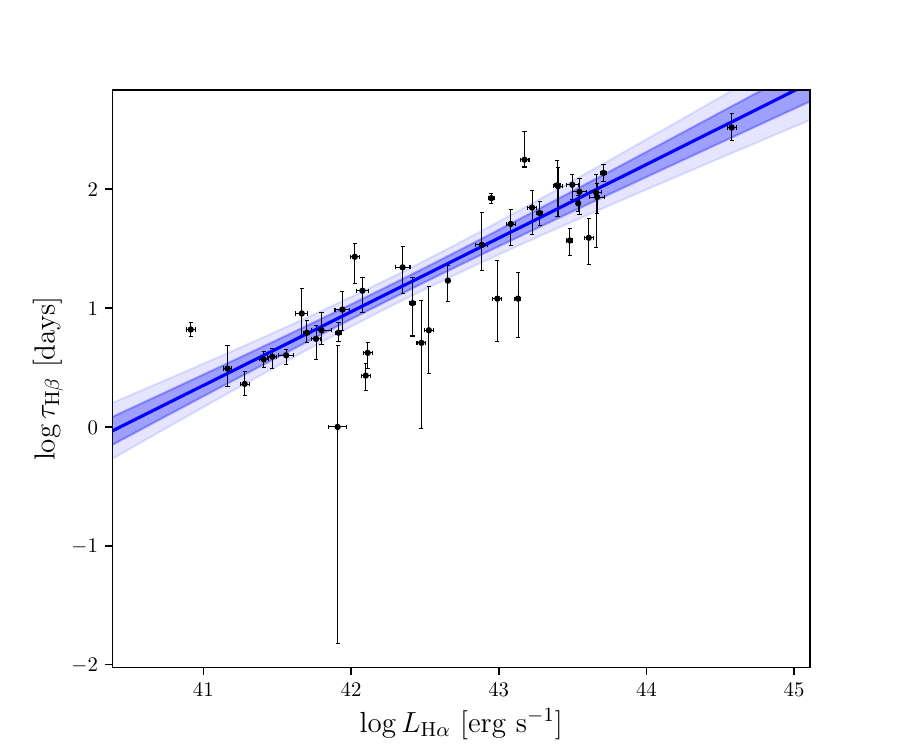}}\\
 \subfloat[Flat \pcdm]{%
    \includegraphics[width=0.45\textwidth,height=0.35\textwidth]{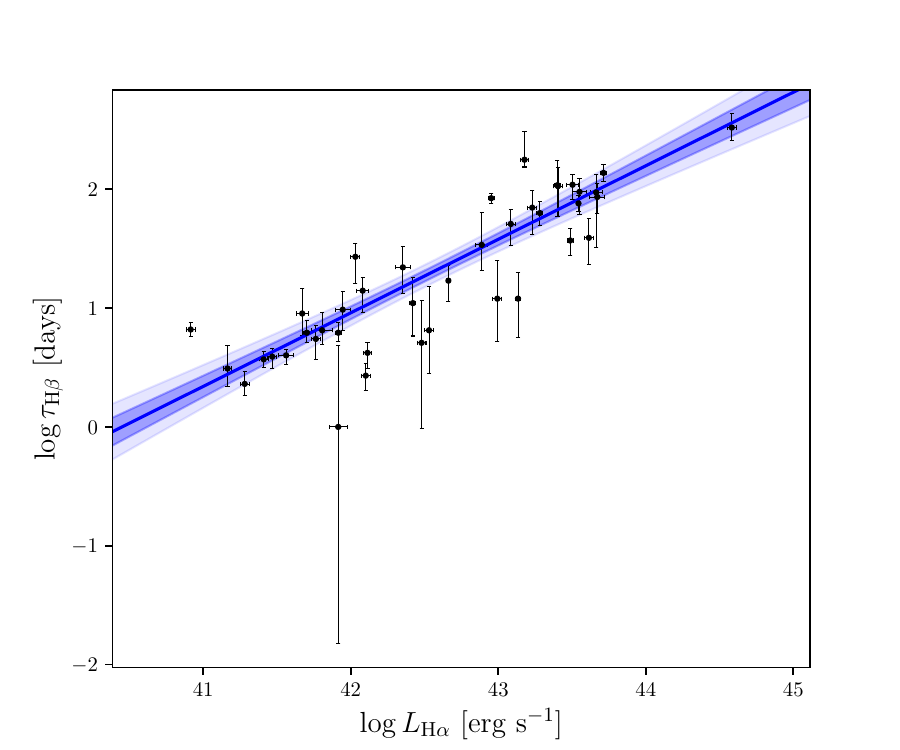}}
 \subfloat[Nonflat \pcdm]{%
    \includegraphics[width=0.45\textwidth,height=0.35\textwidth]{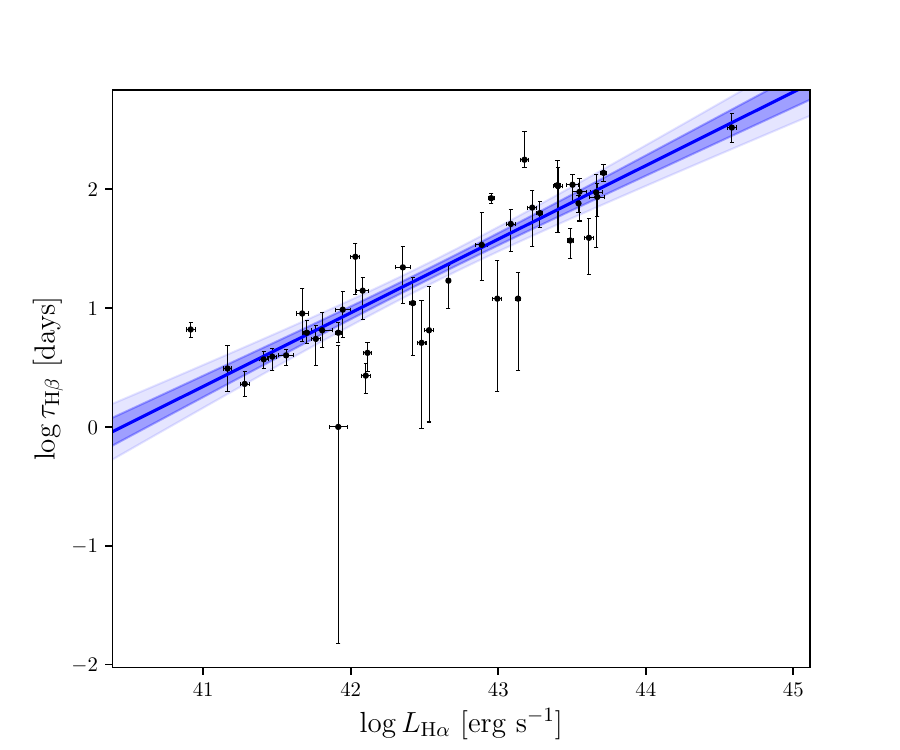}}\\
\caption{Same as Fig.\ \ref{R-L1}, but for H$\beta$ broad.}
\label{R-L3}
\end{figure*}

\begin{figure*}[htbp]
\centering
 \subfloat[Flat \lcdm]{%
    \includegraphics[width=0.45\textwidth,height=0.35\textwidth]{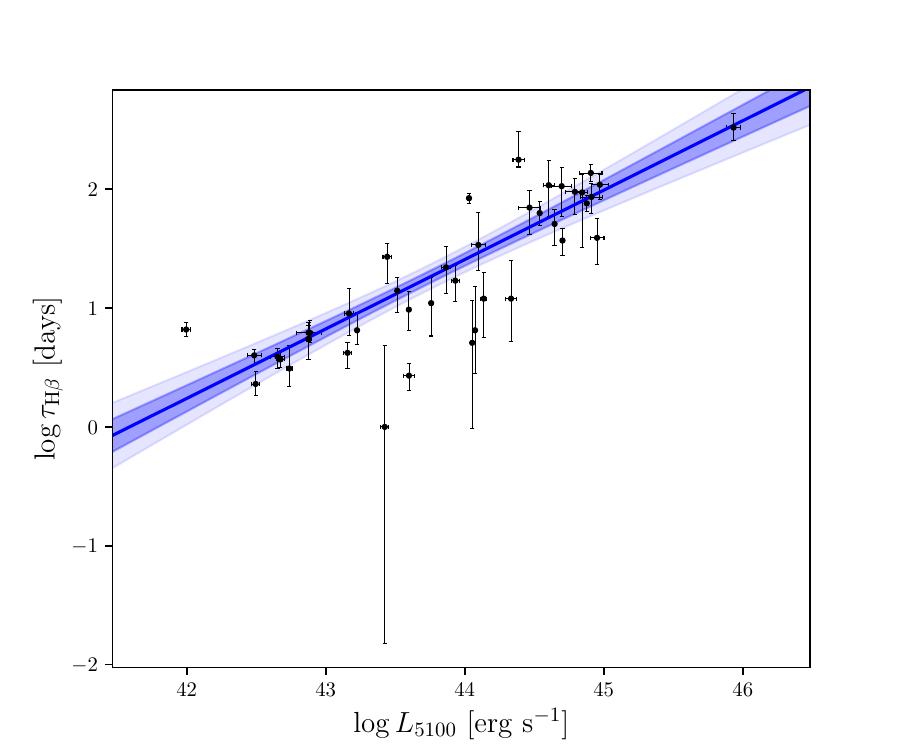}}
 \subfloat[Nonflat \lcdm]{%
    \includegraphics[width=0.45\textwidth,height=0.35\textwidth]{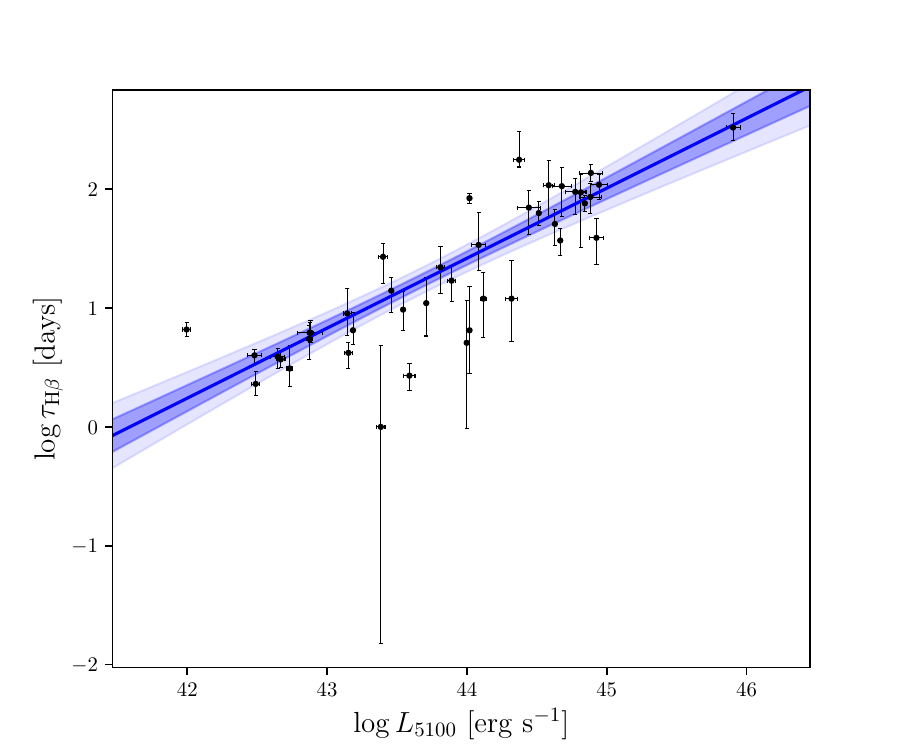}}\\
 \subfloat[Flat XCDM]{%
    \includegraphics[width=0.45\textwidth,height=0.35\textwidth]{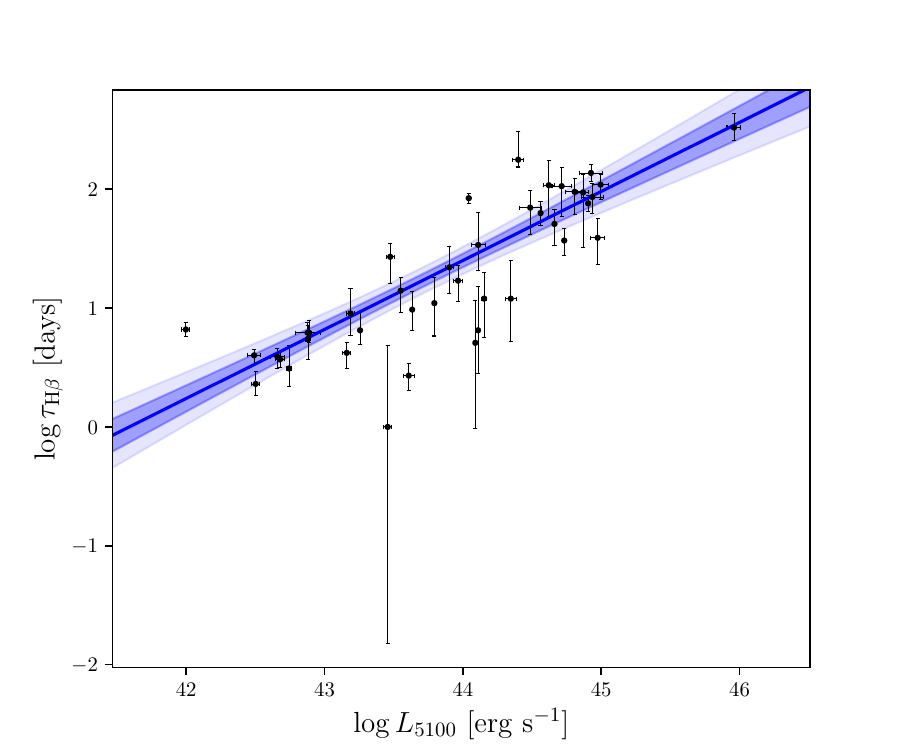}}
 \subfloat[Nonflat XCDM]{%
    \includegraphics[width=0.45\textwidth,height=0.35\textwidth]{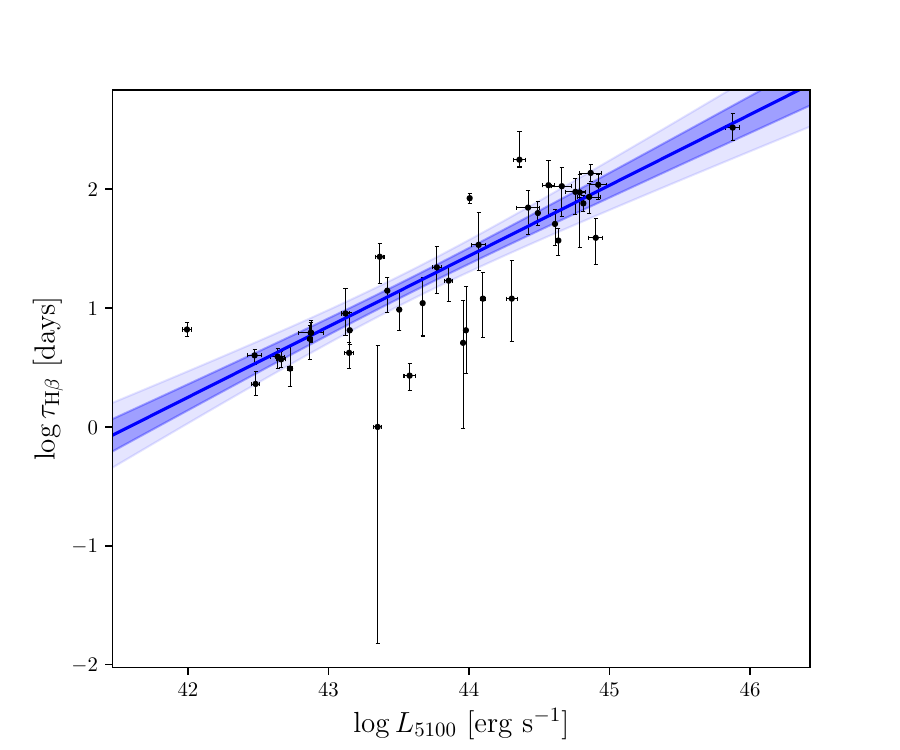}}\\
 \subfloat[Flat \pcdm]{%
    \includegraphics[width=0.45\textwidth,height=0.35\textwidth]{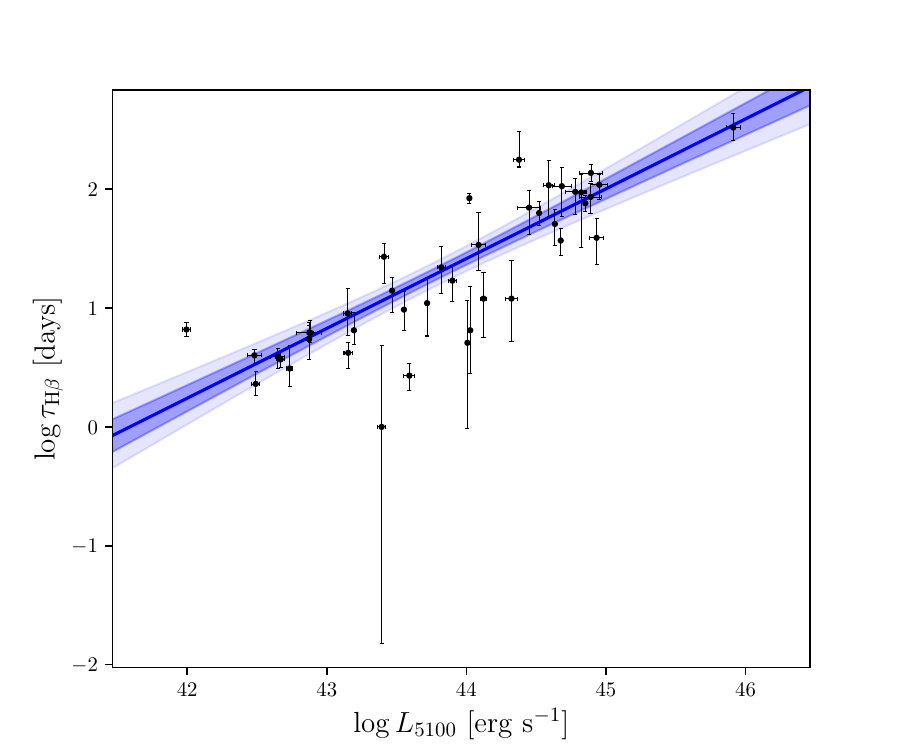}}
 \subfloat[Nonflat \pcdm]{%
    \includegraphics[width=0.45\textwidth,height=0.35\textwidth]{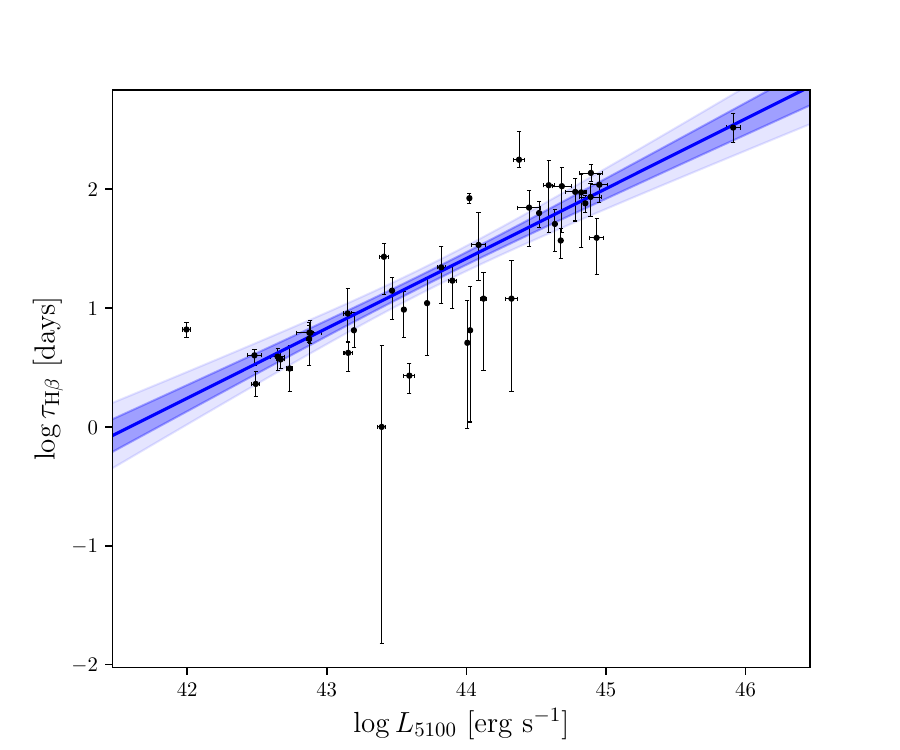}}\\
\caption{Same as Fig.\ \ref{R-L1}, but for H$\beta$ mono.}
\label{R-L4}
\end{figure*}

\begin{figure*}[htbp]
\centering
 \subfloat[Flat \lcdm]{%
    \includegraphics[width=0.45\textwidth,height=0.35\textwidth]{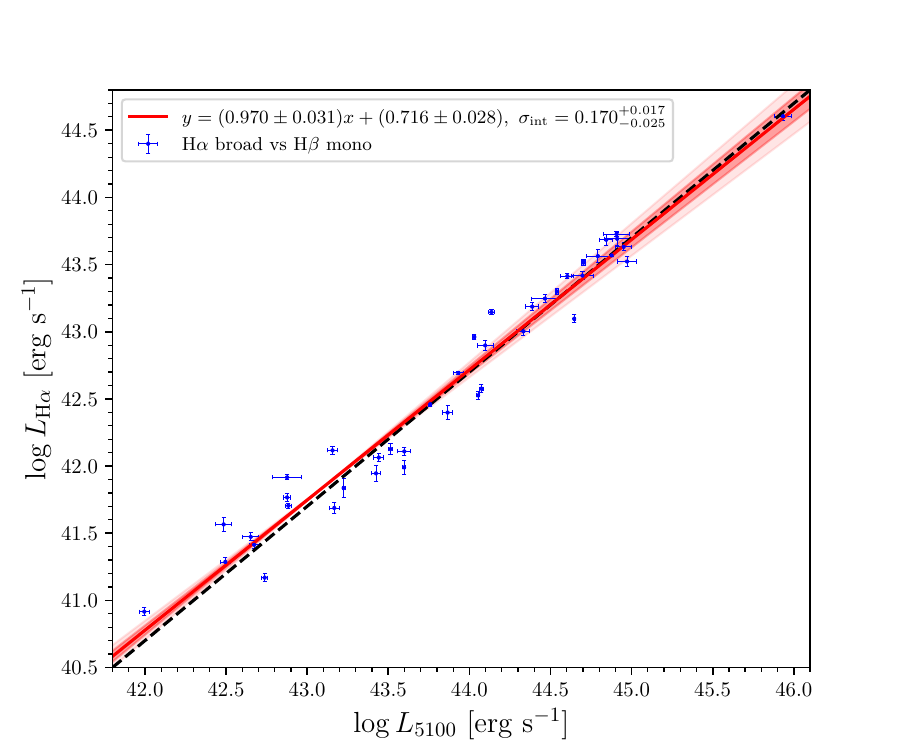}}
 \subfloat[Nonflat \lcdm]{%
    \includegraphics[width=0.45\textwidth,height=0.35\textwidth]{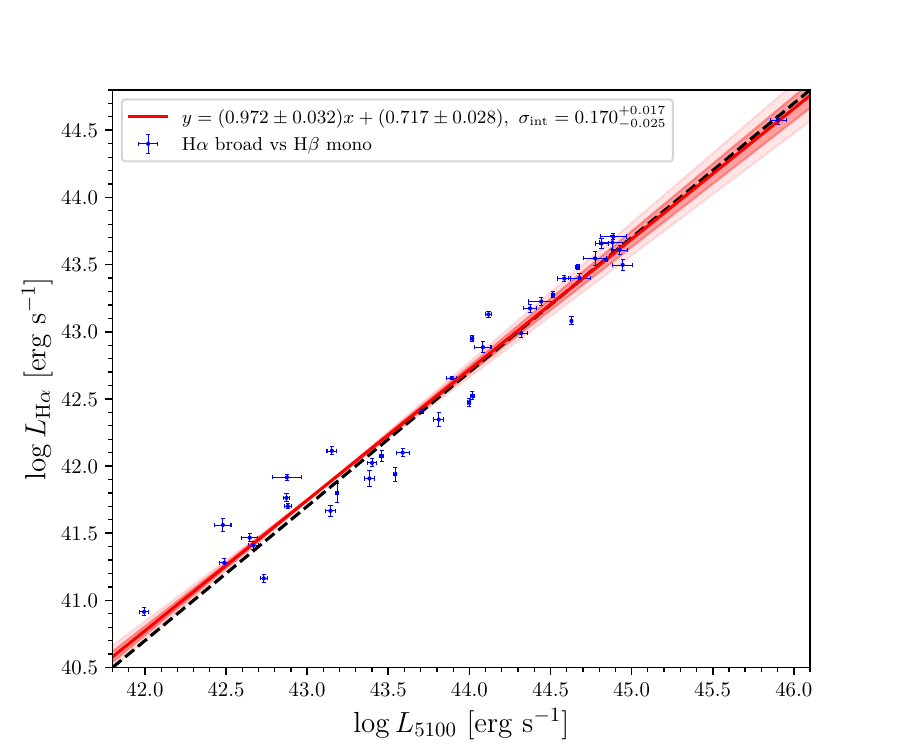}}\\
 \subfloat[Flat XCDM]{%
    \includegraphics[width=0.45\textwidth,height=0.35\textwidth]{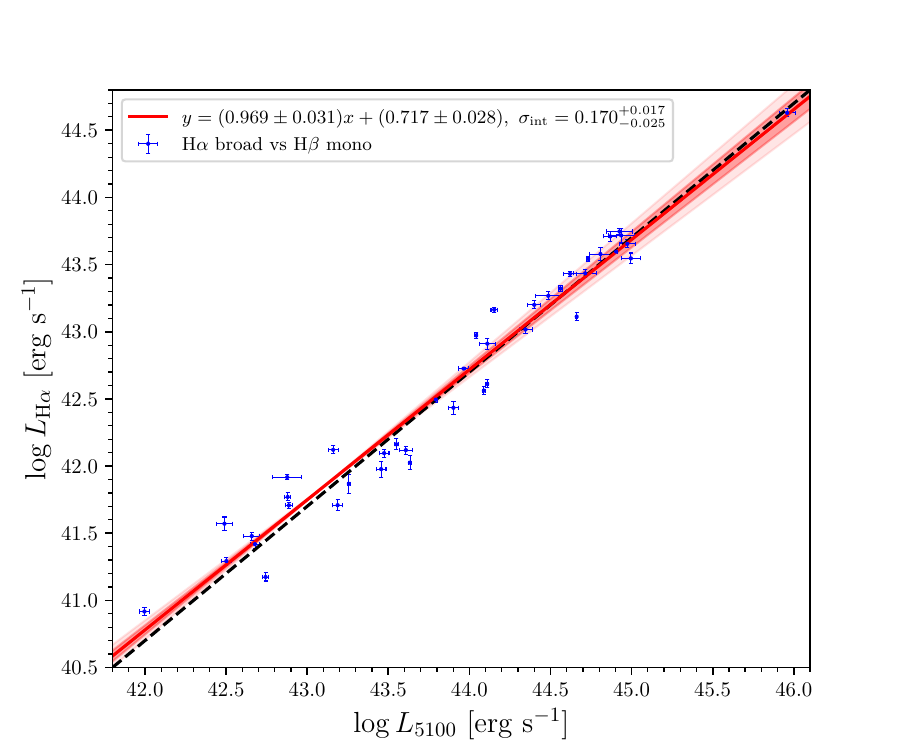}}
 \subfloat[Nonflat XCDM]{%
    \includegraphics[width=0.45\textwidth,height=0.35\textwidth]{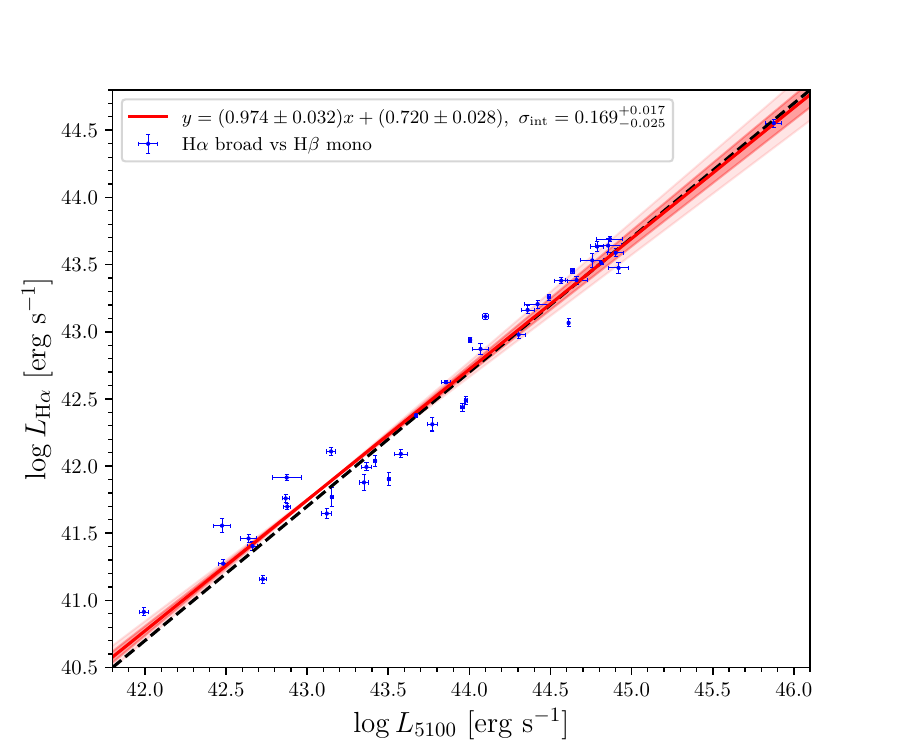}}\\
 \subfloat[Flat \pcdm]{%
    \includegraphics[width=0.45\textwidth,height=0.35\textwidth]{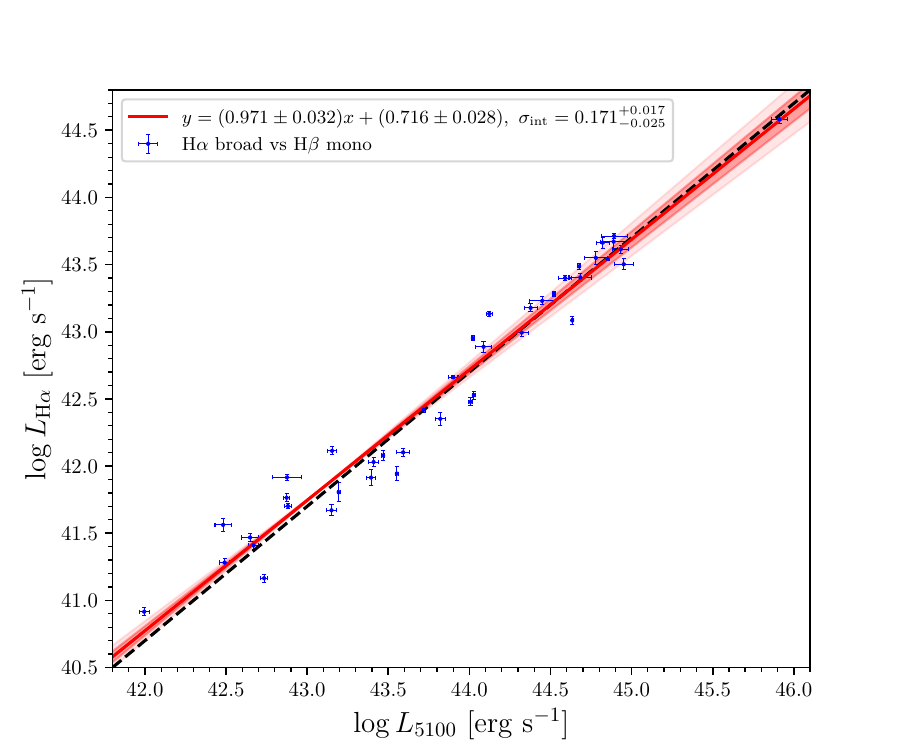}}
 \subfloat[Nonflat \pcdm]{%
    \includegraphics[width=0.45\textwidth,height=0.35\textwidth]{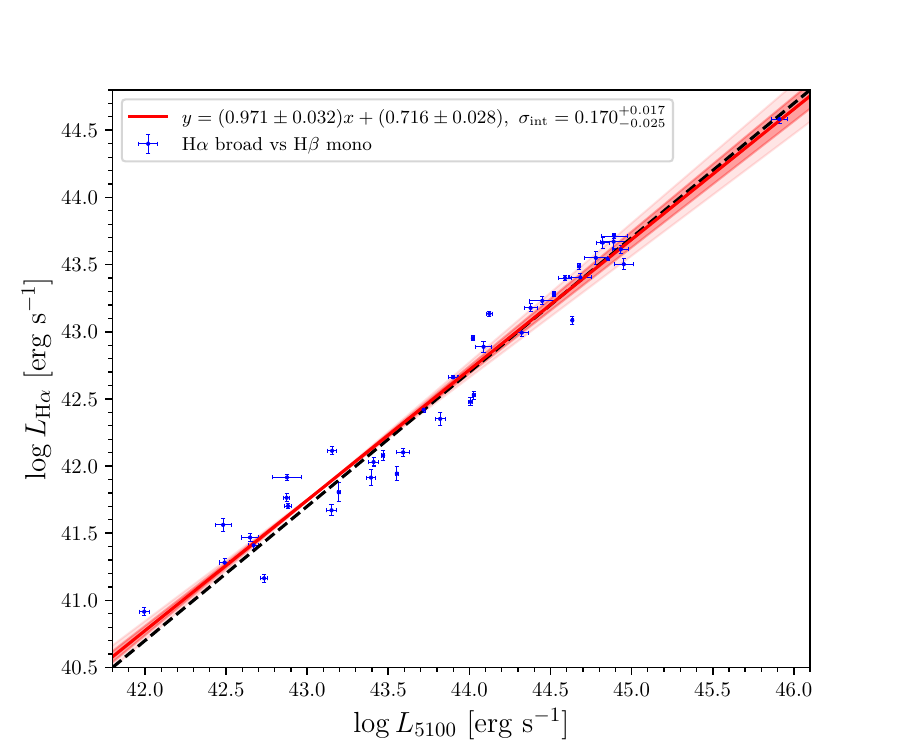}}\\
\caption{Same as Fig.\ \ref{L-L1}, but for H$\alpha$ broad and H$\beta$ mono.}
\label{L-L2}
\end{figure*}

\begin{figure*}[htbp]
\centering
 \subfloat[Flat \lcdm]{%
    \includegraphics[width=0.45\textwidth,height=0.35\textwidth]{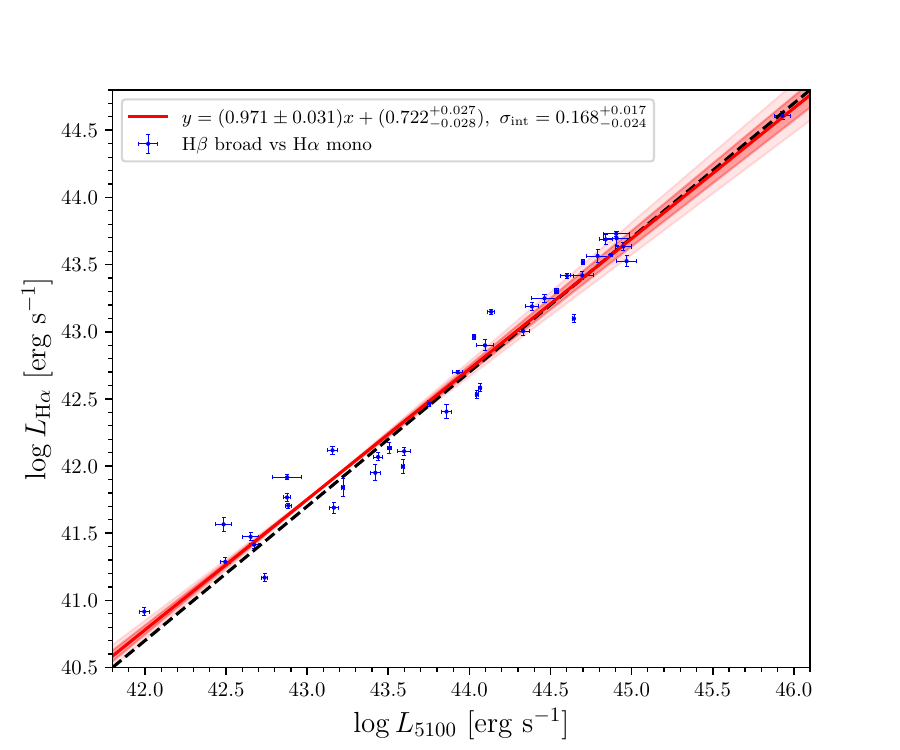}}
 \subfloat[Nonflat \lcdm]{%
    \includegraphics[width=0.45\textwidth,height=0.35\textwidth]{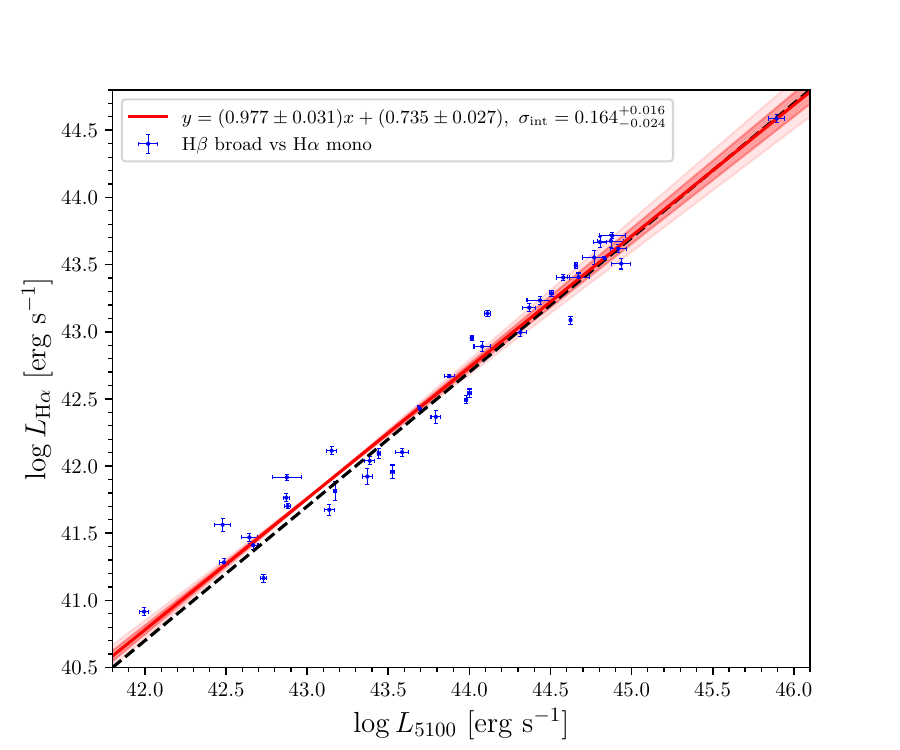}}\\
 \subfloat[Flat XCDM]{%
    \includegraphics[width=0.45\textwidth,height=0.35\textwidth]{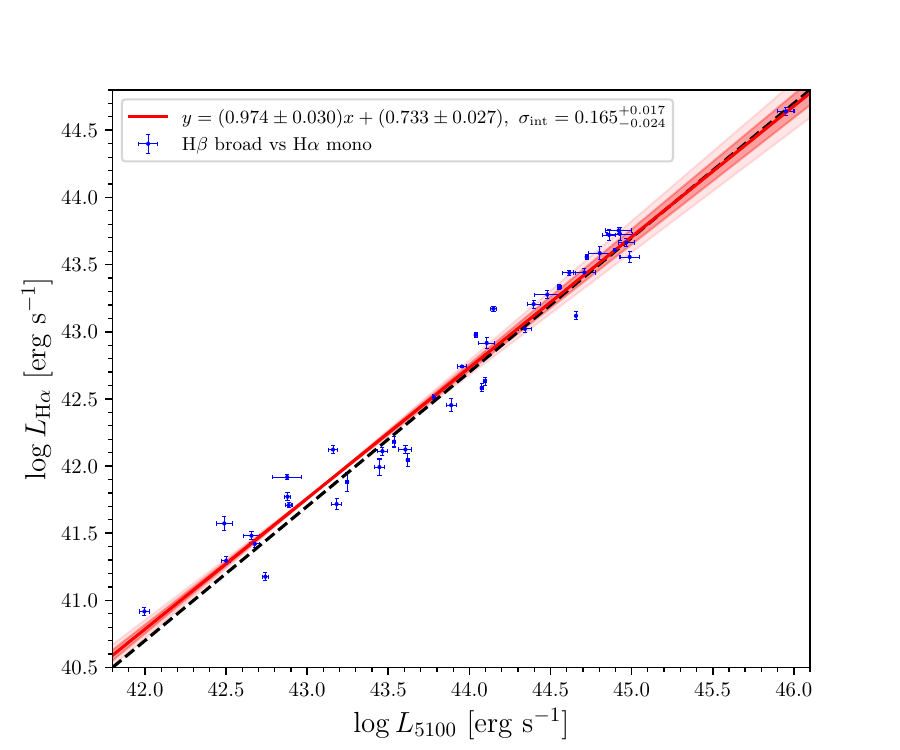}}
 \subfloat[Nonflat XCDM]{%
    \includegraphics[width=0.45\textwidth,height=0.35\textwidth]{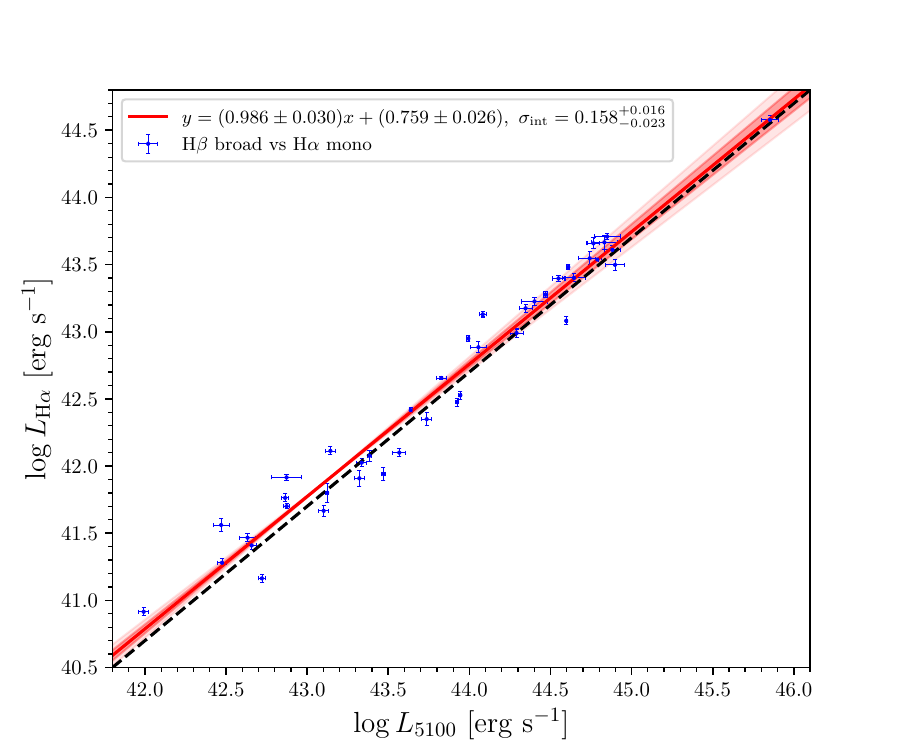}}\\
 \subfloat[Flat \pcdm]{%
    \includegraphics[width=0.45\textwidth,height=0.35\textwidth]{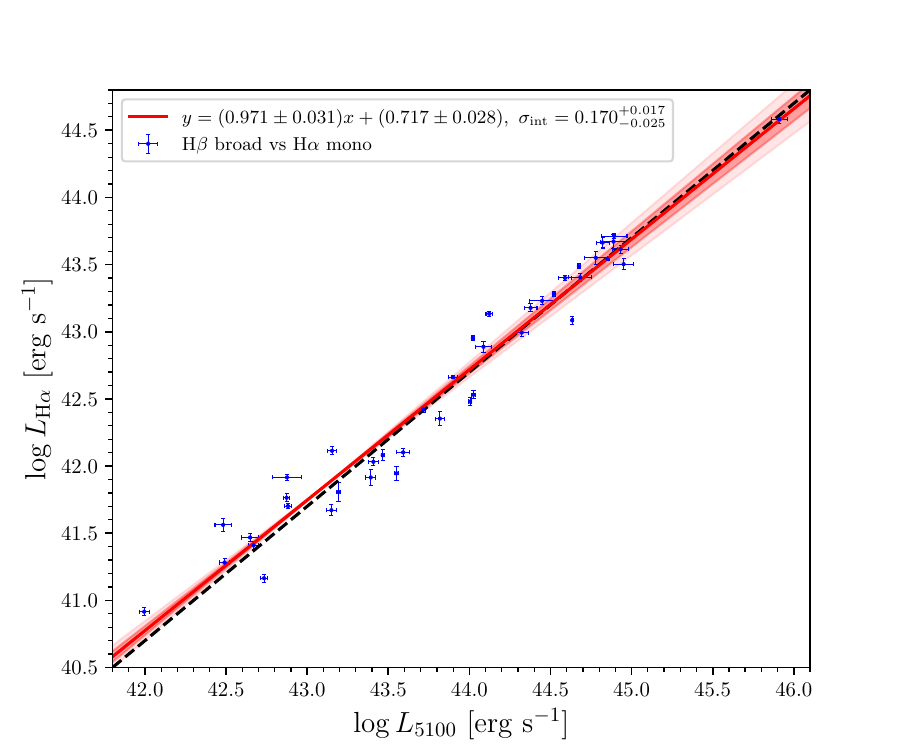}}
 \subfloat[Nonflat \pcdm]{%
    \includegraphics[width=0.45\textwidth,height=0.35\textwidth]{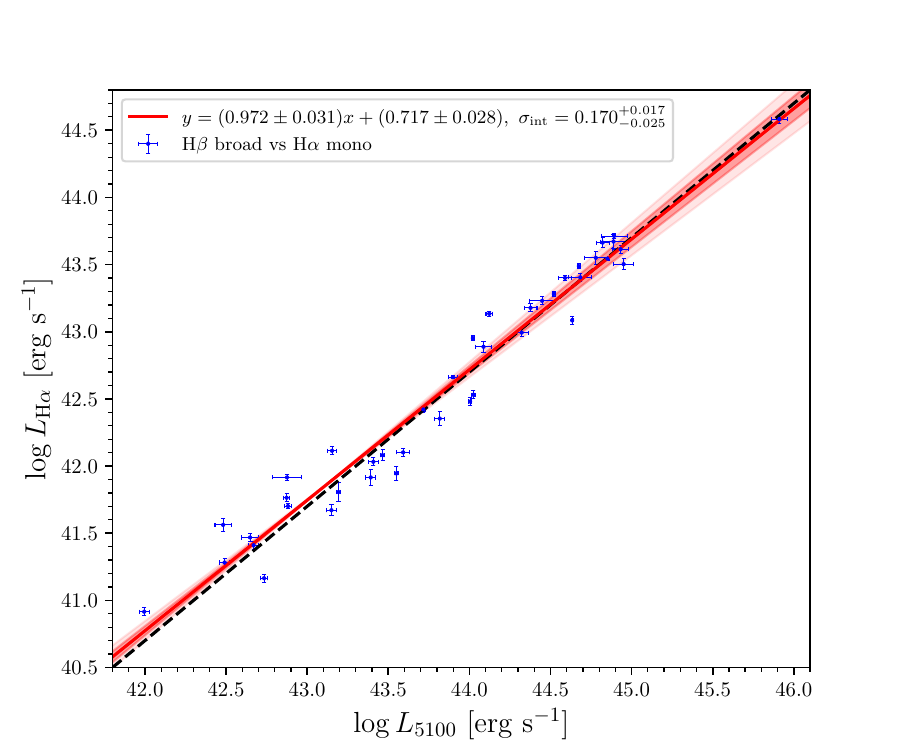}}\\
\caption{Same as Fig.\ \ref{L-L1}, but for H$\beta$ broad and H$\alpha$ mono.}
\label{L-L3}
\end{figure*}

\begin{figure*}[htbp]
\centering
 \subfloat[Flat \lcdm]{%
    \includegraphics[width=0.45\textwidth,height=0.35\textwidth]{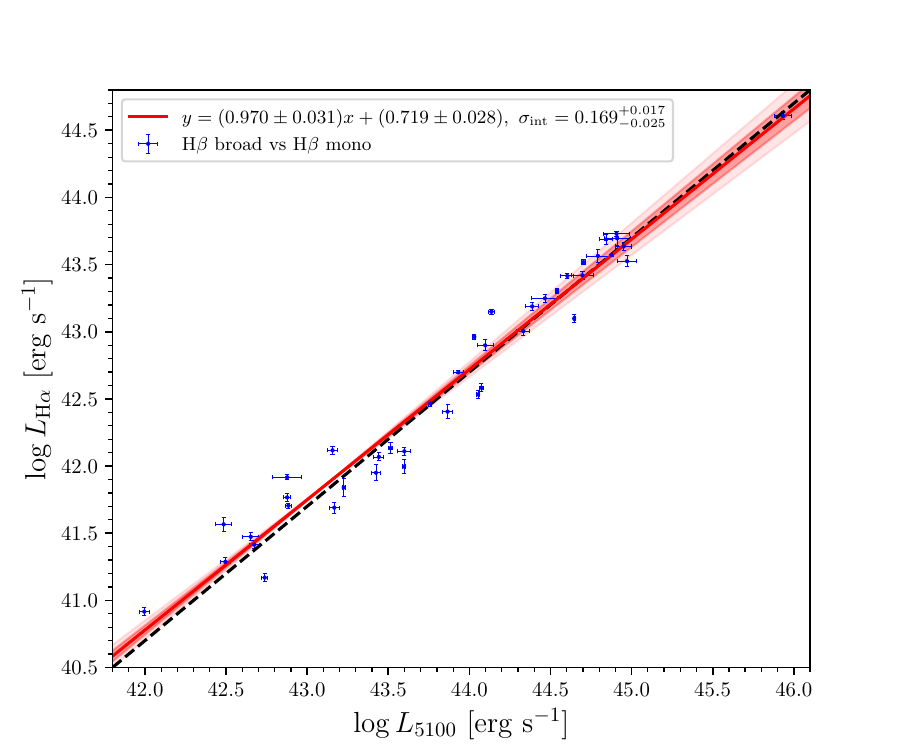}}
 \subfloat[Nonflat \lcdm]{%
    \includegraphics[width=0.45\textwidth,height=0.35\textwidth]{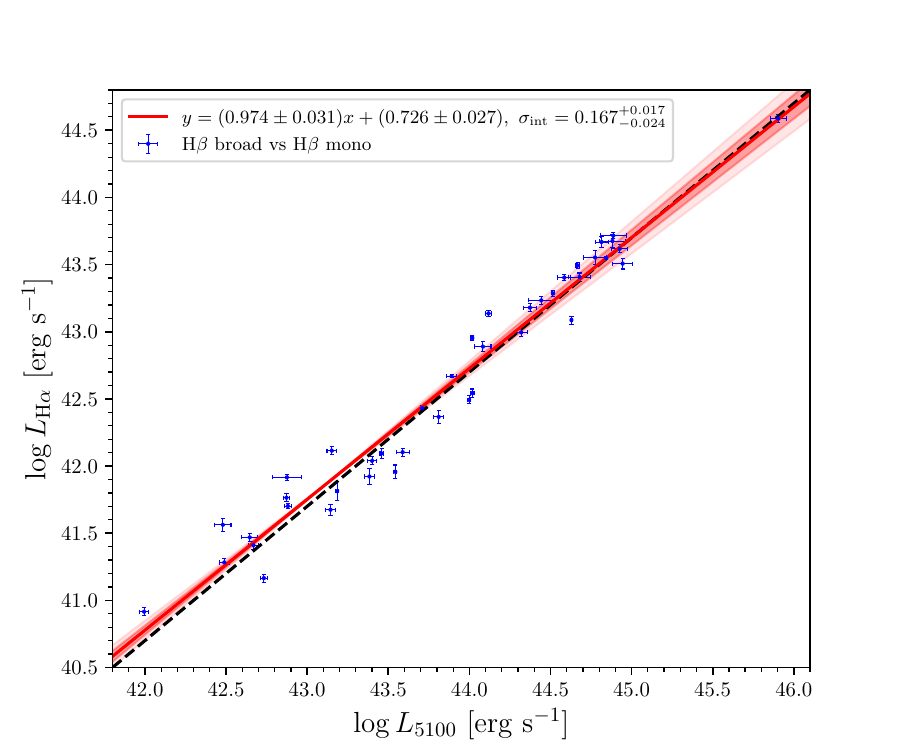}}\\
 \subfloat[Flat XCDM]{%
    \includegraphics[width=0.45\textwidth,height=0.35\textwidth]{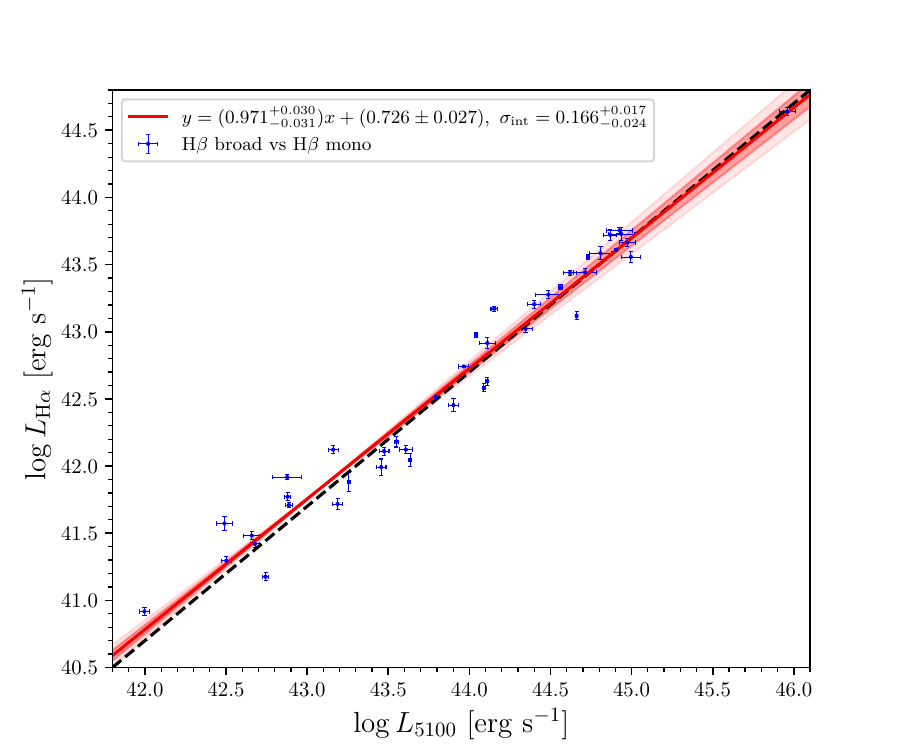}}
 \subfloat[Nonflat XCDM]{%
    \includegraphics[width=0.45\textwidth,height=0.35\textwidth]{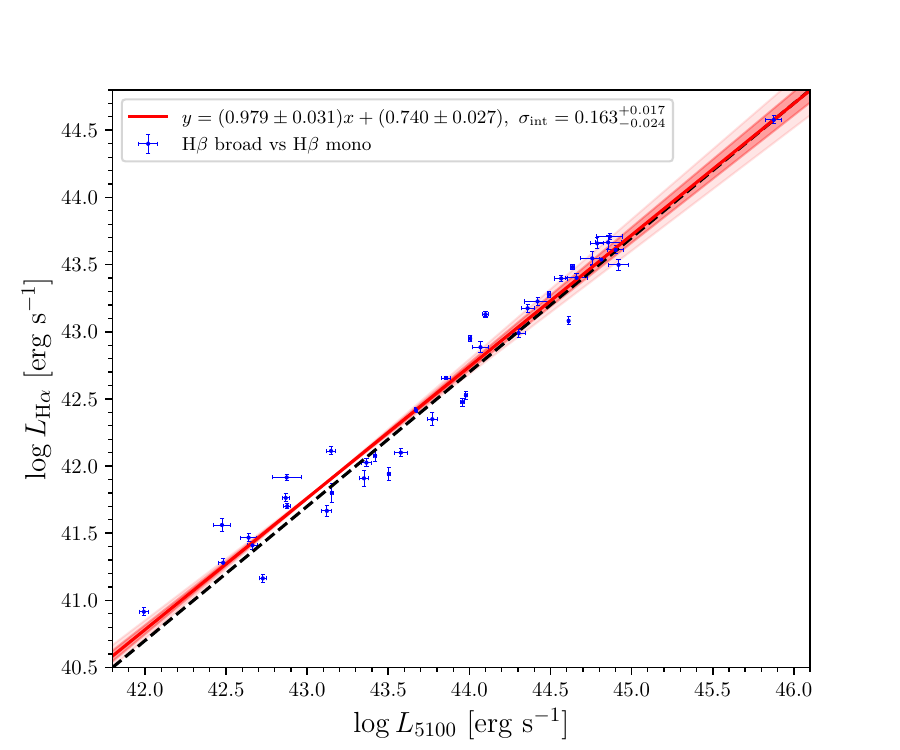}}\\
 \subfloat[Flat \pcdm]{%
    \includegraphics[width=0.45\textwidth,height=0.35\textwidth]{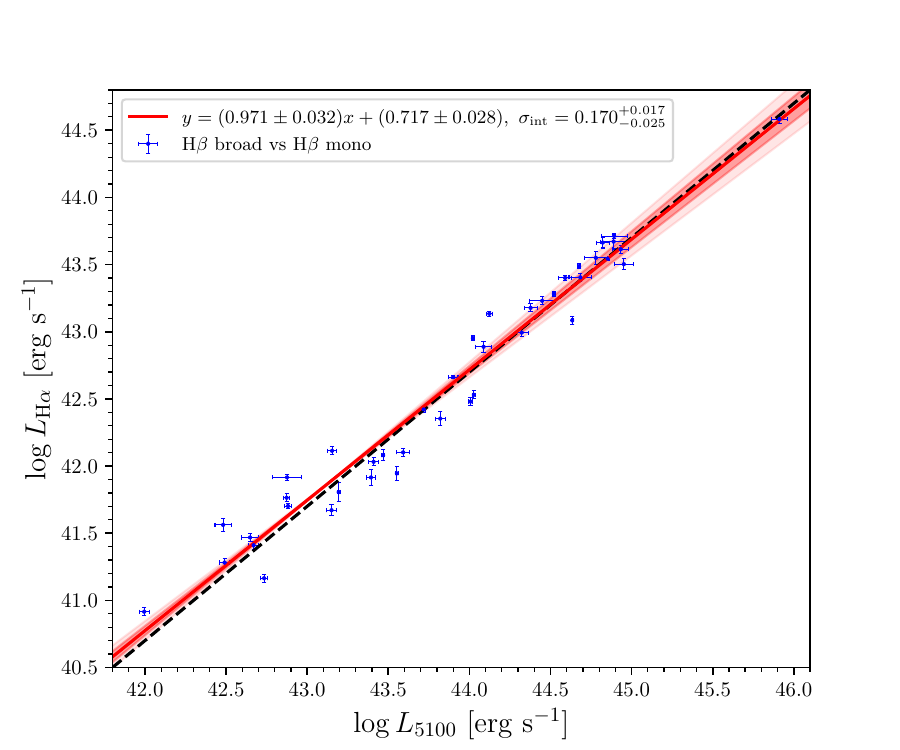}}
 \subfloat[Nonflat \pcdm]{%
    \includegraphics[width=0.45\textwidth,height=0.35\textwidth]{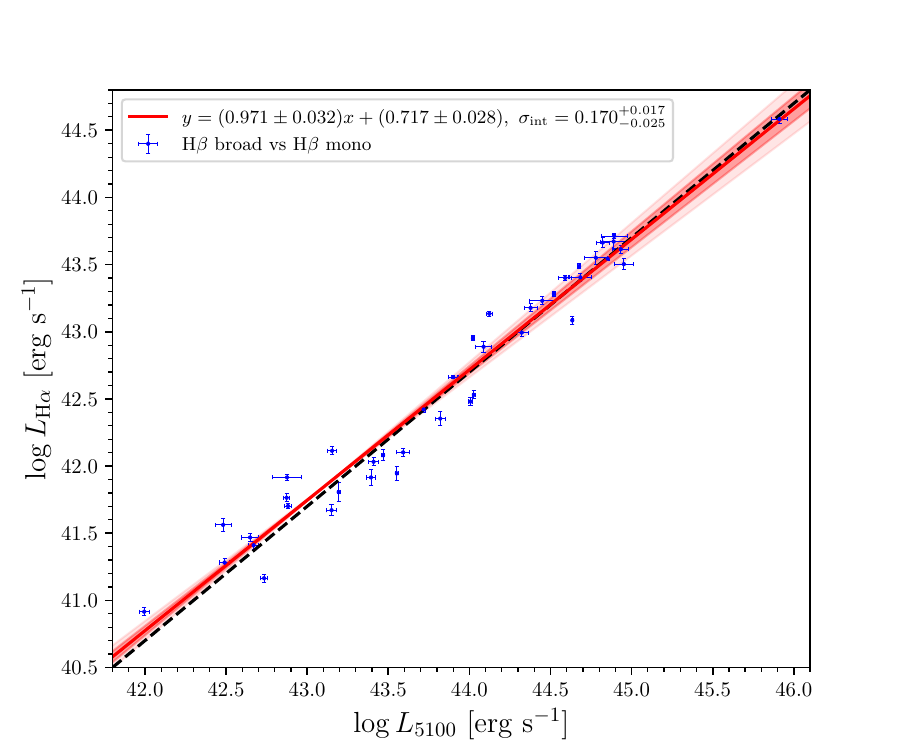}}\\
\caption{Same as Fig.\ \ref{L-L1}, but for H$\beta$ broad and H$\beta$ mono.}
\label{L-L4}
\end{figure*}

The figures presented in this appendix (Figs.~\ref{fig3}--\ref{L-L4}) offer further details and supporting data referenced in the main text.     

Figure \ref{fig3} presents the one-dimensional likelihood distributions and the 1$\sigma$, 2$\sigma$, and 3$\sigma$ two-dimensional confidence contours for the H$\alpha$ mono sample, while the corresponding plots for the cosmological parameters are shown in Fig.~\ref{fig4}. Similarly, Figs.~\ref{fig5}, \ref{fig6}, \ref{fig7}, and \ref{fig8} display the likelihood distributions for the H$\beta$ broad and H$\beta$ mono samples, along with the respective cosmological parameter plots. Figure \ref{R-L2} illustrates the $R-L$ relations for the six different models adopted in this work for the H$\alpha$ mono sample, whereas Figs.~\ref{R-L3} and \ref{R-L4} present the corresponding relations for the H$\beta$ broad and H$\beta$ mono samples, respectively. Figures \ref{L-L2}--\ref{L-L4} compare the broad H$\alpha$ and monochromatic luminosities at 5100{\AA} for different sample pairs: H$\alpha$ broad and H$\beta$ mono, H$\beta$ broad and H$\alpha$ mono, and H$\beta$ broad and H$\beta$ mono, respectively.

These figures collectively provide a comprehensive visualization of the likelihood distributions, cosmological parameter constraints, $R-L$ relations, and luminosity comparisons, supporting the key analyses and findings discussed in the main text.

% \nocite{*}

\newpage
\bibliographystyle{apsrev4-2-author-truncate}
\bibliography{apssamp}% Produces the bibliography via BibTeX.

\end{document}